\documentclass[JHEP,12pt]{article}
\usepackage{graphics,color,soul}
\usepackage{graphicx}
\usepackage{amssymb}
\usepackage{jheppub}
\renewcommand{\afterTocSpace}{\par\nobreak\smallskip\nobreak}
\renewcommand{\afterTocRuleSpace}{\clearpage}
\usepackage{lmodern,mathtools}
\usepackage{caption}
\usepackage{subcaption}
\usepackage{placeins}
\usepackage{booktabs}
\usepackage[normalem]{ulem}
\usepackage[english]{babel}
\usepackage{amsmath,amssymb,amsbsy,amstext, amsthm, simplewick}
\usepackage{hyperref}
\usepackage{tikz}
\usetikzlibrary{decorations.markings}
\usepackage{xcolor}
\usepackage{braket}
\usepackage{caption}

\usepackage{amsfonts}
\usepackage{amssymb}
\usepackage{upgreek}
\usepackage{simplewick}
 \usepackage{exscale,relsize}
\usepackage{mathtools}
\usepackage{comment}

\usepackage[margin=0cm,labelfont={sf,bf,footnotesize},textfont={sf,footnotesize}]{caption}

\def \t {\widetilde}

\renewcommand{\d}{d}

\def\be#1\ee{\begin{align}#1\end{align}}

\definecolor{darkgreen}{RGB}{0,110,0}

\begin{document}

\title{Matrix   scattering and the black hole interior}
\author[a]{Matthew Dodelson,}
\author[b]{Henry W. Lin,}
\author[a]{Haitian Xie,}
\author[c]{and Jessica Yeh}
\affiliation[a]{Center for the Fundamental Laws of Nature, Harvard University, Cambridge, MA, USA}
\affiliation[b]{Joseph Henry Laboratories and Leinweber Forum for Theoretical Physics, Princeton University, Princeton, NJ 08544, USA}
\affiliation[c]{Stanford Leinweber Institute for Theoretical Physics, Stanford University, Stanford, CA 94305, USA}
\abstract{Complex-time singularities of thermal correlators carry imprints of black hole interiors. We investigate their microscopic origin in matrix quantum mechanics at high temperature, where the dynamics becomes semi-classical. We introduce a complex-time scattering kernel that encodes these singularities and makes their finite-$N$ resolution accessible to numerics. Studying both integrable and chaotic models, we find evidence for a smoothing scale $\Delta t\sim N^{-1/2}$, suggesting that the resolution is due to quantum fluctuations of D-branes. This scattering picture provides a starting point for probing the black hole interior beyond a classical geometric description. }
\maketitle
\setcounter{topnumber}{1}
\setcounter{totalnumber}{1}
\section{Introduction}\label{sec:introduction}
 In classical Einstein gravity, nothing can escape a black hole. Nevertheless, the interior leaves subtle imprints on holographic boundary observables. Suitably analytically continued thermal two-point functions become singular at complex time separations determined by null geodesics that bounce off the black hole singularity. These bouncing geodesics probe the geometry behind the horizon \cite{Fidkowski2004, FestucciaLiu2006, Kraus:2002iv, AfkhamiJeddi2025Imprint, Chakravarty:2025ncy, Ceplak2026Fooling, Jia2025ThermalAsymptotics, Jia:2026ryl, Hartnoll:2026vhu, Amado2008ReflectingCavities, Horowitz:2023ury, Grozdanov:2026cut, JiaKulaxizi2026PhaseShift, AliAhmad:2026wem, Araya:2026shz, Grozdanov:2026ktq, ArnaudoWithers2026Bouncing, ArnaudoWithers2026Fourier, Ceplak:2024bja}.

\begin{figure}[t]
\centering
\includegraphics[width=0.94\textwidth]{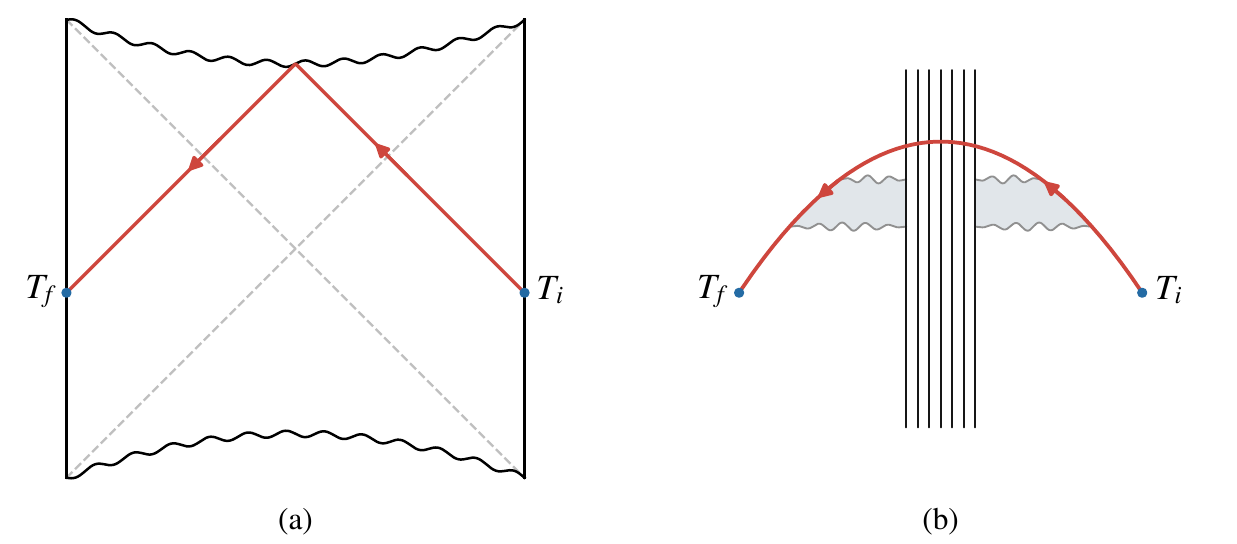}
\begingroup
\phantomsubcaption\label{fig:bouncing_geodesic}
\phantomsubcaption\label{fig:probe_d0_scattering}
\endgroup
\caption{(a) A bouncing geodesic, viewed as a complexified
scattering trajectory from $T_i$ to $T_f$. The wavy curves
represent the black hole singularity, and the dashed lines
indicate the horizons. In our computation, we study an adjoint correlator, so the trajectory would be a folded string. (b) Complex scattering of a probe
D0-brane off a stack of branes. The red curve depicts
the complexified probe trajectory, and the black lines represent the stack.
The interaction is mediated by open strings, shown in gray.}
\label{fig:black_hole_bouncing_scattering}
\end{figure}

One might expect that string theory smooths out the black hole geometry, thereby resolving the bouncing geodesic singularities. The surprise is that this does not happen. In the SYK model \cite{Sachdev:1992fk, Kitaev2015Talks, Polchinski:2016xgd, Maldacena:2016hyu}, higher-derivative bulk corrections shift the singularities of the correlator into the complex plane, rather than resolving them completely \cite{Dodelson2025Stringy, Dodelson:2024atp, DodelsonEtAl2026Temperature, Buric:2026qsp}. The same is true for one-sided correlators in higher dimensions \cite{Hubeny2007BulkCone, Dodelson2024BulkCone,Batra2026NonPrompt}, where bulk tidal effects move the singularities off the real axis \cite{DodelsonOoguri2021}. This suggests that the bulk dual is characterized by some sharply defined geometric information, even beyond the Einstein gravity limit. We can think of the resulting object as a \emph{stringy black hole} \cite{Gesteau2023Ergodic, Ouseph2024Poincare, GesteauLiu2024Horizons, Belin2025Mirage, Herderschee2025Stringy, Dodelson2025Chaos}, whose geometry is determined by the complex-time singularities of the correlator, rather than necessarily by a local spacetime metric. This viewpoint is familiar from inverse scattering theory \cite{Newton1982Scattering}, where a potential is constructed from knowledge of the scattering amplitude.

The SYK example suggests that complex-time singularities may be a generic feature of large-$N$ systems. What microscopic mechanism could generate these singularities in a large-$N$ model, without assuming a weakly curved gravitational dual? A useful clue comes from thinking about the bouncing geodesic contribution in terms of a reflection coefficient at the black hole singularity \cite{AfkhamiJeddi2025Imprint}, as in Figure \ref{fig:bouncing_geodesic}. This reformulates the computation as a scattering problem and leads us to search for an analogous description directly in the boundary dynamics.

The crucial property of large-$N$ quantum systems is that they admit classical descriptions in terms of suitable collective variables~\cite{Yaffe1982, JevickiSakita1980, GopakumarGross1995}. This suggests that we should formulate a scattering matrix for classical dynamical systems, which plays the role of the reflection coefficient off a stringy singularity. For closed systems the phase space at fixed energy is generally compact, and there is no obvious notion of scattering. The catch is that the relevant scattering need not occur in real time: the bouncing geodesic in Figure \ref{fig:bouncing_geodesic} crosses horizons, so its net elapsed time is complex. In the same way, a bounded classical trajectory can reach infinity at a finite complex time. Its behavior near these singularities supplies incoming and outgoing scattering data. With this setup in mind, we introduce a complex-time scattering kernel and investigate how it encodes singularities of correlation functions.

An important advantage of the scattering kernel is that it remains well-defined at finite $N$ as a distribution of complex-time scattering data, which provides a window into the resolution of singularities at finite $N$. By contrast, a naive ensemble average of complexified trajectories need not define an analytic continuation of the correlator beyond its initial strip of analyticity. In holography, finite-$N$ effects correspond to quantum gravity corrections. The scattering description therefore provides a microscopic framework for studying the fate of the black hole singularity in quantum gravity.

In this paper, we develop the scattering kernel picture in detail for the simple case of matrix quantum mechanics at high temperature.  The advantage of this limit is that the relevant classical variables are manifest: quantum corrections are controlled by the inverse temperature~\cite{Kawahara:2007ib, AsplundBerensteinDzienkowski2013, GurAriHanadaShenker2016}. Since the high-temperature limit is very stringy (analogous to the weak 't Hooft coupling limit of $\mathcal{N}=4$ supersymmetric Yang-Mills), it yields a controlled setting for studying complex-time singularities away from the holographic regime. The scattering  trajectories admit a D-brane interpretation: as shown in Figure \ref{fig:probe_d0_scattering}, an eigenvalue behaves like a probe brane interacting with a thermal stack through off-diagonal open strings. At finite $N$, we find that the complex-time singularities are resolved, consistent with general expectations~\cite{Komargodski2026Bouncing, Batra:2026gbq}. The scale $\Delta t\sim N^{-1/2}$ on which the singularities are smoothed out corresponds to fluctuations of the position of the D-brane probe in quantum gravity.

The paper is organized as follows. In Section~\ref{scattsec}, we study complex scattering in general,  and illustrate it in integrable and chaotic systems, relating the scattering kernel to return maps and equilibration. In Section~\ref{sec:one-matrix}, we study the quartic one-matrix model, comparing planar moment calculations with numerical scattering data.  Section~\ref{sec:multi-matrix} extends the analysis to chaotic multi-matrix models and explores the additional structure associated with branching and monodromy, including simplifications on high sheets of the complex-time Riemann surface. Section~\ref{sec:beyondhighT} studies the one- and two-matrix quantum mechanics models beyond the high-temperature limit. Here our methods are either the non-perturbative matrix bootstrap or perturbation theory of the Wigner distribution.

\section{Scattering in complex time}\label{scattsec}

 In this section we introduce the notion of a complex scattering kernel for a classical dynamical system. We then explore some of its basic properties in both chaotic and integrable models.
\subsection{Analytic structure of classical trajectories}
We consider a classical dynamical system in $d$ dimensions, which is a collection of $d$ first-order ordinary differential equations together with a stationary measure on phase space~\cite{Cvitanovic2020Chaos, Gaspard1998}. If the system is Hamiltonian, then we work in either the microcanonical or canonical ensemble,
\begin{align}
\langle \mathcal{O}(p,q)\rangle_{E}&=\frac{1}{\Omega(E)}\int \prod_i dp_i\, dq^i\, \delta(H(p,q)-E)\mathcal{O}(p,q),\\
\langle \mathcal{O}(p,q)\rangle_{\beta}&=\frac{1}{Z(\beta)}\int \prod_i dp_i\, dq^i\, e^{-\beta H(p,q)}\mathcal{O}(p,q),
\end{align}
where the density of states and partition function are given by 
\begin{align}
\Omega(E)&=\int \prod_i dp_i\, dq^i\, \delta(H(p,q)-E),\\
Z(\beta)&=\int \prod_i dp_i\, dq^i\, e^{-\beta H(p,q)}.
\end{align}
For dissipative systems one can instead take the Sinai-Ruelle-Bowen measure defined by time averaging~\cite{BowenRuelle1975}, 
\begin{align}
\langle \mathcal{O}(x)\rangle_{\text{SRB}}=\lim_{T\to\infty}\frac{1}{T}\int_{0}^{T} dt\, \mathcal{O}(x(t)).
\end{align}
\begin{figure}[t]
\centering
\begin{subfigure}{0.48\textwidth}
\centering
\includegraphics[width=\textwidth]{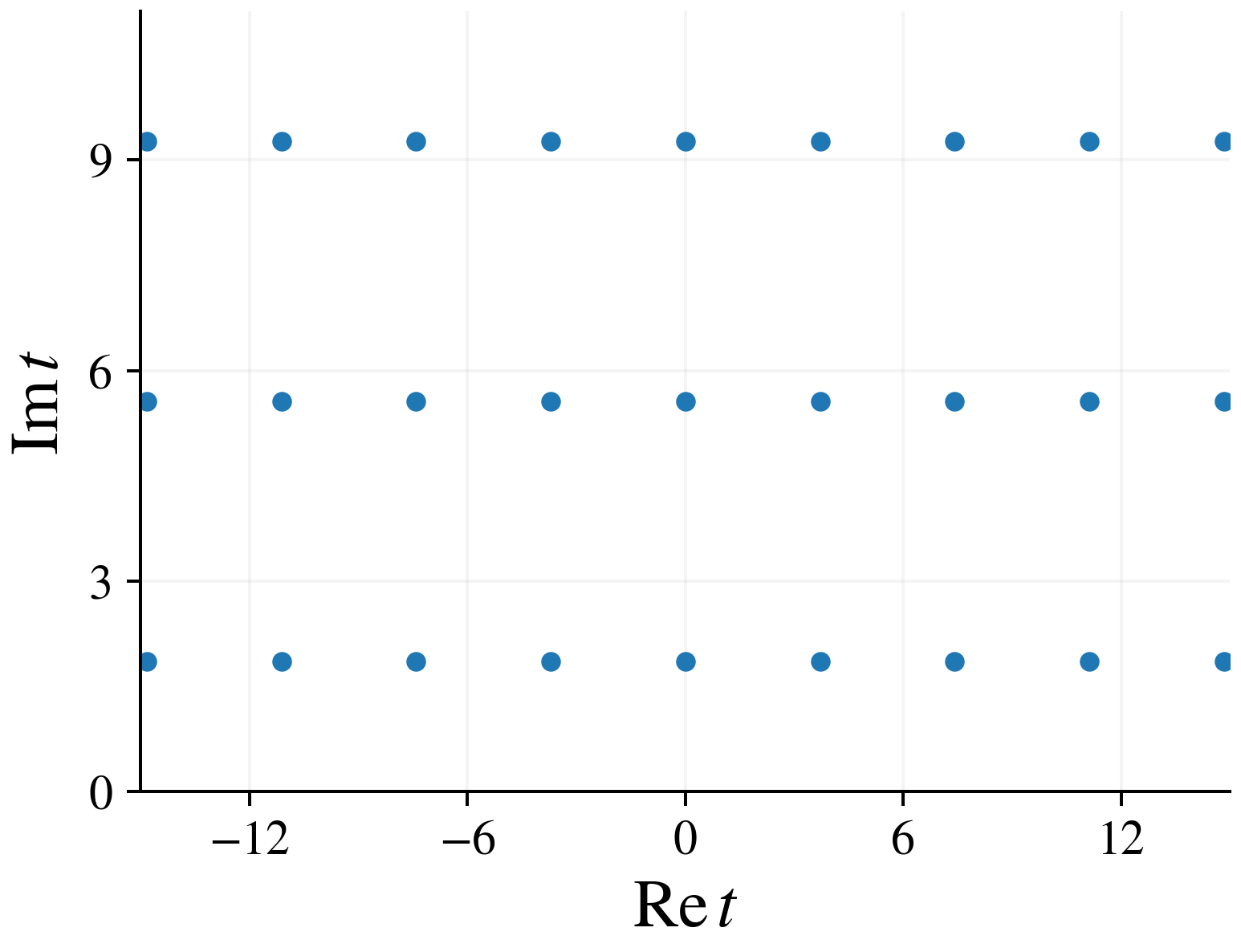}
\caption{\label{fig1a}}
\end{subfigure}
\hfill
\begin{subfigure}{0.48\textwidth}
\centering
\includegraphics[width=\textwidth]{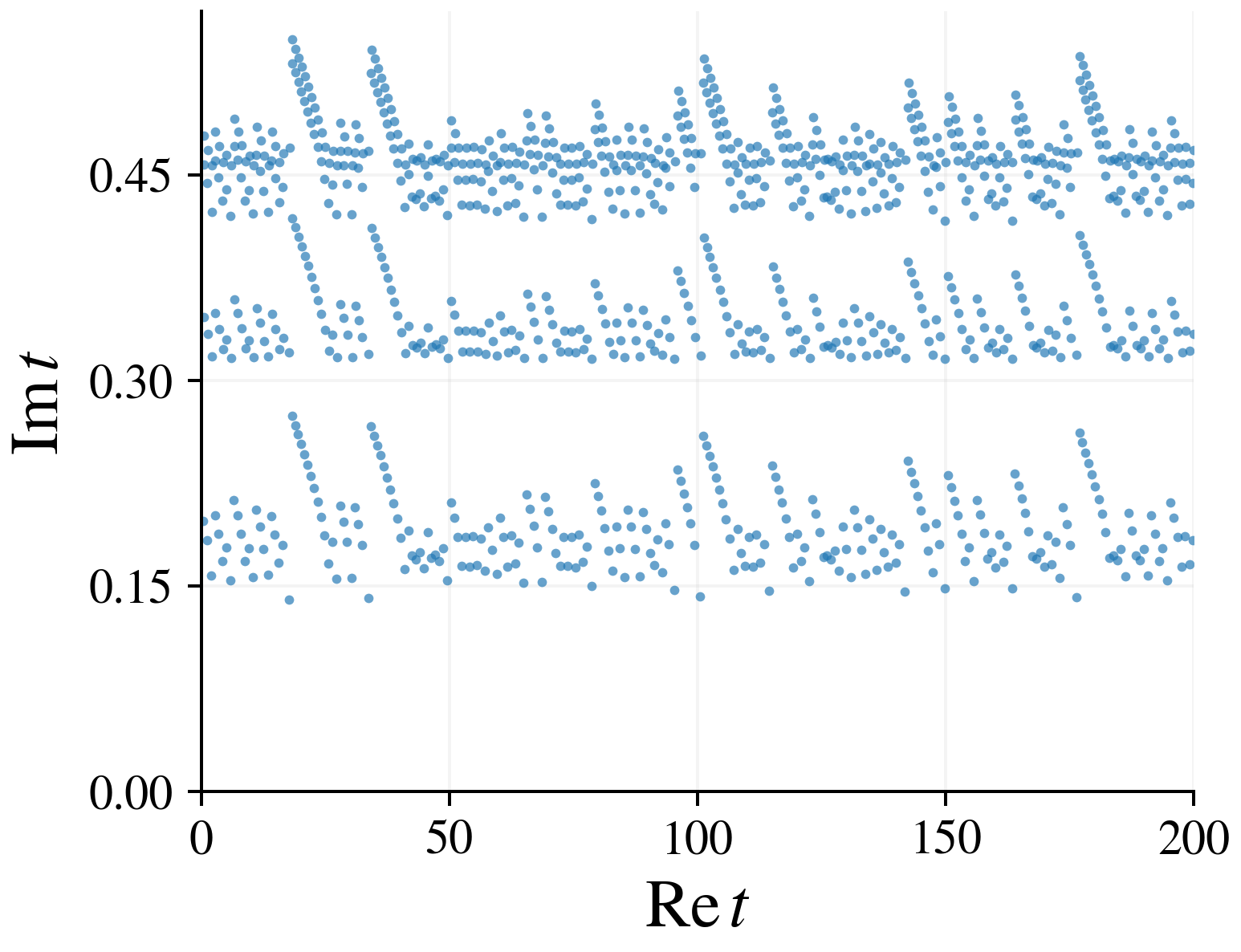}
\caption{\label{fig1b}}
\end{subfigure}
\caption{Singularities of individual complexified classical trajectories. (a) Exact pole lattice for the quartic oscillator, with $q(0)=1$ and $p(0)=0$. (b) First-accessible singularities along a generic Lorenz trajectory. Each singularity is a branch point, and we choose the corresponding branch cut to extend vertically towards $i\infty$. We only include singularities on the first sheet.}
\label{fig:classical_singularity_examples}
\end{figure}

In a compact ergodic system, most trajectories explore all of phase space under real-time evolution. When time is complexified, the picture becomes more interesting. For flows on a compact phase space generated by an analytic vector field, the complexified trajectories are analytic in a strip in complex time~\cite{Hille1976}. Beyond this strip, the solutions may develop singularities, so that the trajectories naturally live on a Riemann surface~\cite{FrischMorf1981}. The local expansion of a trajectory near a singularity should contain $d$ free parameters, since changing the initial conditions modifies the expansion coefficients. These free parameters are known as \emph{resonances}~\cite{Hone2009Painleve}. For a movable singularity, one resonance is the singular location itself.

For example, let us consider a quartic oscillator
\begin{align}
\dot{q}=p,\hspace{10 mm}\dot{p}=-q^3.
\end{align}
The local behavior of a trajectory near a singularity at $t=t_0$ is
\begin{align}
q(t)&\sim \pm\frac{i\sqrt{2}}{t-t_0}\pm \frac{iE}{5\sqrt{2}}(t-t_0)^3+\cdots,\\
p(t)&\sim \mp \frac{i\sqrt{2}}{(t-t_0)^2}\pm \frac{3iE}{5\sqrt{2}}(t-t_0)^2+\cdots.
\end{align}
Here $E$ is the conserved energy along the trajectory. The set of singularities is therefore parameterized by the resonances $t_0$ and $E$. Up to an overall time translation, the full solution to the equation of motion can be written in terms of Jacobi elliptic functions,
\begin{align}
q(t)=(4E)^{1/4}\text{cn}\left[(4E)^{1/4}t,\frac{1}{2}\right].
\end{align}
This is a meromorphic function whose poles form a lattice in the complex plane,
\begin{align}
t_{nm}=\frac{T(E)}{2}\left[n+\left(m+\frac{1}{2}\right)i\right],\hspace{10 mm}T(E)=\frac{4K\left(\frac{1}{2}\right)}{(4E)^{1/4}}.
\end{align} 
Here $T(E)$ is the real period of the solution, and $K(m)$ is the complete elliptic integral of the first kind with parameter $m$. Meromorphy of generic trajectories is a hallmark of integrability, and is known as the Painlev\'e property~\cite{Hone2009Painleve}.

At the opposite end of the ergodic hierarchy, we have globally chaotic systems such as the Lorenz attractor~\cite{Lorenz1963}, 
\begin{align}
\dot{x}=10(y-x),\hspace{10 mm}
\dot{y}=x(28-z)-y,\hspace{10 mm}
\dot{z}=xy-\frac{8z}{3}.
\end{align}
Balancing powers again gives a meromorphic answer to leading order, 
\begin{align}
x(t)\sim \pm \frac{2i}{t-t_0},\hspace{10 mm}y(t)\sim \mp\frac{ i}{5(t-t_0)^2},\hspace{10 mm}z(t)\sim-\frac{1}{5(t-t_0)^2}.
\end{align}
However, in this case logarithms appear at higher orders in the expansion, so the solutions are defined on a complicated multi-sheeted Riemann surface~\cite{TaborWeiss1981, LevineTabor1988}. The singularities of a generic trajectory are depicted in Figure \ref{fig1b}.  We observe an underlying band structure in imaginary time. The resulting band spacing can be thought of as an effective inverse temperature, and will play an important role in this work. Another example of a globally chaotic system with this band structure is the geodesic flow on a hyperbolic manifold, in which case the trajectories are explicitly periodic in imaginary time.

In practice, the systems we consider will be neither fully integrable nor globally chaotic. This means that we expect to find trajectories that look qualitatively somewhere in between Figure \ref{fig1a} and Figure \ref{fig1b}.
\subsection{The scattering kernel}

In chaotic systems, individual trajectories are exponentially sensitive to initial conditions, so cannot be predicted accurately. Instead one should consider coarse-grained observables like correlation functions. We now explain how these correlation functions inherit their singularities from individual trajectories.

Let $\mathcal{O}(x)$ be a classical observable which diverges near the singularities $t_n$ of trajectories. Its leading singular part for a trajectory $x(t)$ takes the form
\begin{align}
\mathcal{O}(x(t))\sim \sum_n \frac{A_n}{(t-t_n)^\Delta}\label{singular},
\end{align}
where $t_n$ and $A_n$ depend on the initial condition and $\Delta$ is an effective scaling dimension. Inserting (\ref{singular}) into the two-point function, we find 
\begin{align}\label{cint0}
C(t)&\equiv \left\langle \overline{\mathcal{O}}\bigl[x(t/2)\bigr]\mathcal{O}\bigl[x(-t/2)\bigr]\right\rangle\notag\\
&\sim \sum_{nm} \left\langle\frac{\overline{A}_nA_m}{\left(\frac{t}{2}-\overline{t}_n\right)^\Delta\left(-\frac{t}{2}-t_m\right)^{\Delta}}\right\rangle.
\end{align}
Here $\overline{\mathcal{O}}$ denotes complex conjugation on real phase space followed by analytic continuation, so that $\overline{\mathcal{O}}(x(t))=\overline{\mathcal{O}(x(\bar t))}$. The problem with this expansion is that it does not by itself provide an analytic continuation into the complex $t$ plane. Indeed, the singularities $\overline{t}_n$ and $t_m$ can cover a two-dimensional region of the complex plane as the initial conditions are varied, so it is not obvious that the sum over $m$ and $n$ is analytic.

In order to express $C(t)$ in a form amenable to analytic continuation, let us rewrite (\ref{cint0}) as
\begin{align}\label{cint}
C(t)\sim \int d^2T_i\,d^2T_f\, \frac{S_{A}(T_i,T_f)}{\left(\frac{t}{2}-T_f\right)^{\Delta}\left(-\frac{t}{2}-T_i\right)^\Delta},
\end{align}
where we have defined
\begin{align}\label{densitydef}
S_A(T_i,T_f)=\sum_{nm}\langle \overline{A}_nA_m\delta^2(T_f-\overline{t}_n)\delta^2(T_i-t_m)\rangle.
\end{align}
 Because the measure is stationary, $S_A$ is independent of the real part of $T_i+T_f$. The integral over $\operatorname{Re}(T_i+T_f)$ is singular when the two singularities in the denominator pinch the integration contour~\cite{Landau1959}. For $\operatorname{Im}t>0$, this fixes $T_i$ and $T_f$ to be in the lower and upper half plane respectively. Performing the integral over $T_i+T_f$ then gives 
\begin{align}
C(t)&\sim \frac{\pi\Gamma(2\Delta-1)}{\Gamma(\Delta)^2}\int d^2 T\, \frac{\tilde{S}_A(T)}{\bigl[i(t-T)\bigr]^{2\Delta-1}},\hspace{10 mm}\operatorname{Im}t>0,
\label{eq:scattering-pinch}
\end{align}
where the reduced kernel is defined by 
\begin{align}
\tilde{S}_A(T)=\int_{-\operatorname{Im}T}^{\operatorname{Im}T}dy\, S_A\left(-\frac{T}{2}+\frac{iy}{2},\frac{T}{2}+\frac{iy}{2}\right).
\end{align}

  For the choice of coefficients $A_n=1$, the density (\ref{densitydef}) represents a joint distribution for trajectories to be singular at both $T_i$ and $T_f$. More generally, we can think of the singularity at $t_n$ as an asymptotic scattering region, with scattering data given by the resonances $\mathbf{r}_n$. Then it is natural to consider the scattering kernel
\begin{align}\label{skernel}
S(T_i,\mathbf{R}_i,T_f,\mathbf{R}_f)=\sum_{nm}\langle \delta^2(T_f-\overline{t}_n)\delta^{2d-2}(\mathbf{R}_f-\overline{\mathbf{r}}_n)\delta^2(T_i-t_m) \delta^{2d-2}(\mathbf{R}_i-\mathbf{r}_m)\rangle.
\end{align}
 In systems where the trajectories have branch cuts, one must include sheet labels for both $T_i$ and $T_f$, with conjugation also acting on the sheet labels, which we leave implicit.

For closed dynamical systems we do not have a notion of scattering in real time, but the complex scattering kernel (\ref{skernel}) will turn out to be perfectly well-defined in the models we study. It contains all information about singularities of two-point functions. Here we have specified $d$ complex variables for the scattering data $(T,\mathbf{R})$, but the physical scattering will be supported on a real section of this data. Note that for any real variable $x$, we have a conjugation relation $\overline{x}(\overline{t})=x(t)$. It follows that there is always a self-pairing contribution to the kernel from $t_m=t_n$, which can be thought of as the elastic piece. 

For holographic systems, the scattering kernel admits a black hole interpretation \cite{AfkhamiJeddi2025Imprint}. Let us think about propagation on an $\text{AdS}_{D}$ black hole as a potential scattering problem, with free propagation given by the pure AdS background. We represent the bulk-to-bulk propagator as~\cite{Newton1982Scattering} 
\begin{align}
G(x,y)=G_0(x,y)+\int d\mu(x_1)\,d\mu(x_2)\, G_0(x,x_1)\mathcal{T}(x_1,x_2)G_0(x_2,y).
\end{align}
Here $G_0$ is the propagator in pure AdS and $d\mu(x)=\sqrt{-g}\, d^{D}x$. The transition kernel $\mathcal{T}$ contains the interaction of the probe with the black hole. The only singularities of $G_0$ are on the light cone, so nontrivial singularities of $G$ can only come from the second term on the right-hand side. Taking $x$ to the boundary gives the bulk-to-boundary propagator
\begin{align}
G_0(x,x_1)\sim \Bigl(\frac{1}{rr_1\bigl[(t-t_1)^2-(\mathbf{x}-\mathbf{x}_1)^2-r_1^{-2}\bigr]}\Bigr)^{\Delta},\hspace{10 mm}r\to\infty,
\end{align}
and similarly for $G_0(x_2,y)$. Near the light cone, this external leg has a singularity $(t-t_0)^{-\Delta}$, so we find that the singularities of $G(x,y)$ are captured by a formula of the form (\ref{cint}). Therefore the scattering kernel $S$ generalizes the transition operator $\mathcal{T}$ to systems without a holographic description.
\subsection{The integrable case}
\begin{figure}[t]
\centering
\begin{subfigure}[t]{0.139137\textwidth}
\centering
\includegraphics[width=\linewidth]{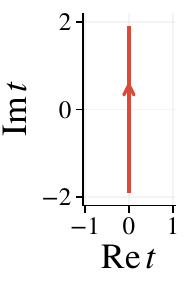}
\caption{}\label{fig:quartic_imaginary_contour}
\end{subfigure}\hfill
\begin{subfigure}[t]{0.266522\textwidth}
\centering
\includegraphics[width=\linewidth]{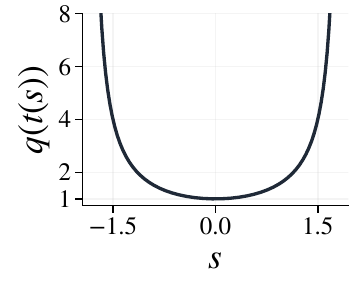}
\caption{}\label{fig:quartic_imaginary_trajectory}
\end{subfigure}\hfill
\begin{subfigure}[t]{0.256923\textwidth}
\centering
\includegraphics[width=\linewidth]{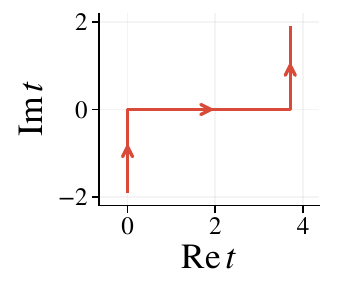}
\caption{}\label{fig:quartic_bounce_contour}
\end{subfigure}\hfill
\begin{subfigure}[t]{0.277418\textwidth}
\centering
\includegraphics[width=\linewidth]{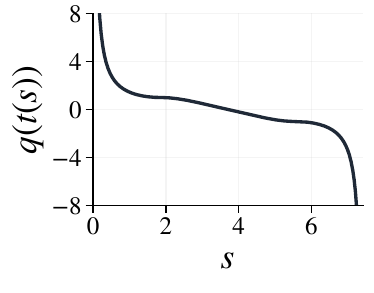}
\caption{}\label{fig:quartic_bounce_trajectory}
\end{subfigure}
\caption{Complex scattering trajectories of the quartic oscillator, in units $E=1/4$. (a,b) The imaginary-time contour from $-iK$ to $iK$ and the corresponding trajectory. (c,d) The contour $-iK\to0\to2K\to2K+iK$ and the corresponding trajectory, including a half-oscillation in real time. }
\label{fig:quartic_scattering_trajectories}
\end{figure}
Let us start by analyzing the scattering kernel in an integrable system, which has $d/2$ conserved charges. It is useful to go to action-angle variables $(I_a,\theta_a)$, where $a=1,\ldots, d/2$. In these coordinates, the actions $I_a$ are conserved, and trajectories live on invariant tori of the form $\boldsymbol{\theta}(t)=\boldsymbol{\theta_0}+\boldsymbol{\omega}(I) t$.  If the level sets $\mathbf{I}(x)=\mathbf{I}_0$ can be parameterized by globally meromorphic functions of $\boldsymbol{\theta}$, then the system is known as algebraically completely integrable, and the Painlevé property holds in the $x$ coordinates~\cite{AdlerVanMoerbeke1980}.

As a first example, let us consider the quartic oscillator in the microcanonical ensemble. The scattering kernel can be computed by averaging over a periodic orbit, 
\begin{align}
S(T_i,E_i,T_f,E_f)&=\delta^2(E_i-E)\delta^2(E_f-E)\\
&\hspace{5 mm}\times\sum_{n_im_in_fm_f}\frac{1}{T(E)}\int_{0}^{T(E)} dt\, \delta^2(T_i-t_{n_im_i}-t)\delta^2(T_f-\overline{t}_{n_fm_f}-t)\notag.
\end{align}
Performing the integral over $t$, we find
\begin{align}\label{scattquart}
S(T_i,E_i,T_f,E_f)&\sim \frac{2\delta^2(E_i-E)\delta^2(E_f-E)}{T(E)} \\
&\hspace{5 mm}\times\sum_{nmk}\delta^2\left(T_f-T_i-\frac{T(E)}{2}(n+mi)\right)\delta\left(\operatorname{Im}T_i-\frac{T(E)}{2}\left(k+\frac{1}{2}\right)\right).\notag
\end{align}
Since the energy $E$ is constant along a trajectory, the scattering kernel is elastic in $E$. The conjugate variable to $E$ is the complex time delay $T_f-T_i$.

The scattering kernel (\ref{scattquart}) predicts singularities in the correlator on the lattice 
\begin{align}\label{singquart}
t=\frac{T(E)}{2}(n+mi),\hspace{10 mm}m\not=0.
\end{align}
We can see the same structure directly in the microcanonical two-point function after analytic continuation. The correlator is obtained by averaging over the real orbit,
\begin{align}
C(t)&=\frac{1}{T(E)}\int_0^{T(E)} ds\, q(s+t)q(s)\\
&=\sqrt{4E}\Bigl[\text{cn}\left(u,\frac{1}{2}\right)-2\frac{\text{dn}\left(u,\frac{1}{2}\right)}{\text{sn}\left(u,\frac{1}{2}\right)}Z\left(u,\frac{1}{2}\right)\Bigr],\qquad u=(4E)^{1/4}t,
\notag
\end{align}
where $Z$ is the Jacobi zeta function. This expression is explicitly meromorphic, with simple poles at the expected points (\ref{singquart}).

In order to picture the scattering process, it is useful to choose a specific contour connecting $T_i$ and $T_f$, consisting of alternating Euclidean and Lorentzian segments. For the quartic oscillator the coordinate $q(t)$ is real along this contour, as shown in Figure \ref{fig:quartic_scattering_trajectories}. The scattering trajectory splits into three segments. The particle first comes in from infinity, then propagates in real time, and finally rolls down the hill back to infinity. This interpretation holds in more general models as well, but one must allow for the coordinates to be complex along the trajectory.
\begin{figure}[t]
\centering
\includegraphics[width=0.90\textwidth]{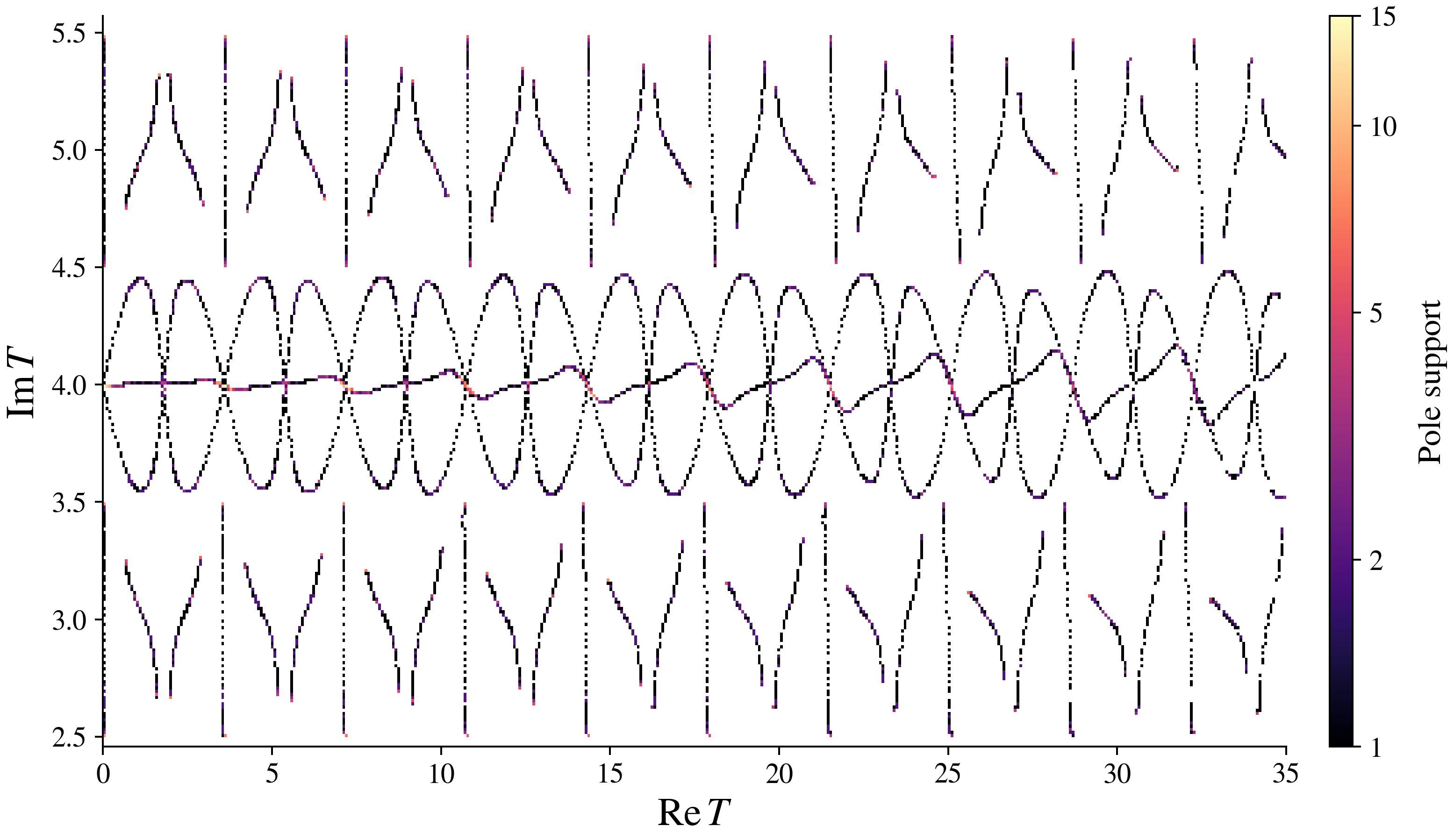}
\caption{The reduced scattering kernel $\tilde{S}(T)$ for the Toda chain, plotted as a function of pole separations $T=T_f-T_i $, where $\operatorname{Im}T_f>0$ and $\operatorname{Im}T_i<0$. The color signifies the number of poles per bin.  Here we have arbitrarily chosen an initial condition. The kernel is quasiperiodic as a function of $\operatorname{Re}T$.}
\label{fig:toda_scattering_kernel}
\end{figure}

More generally, there are interacting integrable systems whose singularities do not lie on a lattice. For example, let us take the three-particle periodic Toda chain~\cite{Toda1967},
\begin{align}
H=\sum_{i=1}^{3}\left(\frac{p_i^2}{2}+e^{q_i-q_{i+1}}\right),\hspace{10 mm}q_{4}=q_1.
\end{align}
This Hamiltonian can be solved exactly~\cite{KacVanMoerbeke1975}, but for our purposes it suffices to integrate the equations of motion numerically. Choosing an arbitrary initial condition, we find the scattering kernel displayed in Figure \ref{fig:toda_scattering_kernel}. In contrast to the quartic oscillator, the kernel has codimension-one support in the complex $T_f-T_i$ plane.

In both of the examples we have studied here, the scattering kernel is a periodic or quasiperiodic function of $\operatorname{Re}(T_f-T_i)$. This follows from the fact that the trajectories live on invariant tori, whose frequencies are generically rationally independent. Therefore scattering in integrable systems with a fixed-action ensemble does not equilibrate at late times. As we will see next, the chaotic case behaves very differently.
\begin{figure}[t]
\centering
\includegraphics[width=0.82\textwidth]{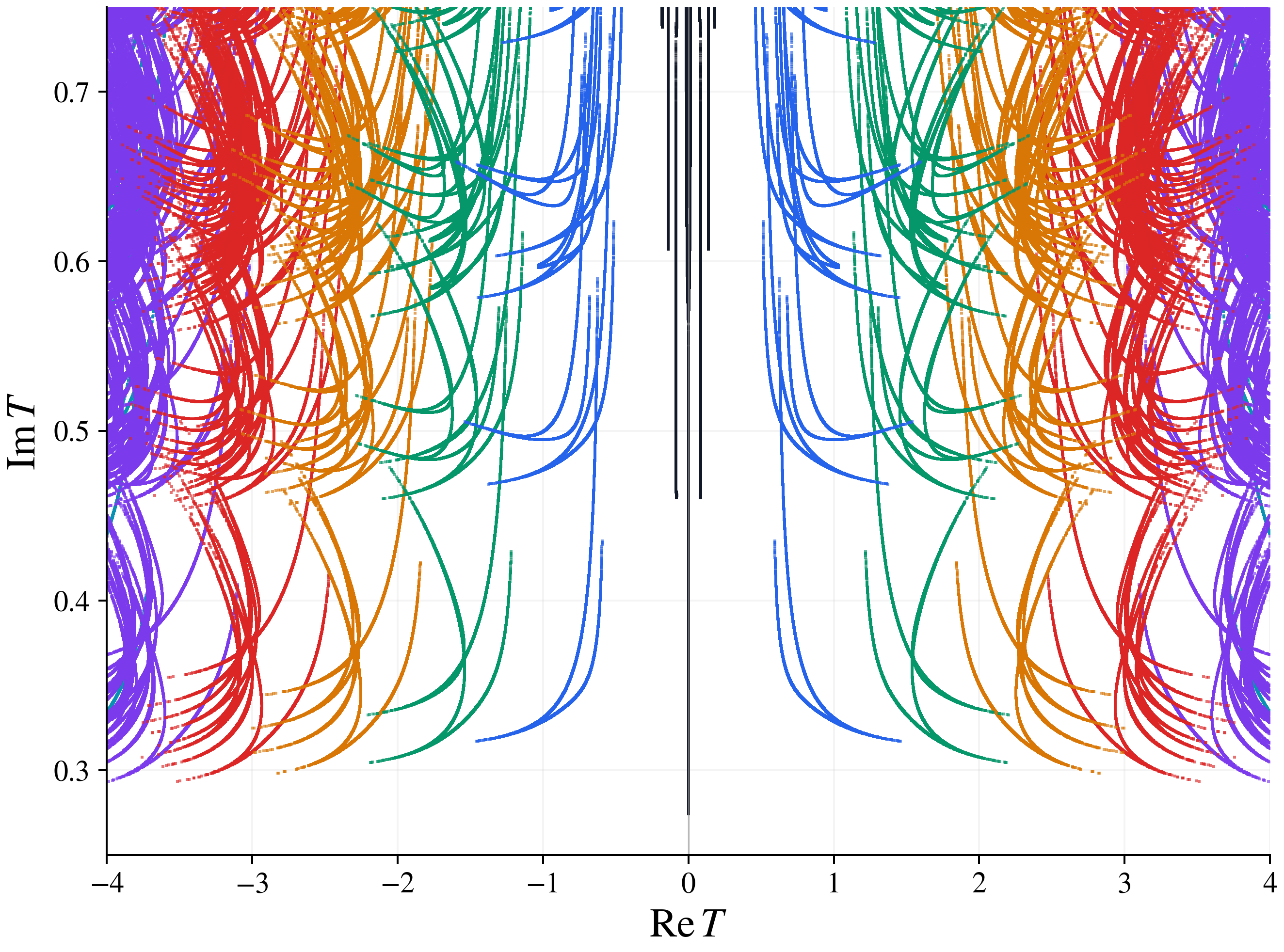}
\caption{The reduced scattering kernel for the Lorenz attractor. The colors distinguish the return separation of the associated Poincar\'e crossings, and the displayed window keeps the first three visible bands.}
\label{fig:lorenz_scattering_kernel}
\end{figure}
\subsection{Chaos and thermalization}
\begin{figure}[t]
\centering
\includegraphics[width=\textwidth]{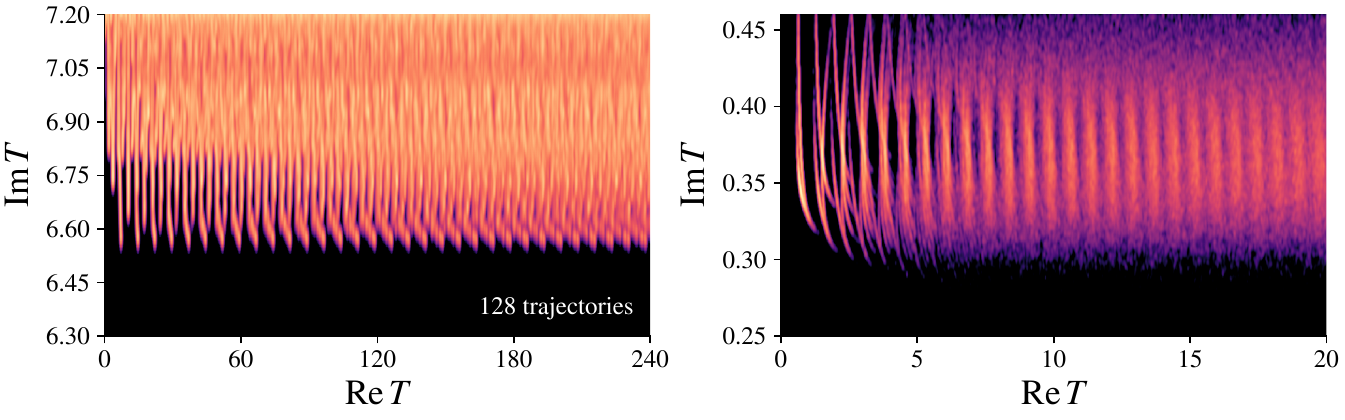}
\caption{The reduced scattering kernel for the H\'enon--Heiles system at microcanonical energy $E=0.12$ (left) and the Lorenz attractor (right), plotted using a Gaussian smoothing function. The Lorenz attractor quickly equilibrates, while the mixed H\'enon--Heiles system exhibits slow relaxation.}
\label{fig:scattering_kernel_late_times}
\end{figure}
We conclude this section with a discussion of scattering in chaotic systems. Let us begin with the Lorenz attractor, whose scattering kernel is depicted in Figure \ref{fig:lorenz_scattering_kernel}. Here we are choosing both $T_i$ and $T_f$ to be on the first sheet of the trajectory displayed in Figure \ref{fig1b}, where the branch cuts are oriented vertically away from the real axis. As is often the case in dynamical systems, it is useful to think of this figure in terms of a return map $P$ acting on a Poincaré section. Let us define the escape section $\Sigma$, consisting of all points $\sigma$ in phase space where $\sigma(it)$ is singular for some real $t$. For a chosen singularity of $\sigma(it)$ with $t>0$, define the height function $h(\sigma)=t$. From our earlier discussion of scattering trajectories, it is clear that the singularities contributing to the kernel are separated by
\begin{align}
T_f-T_i=ih(\sigma)+ih(P^k\sigma)+t_k(\sigma),
\end{align}
where $t_k(\sigma)$ is the total real time elapsed between the points $\sigma$ and $P^k\sigma$. Since the return map is chaotic, it stretches and folds sets of initial conditions on the escape section, leading to the structures displayed in Figure \ref{fig:lorenz_scattering_kernel}.

At late times $\operatorname{Re}(T_f-T_i)\to \infty$, we expect that the scattering process forgets about the initial conditions. This is the statement of thermalization (or equilibration in a non-Hamiltonian system). In a strongly chaotic system like the Lorenz attractor, we expect an exponential approach to equilibrium, with decay rate set by the leading Ruelle resonance~\cite{Pollicott1985, Ruelle1986}. However, almost all real-world systems are mixed, with both chaotic regions and invariant tori in phase space. An example is the Hénon--Heiles Hamiltonian~\cite{HenonHeiles1964},
\begin{align}
H=\frac{1}{2}\left(p_x^2+p_y^2+x^2+y^2\right)+x^2y-\frac{1}{3}y^3.
\end{align}
In such a mixed system, chaotic trajectories may become temporarily trapped near regular islands, leading to slow relaxation~\cite{MacKayMeissPercival1984, MeissOtt1985}. Independently, averaging over invariant tori with different frequencies produces dephasing. Both mechanisms generically yield algebraic tails in the approach to equilibrium. The two scenarios are depicted in Figure \ref{fig:scattering_kernel_late_times}.

\FloatBarrier
\begin{figure}[t]
\centering
\includegraphics[width=0.86\textwidth]{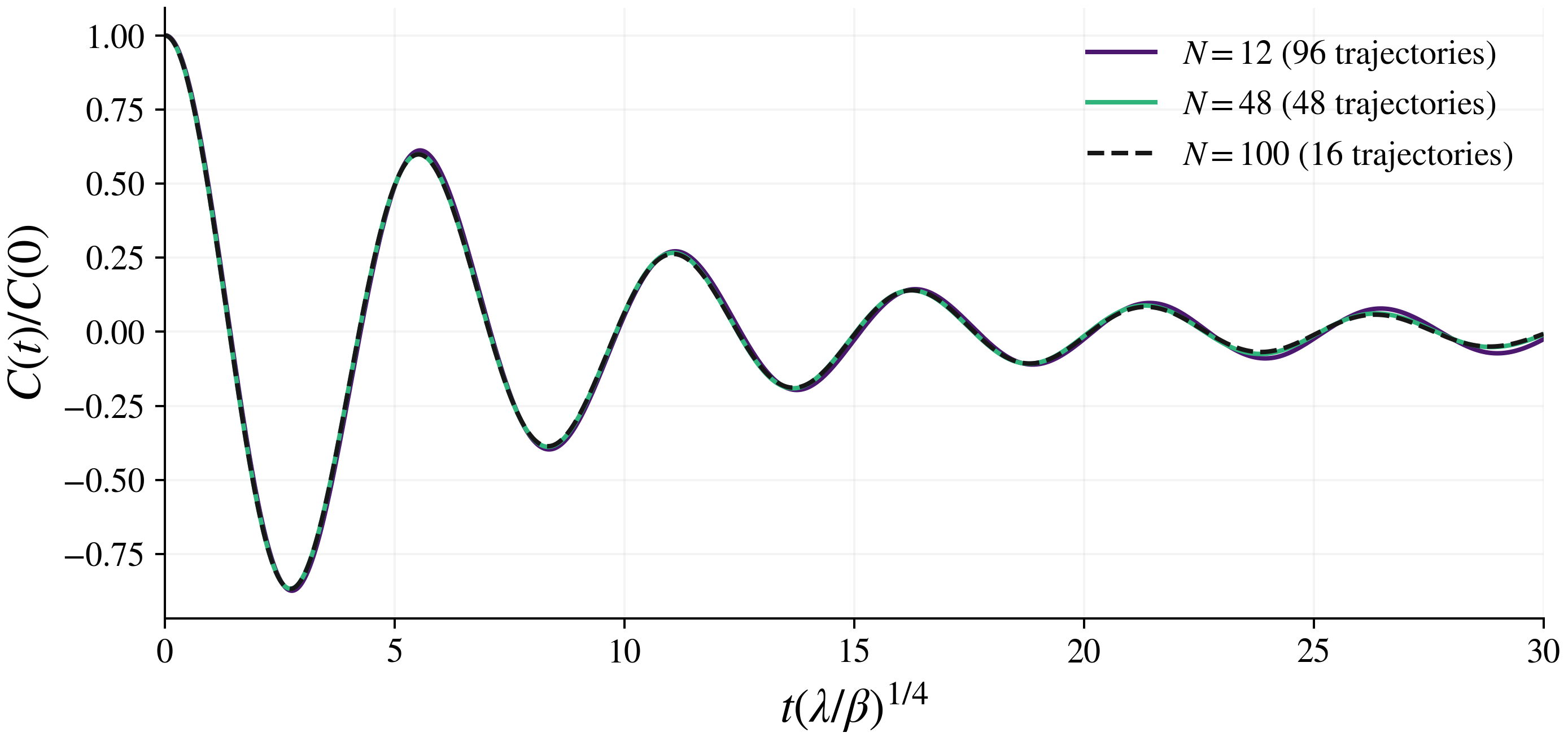}
\caption{Normalized real-time correlator in the classical quartic one-matrix model at $m=0$, plotted against the dimensionless time. The agreement between $N=12$, $N=48$, and $N=100$ illustrates the rapid approach to the planar limit for real-time observables. The curves agree well for early times, and start diverging as the time increases.}
\label{fig:one_matrix_realtime_convergence}
\end{figure}
\section{Classical dynamics of a single matrix}\label{sec:one-matrix}
Having understood the scattering kernel in general classical systems, we now turn to the case of planar theories. We begin with the simplest model of large-$N$ matrix quantum mechanics, 
\begin{align}
\label{eq:mqmhamiltonian}
H=\frac{1}{2}\text{Tr}(P^2)+\frac{m^2}{2}\text{Tr}(X^2)+\frac{\lambda}{4N}\text{Tr}(X^4).
\end{align}
Here $X$ and $P$ are Hermitian $N\times N$ matrices, and the 't Hooft limit is defined by taking $N\to \infty$ with $\lambda$ held fixed~\cite{tHooft1974}. This Hamiltonian is invariant under the $SU(N)$ transformation
\begin{align}
(X,P)\to (U^\dagger XU,U^\dagger PU),\hspace{10 mm}U^\dagger U=1. 
\end{align}
A well-known property of this model is that if one gauges the $SU(N)$ symmetry, the theory can be solved in terms of fermions in a potential~\cite{bipz}. Here we are interested in the ungauged system, for which much less is known.

In order to make progress, we will focus on the high-temperature limit $m \beta, \lambda \beta^3 \to 0$ with the ratio $\lambda/(\beta m^4)$ held fixed. In this weakly coupled regime, quantum corrections are suppressed, and observables can be computed using classical statistical mechanics~\cite{AsplundBerensteinDzienkowski2013, GurAriHanadaShenker2016}. The main object of interest is the Lorentzian thermal two-point function,
\begin{align}
C(t)=\frac{1}{N^2}\langle \text{Tr}(X(t)X(0))\rangle_{\beta}.
\end{align}
This can be computed to high accuracy using Monte Carlo simulation, as shown in Figure \ref{fig:one_matrix_realtime_convergence}. We observe an overall power-law decay in time with an oscillating envelope, which is characteristic of integrable systems in the thermal ensemble. The power-law decay comes from the integral over the conserved actions $I_a$, which leads to algebraic dephasing at late times. This model indeed turns out to be classically integrable~\cite{BruschiCalogero2000Matrix, BruschiCalogero2004Quartic}, but extracting the planar correlator from integrability is not straightforward.

\subsection{Complex singularities from the moment expansion}
\label{sec:momentexpansion}

Although the real-time correlator can be accurately computed using Monte Carlo, the analytic structure in the complex time plane is less amenable to direct numerical integration. In order to avoid accumulating finite-$N$ errors, we now introduce an analytic method that works directly in the planar limit. The strategy is to dimensionally reduce the quantum-mechanical problem to a planar matrix integral, for which the exact solution is known.

Consider the series expansion of $C(t)$ around the origin,
\begin{align} \label{ct_moment_expansion}
C(t)=\sum_{n=0}^{\infty}\mu_{2n}\frac{(it)^{2n}}{(2n)!}.
\end{align}
The moments $\mu_{2n}$ can be expressed as one-point functions of operators with increasing size, 
\begin{align}
\mu_{2n}=\frac{(-1)^{n}}{N^2}\langle \mathcal{O}_{2n}(X,P)\rangle_\beta,\hspace{10 mm}\mathcal{O}_{2n}(X,P)=\text{Tr}(X\mathcal{L}^{2n}X),
\end{align}
where $\mathcal{L}=\{H,\cdot\}$ is the classical Liouvillian. Working in the massless case for simplicity, the first few operators take the form
\begin{align}
\mathcal{O}_0&=\text{Tr}(X^2),\\
\mathcal{O}_2&=-\frac{\lambda}{N}\text{Tr}(X^4),\\
\mathcal{O}_4&=\frac{3\lambda^2}{N^2}\text{Tr}(X^6)-\frac{4\lambda}{N}\text{Tr}(P^2X^2)-\frac{2\lambda}{N}\text{Tr}(PXPX).
\end{align}
Since the number of words with length $L$ is exponential in $L$, we expect that the time it takes to generate up to $\mathcal{O}_{2n}$ scales exponentially with $n$. Using symbolic algebra on a computer, we were able to generate up to $\mathcal{O}_{60}$, at which point exponential slowdown kicks in.  This should be contrasted with the case of the SYK model, for which there is a polynomial-time algorithm for computing the moments~\cite{ParkerEtAl2019}.

\begin{figure}[t]
\centering
\includegraphics[width=0.72\textwidth]{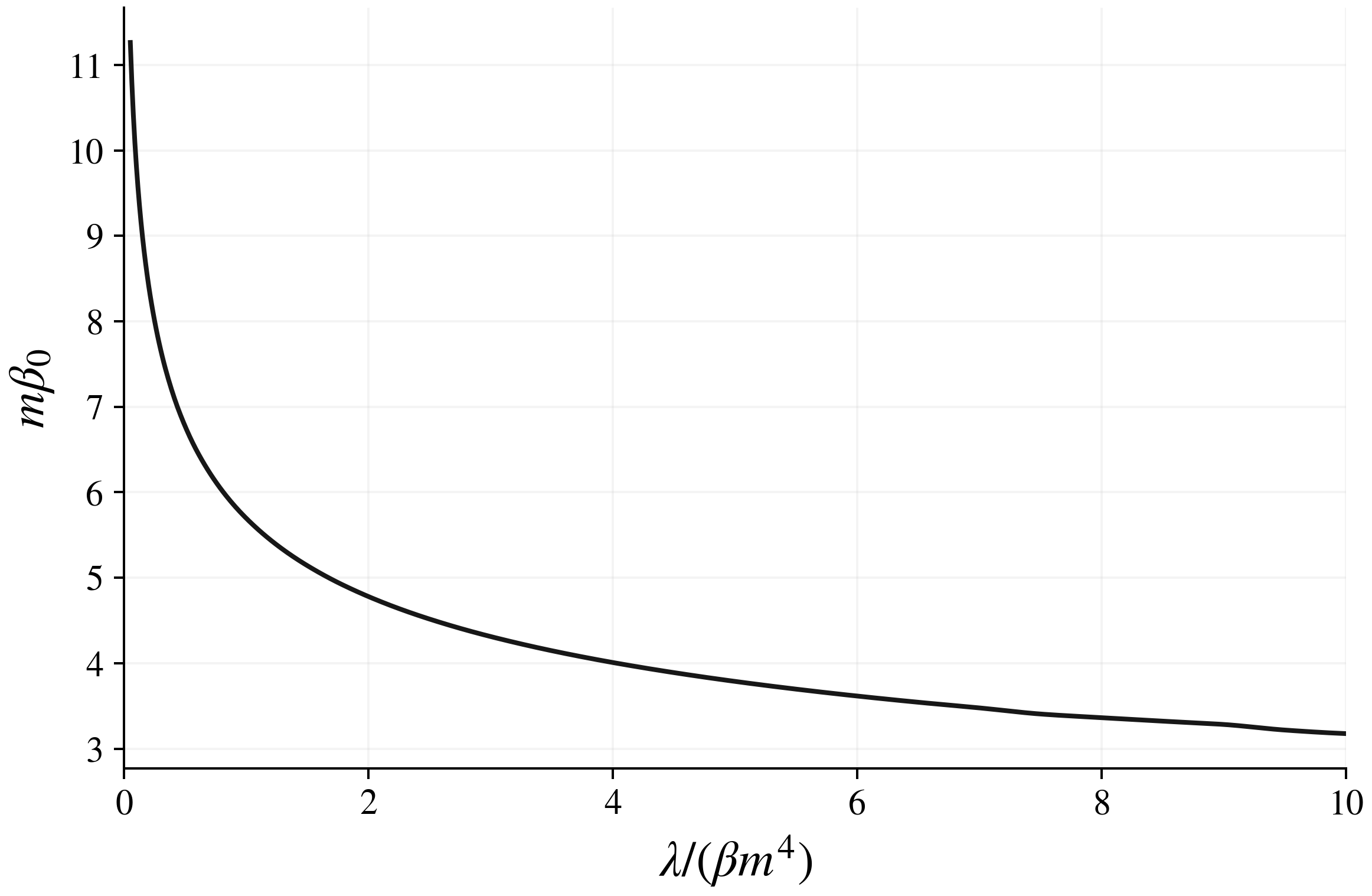}
\caption{Dimensionless singularity scale $m\beta_0$ as a function of the dimensionless coupling $\lambda/(\beta m^4)$ in the classical planar one-matrix model. The values are extracted from the imaginary-axis pole of diagonal Pad\'e approximants to the moment expansion.}
\label{fig:one_matrix_beta0_lambda}   
\end{figure}

In order to evaluate the moments, we first perform the Gaussian integral over $P$ to obtain operators with only $X$-dependence, 
\begin{align}\label{MQMwick}
\overline{\mathcal{O}}(X)=\frac{\int dP\, \mathcal{O}(X,P)e^{-\beta\text{Tr}(P^2)/2}}{\int dP\, e^{-\beta\text{Tr}(P^2)/2}}.
\end{align}
The problem is now reduced to a matrix integral,
\begin{align}\label{MQMbipz}
\langle \mathcal{O}(X,P)\rangle_\beta=\frac{\int dX\, \overline{\mathcal{O}}(X)e^{-\beta\lambda\text{Tr}(X^4)/(4N)}}{\int dX\, e^{-\beta\lambda\text{Tr}(X^4)/(4N)}}, 
\end{align}
where we set $m=0$ as above. At infinite $N$, we may then evaluate this integral using saddle-point methods, which give the exact solution for the eigenvalue density~\cite{bipz}
\begin{align}\label{eigensingle}
\rho(x)=\frac{\beta\lambda}{2\pi N}\left(x^2 +\frac{a^2}{2}\right)\sqrt{a^2-x^2},\hspace{10 mm}a^4=\frac{16N^2}{3\beta\lambda}.
\end{align}
The density is supported in a finite range $|x|<a$, and is normalized as
\begin{align}
\int_{-a}^a dx\, \rho(x)=N.
\end{align}
This leads to the expectation values
\begin{align}
\langle\text{Tr}(X^{2k})\rangle_\beta=\frac{(2k)!(2k+2)N^{1+k}}{k!(k+2)!(3\beta\lambda)^{k/2}}.
\end{align}

\begin{figure}[t]
\centering
\includegraphics[width=0.72\textwidth ]{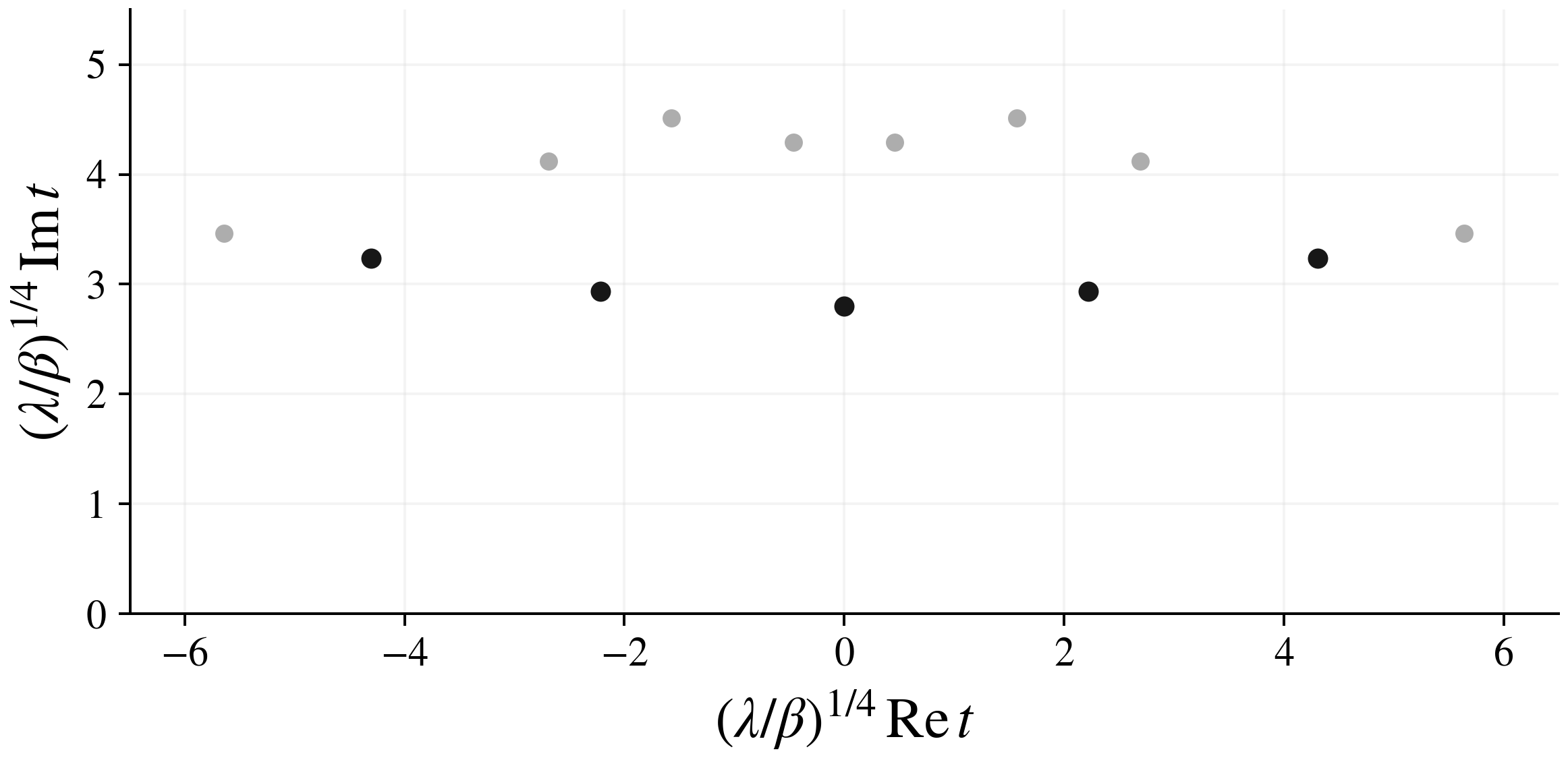}
\caption{Singularities of the classical planar one-matrix correlator at $m=0$ in the upper half of the complex time plane, extracted from a diagonal Pad\'e approximant to the moment expansion. The moments are computed using the large-$N$ solution of the matrix model.  Stable poles are shown in black, while unstable Pad\'e poles are shown in gray.}
\label{fig:one_matrix_pade_singularities}
\end{figure}
Using this methodology, we were able to carry out the moment expansion up to order $t^{60}$. At large $N$, we found the asymptotic behavior 
\begin{align}\label{muasympt}
\mu_{2n}\sim \left(\frac{4n}{e\beta_0}\right)^{2n},\hspace{10 mm}n\to \infty,
\end{align}
which is consistent with the universal operator growth hypothesis~\cite{ParkerEtAl2019}.  As shown in Figure \ref{fig:one_matrix_beta0_lambda}, the parameter $\beta_0$ diverges in the free limit $\lambda=0$, where the correlator is entire, and decreases as $\lambda$ is dialed up at fixed mass. For $m=0$ we find that $\beta_0/\beta\approx 5.6(\beta^3\lambda)^{-1/4}$ has converged to over 10 digits of precision, which demonstrates the computational power of our analytic approach. {In the SYK context, $\beta_0$ is sometimes referred to as the ``circumference of the fake disk'' \cite{Lin:2023trc}.} In general, $\beta_0$ satisfies the chaos bound~\cite{ParkerEtAl2019, AvdoshkinDymarsky2020, GuKitaevZhang2022} 
\begin{align}\label{chaosbound}
\lambda_L\le \frac{2\pi}{\beta_0},
\end{align}
where $\lambda_L$ is the maximal Lyapunov exponent. In this case the Lyapunov exponent vanishes and the bound is trivially satisfied.

\begin{figure}[t]
\centering
\includegraphics[width=0.78\textwidth]{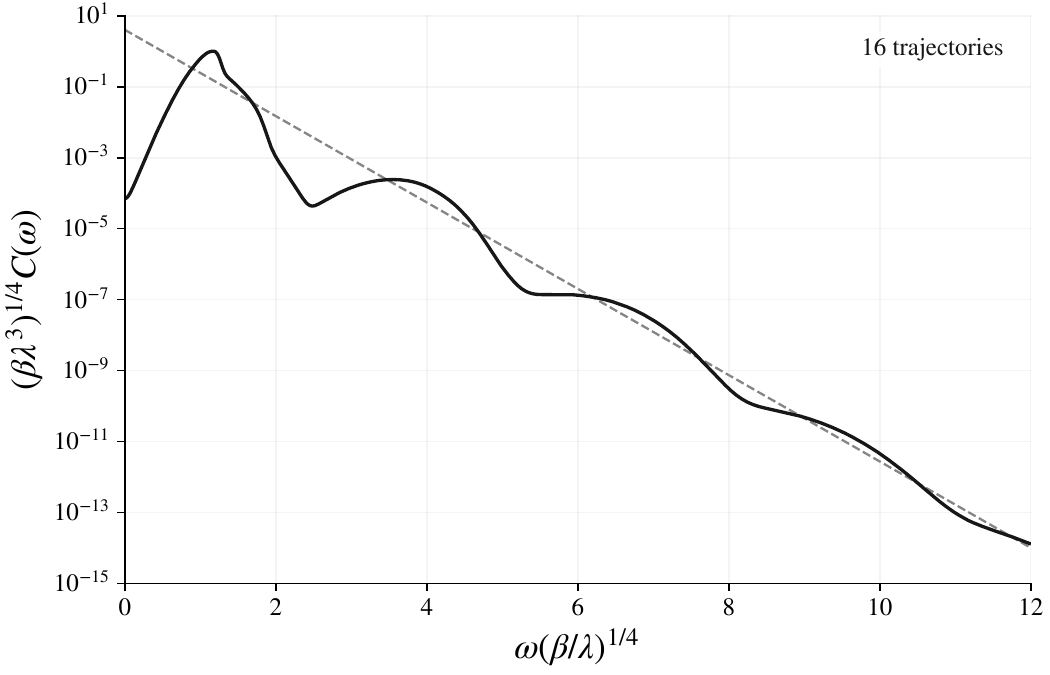}
\caption{Power spectrum of the classical quartic one-matrix model at $N=100$ and $m=0$, on a logarithmic vertical scale. The spectral density exhibits damped oscillations around an exponentially decreasing envelope. The dashed line has the independently determined planar slope $\beta_0(\lambda/\beta)^{1/4}/2\simeq2.8$.}
\label{fig:one_matrix_power_spectrum}
\end{figure}

The asymptotics (\ref{muasympt}) imply that the nearest singularity to the real axis is at 
\begin{align}
\label{eq:beta0singularity}
t=\pm \frac{i\beta_0}{2}.
\end{align}
In order to diagnose the power of this singularity, it is useful to Padé approximate the quantity $d\log C(t)/dt$ and then take the residue of the pole. We find that the singularity in $C(t)$ is very close to a simple pole.

We also obtained two additional stable pairs of simple poles in the upper half plane, as depicted in Figure \ref{fig:one_matrix_pade_singularities}. Note that the singularities here are approximately evenly spaced in real time. This has an interesting dynamical consequence for the power spectrum $C(\omega)$. At large real frequency, $C(\omega)$ is controlled by the singularities of $C(t)$ which are closest to the real axis. Inspired by Figure \ref{fig:one_matrix_pade_singularities}, let us introduce a toy model for these singularities,
\begin{align}
C(t)=-i\sum_{n=-\infty}^{\infty}\frac{c_n}{t-\frac{i\beta_0}{2}-nT-i\tau_n}+(t\to-t),\hspace{10 mm}\tau_0=0,
\end{align}
where $\tau_n=\tau_{-n}>0$ for $n>0$ is an upward shift in complex time, and $c_n=c_{-n}^*=|c_n|e^{i\phi_n}$. Fourier transforming, we find that the asymptotics of the power spectrum are given by
\begin{align}\label{comegaosc}
C(\omega)\sim  e^{-\beta_0\omega/2}\left[c_0+2\sum_{n=1}^{\infty}|c_n| \cos(n\omega T+\phi_n)e^{-\tau_n\omega}\right],\hspace{10 mm}\omega\to \infty.
\end{align}
We see that there are damped oscillations in $C(\omega)$ with period $2\pi/T$, which were first observed in~\cite{AsplundBerensteinDzienkowski2013}. In Figure \ref{fig:one_matrix_power_spectrum} we confirm these oscillations using Monte Carlo simulation.

Incidentally, there is a simple argument that the $\beta_0$ singularity must be the closest singularity to the real axis. Suppose that a complex pair has imaginary part less than $\beta_0/2$, so that $\tau_n<0$ for some $n$. Then at large real frequency, (\ref{comegaosc}) is oscillatory. In a unitary theory $C(\omega)$ must be nonnegative, so this scenario is ruled out.

\subsection{The scattering data}
\label{sec:The scattering data}
We now turn to the complex scattering problem in this model, with the eventual aim of assigning classical trajectories to the singularities in Figure \ref{fig:one_matrix_pade_singularities}. To this end, we consider the equation of motion, 
\begin{align}
\ddot{X}=-m^2X-\frac{\lambda}{N}X^3.
\end{align}
To analyze a singularity at $t=t_0$, we make the ansatz
\begin{align}
X\sim \frac{A}{\epsilon^{\Delta }},\hspace{10 mm}t=t_0+\epsilon.
\end{align}
Balancing $\ddot X$ against $X^3$ gives $\Delta=1$, and the residue matrix obeys
\begin{align}
2A+\frac{\lambda}{N}A^3=0.
\end{align}
In a basis where $A_{ij}=a_{i}\delta_{ij}$ is diagonal, we have $a_i=0$ or $a_i=\pm i\sqrt{2N/\lambda}$. In particular, the leading singularity is independent of the mass $m$.

Now let us analyze the perturbations to a given singularity. Since the dimension of phase space is $2N^2$, a general solution should depend on $2N^2$ resonances, which appear as perturbations to the pole at order $\epsilon ^{r-1}$. Linearizing around the pole,
\begin{align}
X=\frac{1}{\epsilon}(A+B \epsilon^{r}),
\end{align}
one finds that the mass term again enters only at order $\epsilon^{r-1}$ rather than $\epsilon^{r-3}$, so it does not affect the resonance equation. After diagonalizing $A$ as above, the leading linearized equation is therefore
\begin{align}
\left[(r-1)(r-2)+\frac{\lambda}{N}\left(a_i^2+a_i a_j+a_j^2\right)\right]B_{ij}=0.
\end{align}
There are then three cases,
\begin{align}
a_i=a_j=\pm i\sqrt{\tfrac{2N}{\lambda}}
&\qquad \Longrightarrow\qquad r=-1,\ 4,\\
(a_i,a_j)=\left(0,\pm i\sqrt{\tfrac{2N}{\lambda}}\right)\ \text{or}\ \left(\pm i\sqrt{\tfrac{2N}{\lambda}},\mp i\sqrt{\tfrac{2N}{\lambda}}\right)
&\qquad \Longrightarrow\qquad r=0,\ 3,\\
(a_i,a_j)=(0,0)
&\qquad \Longrightarrow\qquad r=1,2.
\end{align}

The $r=-1$ resonance is of particular interest, since it represents a potential instability. Near the pole, the perturbation can be resummed as 
\begin{align}
X_{ii}&=\frac{a_i}{\epsilon}+\frac{B_{ii}}{\epsilon^2}\approx \frac{a_i}{\epsilon-\frac{B_{ii}}{a_i}},\hspace{10 mm}|B_{ii}/a_i|\ll \epsilon.
\end{align}
In the case where $B$ is proportional to $A$, this is the expected time-translation mode, corresponding to a shift in the position of the singularity in the complex plane. When there are two eigenvalues $a_i$ and $a_j$ with nonvanishing residues at the pole, a perturbation with $B_{ii}/a_i\not=B_{jj}/a_j$ splits the pole into two. Since the perturbation grows near $\epsilon=0$, this is an instability of the solution. It follows that unless the initial data are fine-tuned, singularities are characterized by a single eigenvalue running off to infinity.

The remainder of the scattering data is characterized by the other resonances.  The resonance $r=0$ corresponds to a complexified $SU(N)$ rotation of the residue matrix $A$, and $r=1,2$ are perturbations in the kernel of $A$. Finally, $r=3$ and $r=4$ shift the values of the $SU(N)$ conserved charge $[X,P]$ and the Hamiltonian $H$, which scale dimensionally as $\epsilon ^{-3}$ and $\epsilon ^{-4}$ respectively. It is straightforward to check that the sum of the multiplicities of the resonances adds up to $2N^2$, as expected.

Since the resonances are all integers, the expansion around the local singularity is a pure Laurent expansion rather than branched. In general logarithms could appear in the expansion at the order of the resonances, but it is straightforward to check that this does not happen in this example.  Therefore the model passes the Painlev\'e test.

\begin{figure}[t]
\centering
\includegraphics[width=\textwidth]{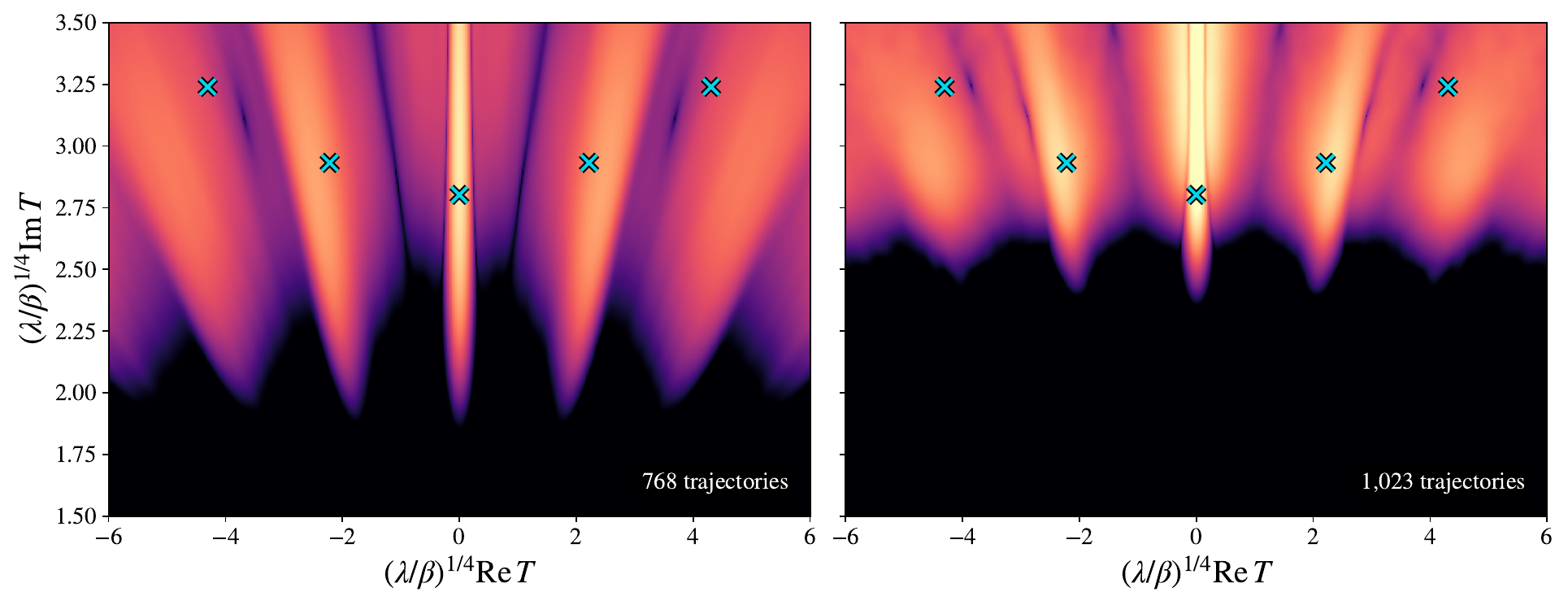}
\caption{Residue-weighted reduced scattering kernel $\tilde{S}_A(T)$ for the correlator $\langle\text{Tr}(X(t)X(0))\rangle_\beta$ in the massless one-matrix model, with $N=4$ (left) and $N=32$ (right). Colors show the magnitude after summing complex weights, on a common logarithmic scale. Cyan crosses mark planar Pad\'e poles; the outermost pair is less well converged.}
\label{fig:one_matrix_weighted_scattering_kernel}
\end{figure}

\subsection{D-brane scattering and quantum gravity}
The pole analysis above shows that a generic movable singularity has rank-one residue. In eigenvalue language this means that one eigenvalue runs off to infinity while the remaining degrees of freedom remain finite. Using the D-brane picture of matrix quantum mechanics~\cite{Witten1996}, we can interpret the escaping eigenvalue as a probe brane propagating on a background formed out of the stack of remaining $N-1$ eigenvalues. The probe brane is coupled to the stack via off-diagonal open-string interactions~\cite{Witten1996, Bachas1996}.  We can then think of the singularities in Figure \ref{fig:one_matrix_pade_singularities} as an approximate quartic oscillator lattice, with deformations mediated by probe-bath interactions. Although we do not expect a local bulk description in this model, we can schematically identify the probe eigenvalue as an emergent radial direction, so that the singularities in complex time describe boundary-to-boundary scattering\footnote{Another context where scattering trajectories arise is the Hawking evaporation of D0 branes from the IIA black hole. At low temperatures/energies, one can follow Hawking's calculation and show that the decay rate is exponentially suppressed as $e^{-N I(E)}$, see Appendix A of \cite{Lin:2013jra}. This is consistent with the fact that at fixed energy, the change in entropy of the black hole is $\Delta S \propto N $ when one emits a D0-brane, sending $N \to N-1$. At high temperatures, one can ask whether there is some instanton effect in the high $T$ matrix integral that represents the metastability of the thermal ensemble.}.

What are the dynamical consequences of the probe approximation? Consider the planar eigenvalue density (\ref{eigensingle}), which is compactly supported for $|x|<a$. Under Euclidean time evolution, the fastest way for the trajectory to reach infinity is if the maximal eigenvalue $x=\pm a$ rolls down the potential. The singularity at $t=i\beta_0/2$ then corresponds to the scattering trajectory in Figure \ref{fig:quartic_imaginary_contour}, where an eigenvalue propagates from infinity, grazes the edge $|x|=a$, and then returns to infinity. Meanwhile, the off-axis singularities in Figure \ref{fig:one_matrix_pade_singularities} correspond to the scattering trajectory in Figure \ref{fig:quartic_bounce_contour}. The longer the trajectory spends propagating through the bath, the more off-diagonal strings are excited. This mixing process increases the Euclidean time it takes to reach infinity, so we expect the off-axis singularities to be shifted upwards in the complex plane. This gives a qualitative explanation of Figure \ref{fig:one_matrix_pade_singularities}.

\begin{figure}[t]
\centering
\includegraphics[width=0.78\textwidth]{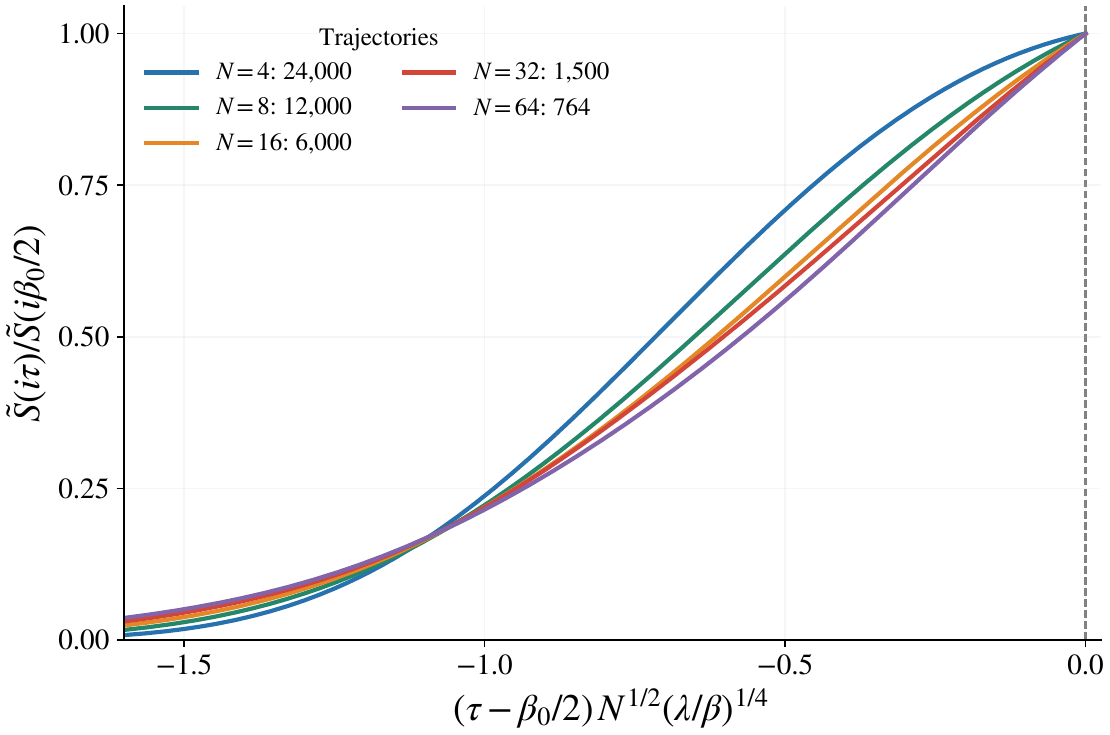}
\caption{
The scattering kernel for the massless one-matrix model at finite $N$ near the planar $\beta_0$ singularity. We plot the diagonal part of the reduced scattering kernel normalized at $T=i\beta_0/2$.  At large $N$, we see that the smoothing timescale is $\Delta T\sim N^{-1/2} (\beta/\lambda)^{1/4}$.}
\label{fig:one_matrix_diagonal_onset}
\end{figure}

At finite $N$, the probe approximation breaks down, and we expect the singularities to be resolved. This is a quantum effect in the bulk description. It is straightforward to see this via Monte Carlo simulation of the finite-$N$ system. In Figure \ref{fig:one_matrix_weighted_scattering_kernel}, we display the scattering kernel for $N=4$ and $N=32$, observing large-$N$ collapse to a localized distribution supported at the singularities in Figure \ref{fig:one_matrix_pade_singularities}. Figure \ref{fig:one_matrix_diagonal_onset} then depicts this collapse in the case of the singularity at $t=i\beta_0/2$. Note that it is essential to use the scattering kernel for this computation; as explained in Section \ref{scattsec}, the complex-time correlator cannot be simulated directly. Since a singularity in a single eigenvalue contributes $1/N$ to the correlator, the planar singularity in the correlator is explained by not just one eigenvalue diverging, but a finite fraction of the eigenvalues rolling to infinity. However, as these eigenvalues approach infinity, the density becomes sufficiently low that the ``atoms'' of the eigenvalue fluid are revealed and we can treat each one in the probe approximation.

\begin{figure}[t]
\centering
\includegraphics[width=\textwidth]{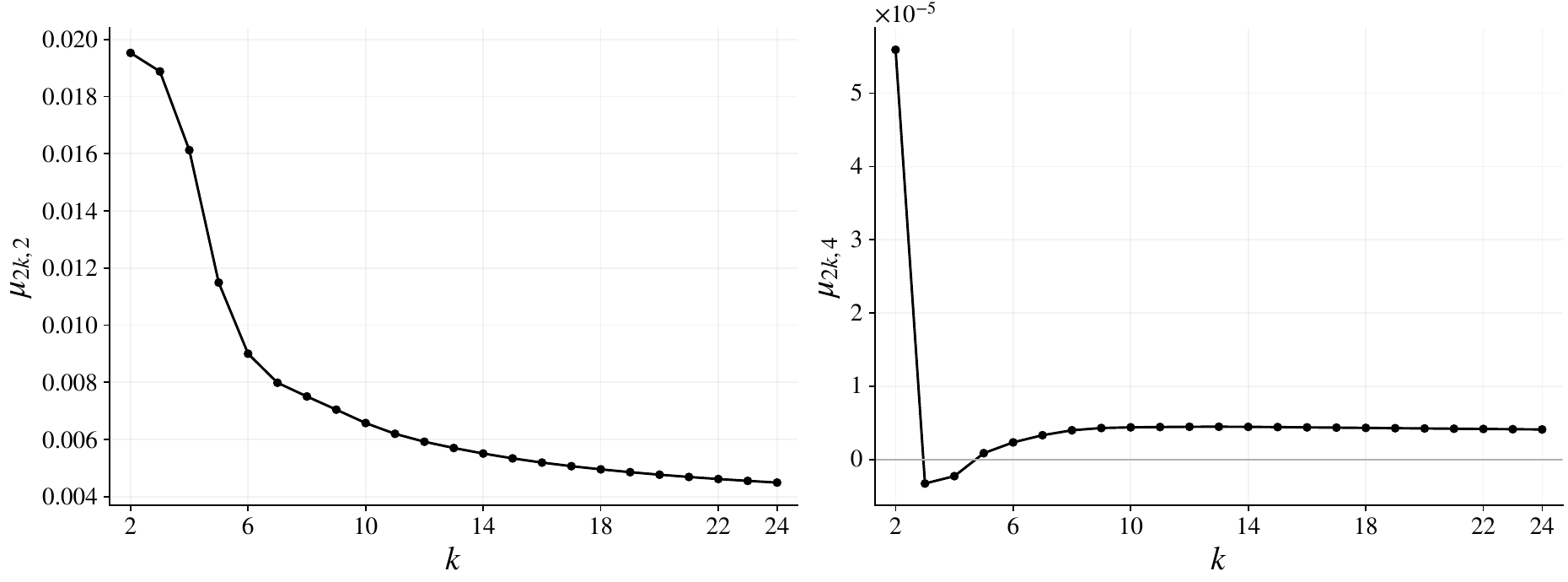}
\caption{Nonplanar coefficients $\mu_{2k,2}$ (left) and $\mu_{2k,4}$ (right) in the massless one-matrix model, computed exactly through $k=24$. Their flattening at large $k$ is consistent with an effective expansion parameter $k^2/N$.}
\label{fig:one_matrix_nonplanar_moments}
\end{figure}

Since the probe D-brane action scales like $N$, the smoothing scale $\Delta t\sim N^{-1/2}$ represents the scale on which the D-brane coordinates fluctuate. This is a natural expectation, since the generic singularity of a trajectory consists of a single D-brane escaping to infinity. In order to better understand this result, note that the $\beta_0$ singularity is dominated by large moments,
\begin{align}
\frac{1}{t-\frac{i\beta_0}{2}}&\sim \frac{4i}{\beta_0}\sum_{k=0}^{\infty}\left(\frac{2it}{\beta_0}\right)^{2k}\notag\\
&\sim \frac{4i}{\beta_0}\sum_{k=0}^{\infty}\exp\left(-\frac{4ik}{\beta_0}\left(t-\frac{i\beta_0}{2}\right)\right),
\end{align}
where we expanded around $t=i\beta_0/2$ in the last equation. The scale $\Delta t\sim N^{-1/2}$ therefore translates to a breakdown of the planar expansion at $k\sim N^{1/2}$. This means that corrections to the large-$k$ moments should be controlled by the effective expansion parameter $k^2/N$,
\begin{align}\label{muplanar}
\mu_{2k}=\mu_{2k,0}\Bigl[1+\left(\frac{k^2}{N}\right)^2\mu_{2k,2}+\left(\frac{k^2}{N}\right)^4\mu_{2k,4}+\cdots\Bigr],
\end{align}
where $\mu_{2k,2j}$ for $j>0$ is expected to plateau at large $k$. In Figure \ref{fig:one_matrix_nonplanar_moments}, we display the leading nonplanar corrections to the moments computed using the method of Section \ref{sec:momentexpansion}, finding that $\mu_{2k,2}$ and $\mu_{2k,4}$ indeed flatten out at large $k$.  Note that the resolution occurs on a wider time scale than fluctuations of the edge of the eigenvalue spectrum, which are controlled by $N^{-2/3}$~\cite{TracyWidom1994, DeiftGioev2007}.

In fact, there is some precedent for the expansion parameter $k^2/N$. In the large-charge sector of $\mathcal{N}=4$ SYM \cite{BerensteinMaldacenaNastase2002}, the expansion parameter for the two-point function of $\text{Tr}(Z^J)$ is $J^2/N$, so perturbation theory breaks down for operators of length $N^{1/2}$. The argument is combinatorial: the first nonplanar correction involves two Wick contractions, and therefore four insertions. Each of these insertions can occur anywhere along the long operator, so the correction scales like $J^4/N^2$~\cite{KristjansenEtAl2002, ConstableEtAl2002}. This result can also be interpreted in terms of fluctuations of a D-brane collective coordinate~\cite{Berenstein2013Collective}. In our case the moment $\mu_{2k}$ computes the trace of an operator of length proportional to $k$, so the breakdown occurs at the same scale.

The large-$k$ asymptotics of the moments lead to a remarkable conjecture which holds to all orders in the genus expansion. Plugging (\ref{muasympt}) into the expansion 
\begin{align}
   C(t) &= \sum_{g=0}^\infty \frac{1}{N^{2g}}C_{g}(t),
\end{align}
we find
\begin{align}
C_g(t)\sim \frac{\mathcal{N}_g}{\left(t-\frac{i\beta_0}{2}\right)^{4g+1}},\hspace{10 mm}t\sim \frac{i\beta_0}{2}.
\end{align}
Note that the power of the singularity increases with genus. This is a striking prediction that should hold for all nonzero values of the coupling $\lambda$ in the matrix quantum mechanics. It predicts the same radius of convergence for each term in the genus expansion, with a  power-law divergence. The time $t_{\text{breakdown}}$ at which planar perturbation theory fails is obtained by comparing the first nonplanar correction with the genus-zero result,
\begin{align}
    C_0\left(\frac{i\beta_0}{2} + t_\text{breakdown}\right) \sim \frac{1}{N^2} C_1\left(\frac{i\beta_0}{2}+ t_{\text{breakdown}}\right) \Rightarrow t_\text{breakdown} \propto \frac{1}{\sqrt{N}}. 
\end{align}

In the strict $k\to \infty$ limit with $N$ held fixed, we found that the ratio $\mu_{2k}/(2k)!$ grows factorially. This implies that the moment expansion (\ref{ct_moment_expansion}) has zero radius of convergence, so that the correlator is nonanalytic at the origin. An equivalent statement is that the bright ridges in Figure \ref{fig:one_matrix_weighted_scattering_kernel} extend down to $T=0$; the fact that they appear to end at some finite value of $\operatorname{Im}T$ is an artifact of the scale used in the plot. We can think of this singularity in terms of exponentially rare high-energy initial conditions. Indeed, the time to reach infinity shrinks to zero as one increases the energy. In the microcanonical ensemble the correlator should therefore be analytic at $t=0$, since the energy is bounded.

Since the one-matrix model is classically integrable, it does not describe a black hole. Nevertheless the probe limit is a robust feature of matrix mechanics~\cite{Ferrari2014GaugeBranes, Ferrari2014MatrixProbes}, so the analysis should generalize to chaotic systems as well. In the next section we will see that this is indeed the case.
\subsection{Zero-temperature singularities \label{zero-temp}}
It is instructive to identify the relevant smoothing scales in other models, where the scattering trajectories need not correspond to escaping branes. One regime which is especially numerically accessible is zero temperature. In the SYK$_p$ model, one can solve the zero-temperature Schwinger-Dyson equations near a singularity, finding \cite{Buric:2026qsp}
\begin{align}
    \langle \psi_1(-i\tau) \psi_1(0) \rangle \propto \frac{1}{(\tau+\tau_0)^{2/(p-2)}},\hspace{10 mm}\tau\sim -\tau_0.
\end{align}
\begin{figure}[t]
\centering
\includegraphics[width=\textwidth]{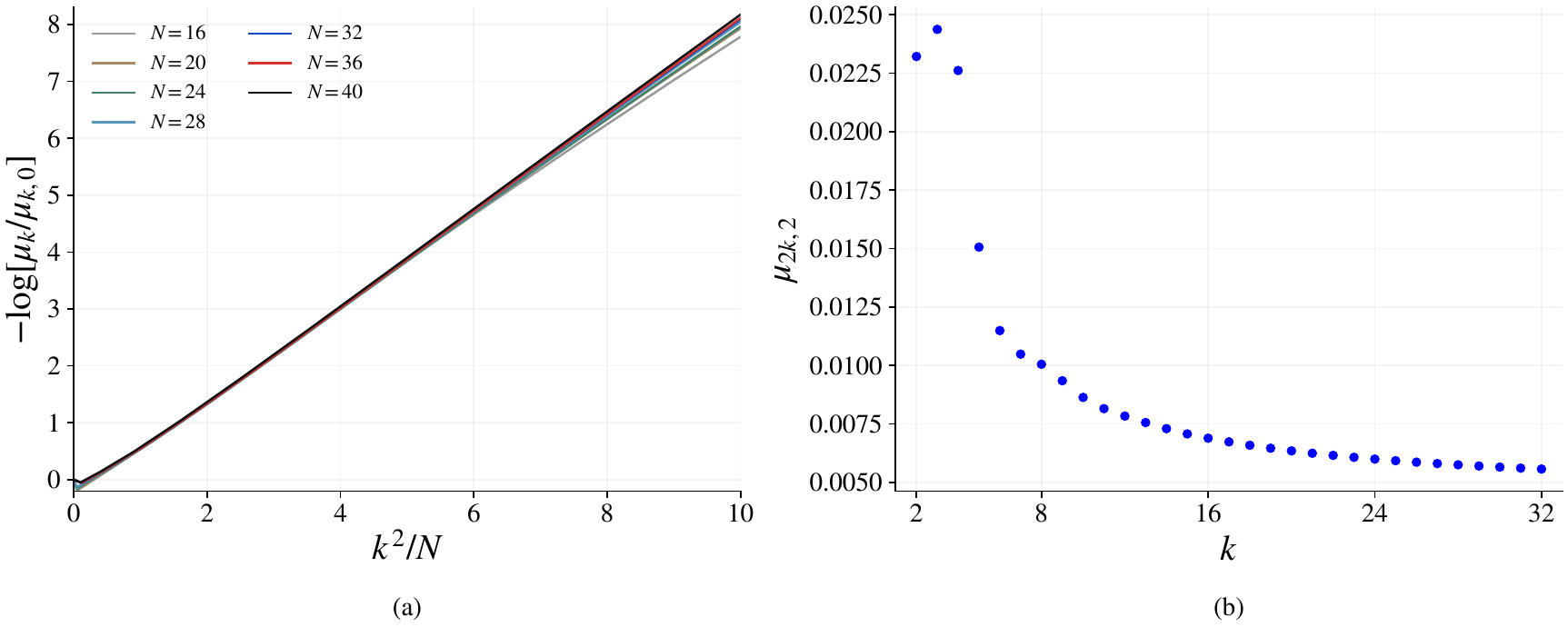}
\begingroup
\phantomsubcaption\label{fig:syk_zero_temperature_collapse}
\phantomsubcaption\label{fig:mqm_zero_temperature_one_loop}
\endgroup
\caption{(a) Moment ratios computed using exact diagonalization in the zero-temperature SYK$_4$ model, with points joined by a solid curve. The approximate collapse as a function of $k^2/N$ is consistent with a breakdown of the large-$N$ expansion at $k\sim N^{1/2}$. (b) The coefficient $\mu_{2k,2}$ in the zero-temperature one-matrix model at $\lambda/m^3=1$, computed through $k=32$. We compute this correction using the one-loop extension of the Marchesini–Onofri equation \cite{Marchesini:1979yq, Karczmarek2009Adjoint}, including corrections to the equilibrium density and inner product, and mixing with collective density excitations.}
\label{fig:syk_zero_temperature_moment_collapse}
\end{figure}%
Note that this is the same power that appears at any temperature \cite{Dodelson:2024atp,Buric:2026qsp}. Even at zero temperature, the singularity has a larger power than the conformal dimension $2 \Delta = 2/p$. The large-$N$ classical description involves the propagator $G$ and the self-energy $\Sigma$, with an action proportional to $N$. This scaling is due to the vector-like nature of the system, rather than an underlying D-brane description. We therefore again expect a smoothing scale $N^{-1/2}$; this is confirmed by a numerical computation of the moments, displayed in Figure~\ref{fig:syk_zero_temperature_collapse}, or by computing $1/N$ corrections to the two-point function by summing non-melonic diagrams~\cite{BenedettiGurau2018}\footnote{For example, the  leading correction to the two-point function is $\propto -N^{-1} (\tau+\tau_0)^{-2/(p-2)-2}$. Note that the first correction is negative and has a higher power; these facts are consistent with the singularity getting smoothed out at finite $N$. To obtain this correction, one can take a propagator inside a melon and replace it with a sum over ladders. The other class of diagrams comes from taking the four-point function (a sum over ladders) and closing its rails to make a vacuum diagram. Then one cuts open either a ``rung'' or a ``rail'' to define a contribution to the two-point function. At large $p$, one can also check this formula by expanding the finite $\lambda$ two-point function in DSSYK.}.

As the parameter $p$ increases, one eventually enters the double-scaling regime where $p^2 \to \infty, N \to \infty$, with the double-scaling parameter $2p^2/N = \lambda_\text{DSSYK}$ held fixed. In this limit, the correlator is under analytic control \cite{Berkooz:2018jqr} and the spectrum of the model is compact \cite{Cotler:2016fpe}. The asymptotic behavior of the moments is  
\begin{align}
\mu_k\propto k^{-3/2}(1-e^{-\lambda_\text{DSSYK}})^{-k/2},\hspace{10 mm}k\to \infty.%
\end{align}
Here the large-$N$ limit is taken before the large-$k$ limit. This expression has a natural interpretation in terms of chord diagrams \cite{Berkooz:2018jqr}, where the penalty factor for chords to cross is $e^{-\lambda_\text{DSSYK}}$.  %
Although these moments grow exponentially, dividing by the factorial of $k$ gives a convergent sum, so that the correlator $C(t)$ is entire. This is also the case at finite $N$, where the correlator is a finite sum over exponentials. For bosons this is no longer true, as we pointed out above.

In Appendix \ref{appb}, we show that for the one-matrix quantum mechanics, one again obtains a power-law divergence at zero temperature: %
\begin{align}
    \langle \text{Tr} ( X^J(- i \tau) X^J(0)) \rangle \sim  \frac{\mathcal{N}_J}{(\tau + \tau_0 )^{2J-1}}, \hspace{10 mm}\tau\sim -\tau_0,
\end{align}
where $\tau_0$ is given in \eqref{tau0general}; for the critical coupling we find $\tau_0 = \sqrt{2} \pi$. The coupling-independent exponent $2J-1$ is the same exponent obtained at high temperature from a delta-function scattering kernel, see \eqref{eq:scattering-pinch}.  The smoothing scale again turns out to be $\Delta \tau\sim N^{-1/2}$, as shown in Figure \ref{fig:mqm_zero_temperature_one_loop}. Near the critical coupling, one could hope to  understand this scale from the point of view of the string worldsheet. As discussed in Appendix \ref{appb}, the zero-temperature divergence near $\tau_0$ for the critical matrix quantum mechanics should be interpreted as the folded string becoming long; it seems plausible that the radius of convergence $\tau_0$ is independent of the topology of the worldsheet. Note that the analysis applies even when $\lambda<0$, in which case the model is ill-defined at finite $N$ due to metastability. In other words, we can use string perturbation theory to define $\tau_\text{breakdown}$ even in situations where the two-point function and moments are not defined at finite $N$.

Presumably the fact that the power of the divergence agrees at zero and high temperatures is related to the fact that the $\tau_0$ singularity probes very high energies. This can also be seen in the AdS black hole context, where the power of bouncing and bulk-cone singularities only depends on the conformal dimension of the external operator.

\FloatBarrier
\section{Multi-matrix models and classical chaos}\label{sec:multi-matrix}
We now turn to the bosonic $d$-matrix commutator model~\cite{Savvidy1984, MandalMahatoMorita2010}
\begin{align}
\label{eqn:2matrix-Hamiltonian}
H=\frac{1}{2}\sum_{a=1}^d \text{Tr}\left(P_a^2+m^2 X_a^2\right)-\frac{\lambda}{4N}\sum_{1\le a<b\le d}\text{Tr}\left([X_a,X_b]^2\right).
\end{align}
For $d=9$ and $m=0$ this is the bosonic sector of the ungauged BFSS matrix model~\cite{BanksEtAl1997, MaldacenaMilekhin2018}, which at low temperatures describes a black hole in Einstein gravity~\cite{ItzhakiEtAl1998, BiggsMaldacena2023, Maldacena2023BlackHole}. The gauge field is nondynamical and will not substantially change the analysis.  We are interested in the opposite high-temperature limit, which we can think of as a black hole with order-one string scale. In contrast to the single-matrix model, this system is chaotic. Our task in this section is to understand the novel elements of scattering introduced by classical chaos.

\subsection{Scattering states}
In the one-matrix case, the only singular configuration with no growing perturbations consists of a single eigenvalue running off to infinity. More general models admit a wider variety of stable scattering states. The equations of motion for the Hamiltonian (\ref{eqn:2matrix-Hamiltonian}) are 
\begin{align}
\ddot{X}_a=-m^2X_a-\frac{\lambda}{2N} [X_b,[X_b,X_a]].
\end{align}
Near a singularity, we can drop the mass term and use a simple pole ansatz to obtain an algebraic relation for the matrix-valued residue $A_a$,
\begin{align}\label{resequationsd}
X_a=\frac{A_a}{\epsilon},\hspace{10 mm}A_a=-\frac{\lambda}{4N} [A_b,[A_b,A_a]].
\end{align}
In particular, the matrices cannot commute near the pole.

Let us first specialize to the simplest case $d=2$, where the residue equations become
\begin{align}
A_1&=-\frac{\lambda}{4N}[A_2,[A_2,A_1]],\\
 A_2&=-\frac{\lambda}{4N}[A_1,[A_1,A_2]].
\end{align}
The solution is an embedded $SU(2)$ triple, 
\begin{align}\label{deq2res}
A_a=2i\sqrt{\frac{N}{\lambda}}T_a,\hspace{10 mm}[T_a,T_b]=i\epsilon_{abc}T_c.
\end{align}
It is straightforward to check that the spin-$j$ representation contains a growing resonance for $j\ge 1$, so that the only stable configuration is the spin-$1/2$ representation. We can think of this two-by-two block as a composite particle formed out of two D-branes connected by open strings.

In contrast to a single eigenvalue, this bound state has internal shape modes, which are characterized by the shape tensor $\text{Tr}(X_aX_b)$. One traceless deformation of the shape tensor is given by
\begin{align}
\delta X_1=\epsilon^{r-1}BT_1,\hspace{10 mm}\delta X_2=-\epsilon^{r-1}BT_2.
\end{align}
The linearized equation of motion gives a pair of complex resonances
\begin{align}
r_{\pm}=\frac{3}{2}\pm \frac{i\sqrt{7}}{2}.
\end{align}
Complex resonances cannot appear in a system obeying the Painlevé property, but are generically present in chaotic models.

We now extend the discussion to higher dimensions. For $d=3$ the configuration (\ref{deq2res}) remains a solution, with the residue $A_3$ set to zero. In order to analyze its stability, we look at the equations of motion for a small perturbation aligned with $T_3$,
\begin{align}
\ddot{B}\sim \frac{4}{\epsilon^2}B,\hspace{10 mm}X_3(t)=B(t)T_3.
\end{align}
This admits a growing solution $B\sim (t-t_0)^{-(\sqrt{17}-1)/2}$, signifying an instability. The natural endpoint of this instability is the Nahm pole solution~\cite{Nahm1980, GaiottoWitten2009Boundary} 
\begin{align}
X_a\sim i\sqrt{\frac{2N}{\lambda}}\frac{T_a}{\epsilon},\hspace{10 mm}a=1,2,3,
\end{align}
where the $T_a$'s satisfy the $SU(2)$ commutation relations as above. In fact, the Nahm pole in the spin-$1/2$ representation is stable for any $d\ge 3$. Numerical investigations suggest that this is the generic behavior of a solution near a pole.

The Nahm pole does not admit complex resonances, but fractional powers and logarithms do appear. Consider a deformation mixing the probe and the bath,
\begin{align}
X_a(t)
&=
\begin{pmatrix}
i\sqrt{\frac{2N}{\lambda}}\frac{T_a}{\epsilon}&
B_a(t)\\
C_a(t)&
0
\end{pmatrix}.
\end{align}
Since the solution is complex, we need not have $B_a^\dagger=C_a$. Plugging this ansatz into the equations of motion gives 
\begin{align}
\ddot{B}_a&=\frac{1}{\epsilon^2}\left(-\frac{B_a}{4}+3T_aT_bB_b\right),\\
\ddot{C}_a&=\frac{1}{\epsilon^2}\left(-\frac{C_a}{4}+3C_bT_bT_a\right).
\end{align}
In particular, imposing the polarization conditions 
\begin{align}
T_bB_b=C_bT_b=0,
\end{align}
we find the branched solutions
\begin{align}\label{nahmdef}
B_a(t)=\sqrt{\epsilon}b_a\log \left(\frac{\epsilon}{\epsilon_b}\right),\hspace{10 mm}C_a(t)=\sqrt{\epsilon}c_a\log \left(\frac{\epsilon}{\epsilon_c}\right).
\end{align}
The logarithm leads to infinitely many branches, which is what we found in the $d=2$ case. We will explore the implications of this infinite branching later in this section.

\begin{figure}[t]
\centering
\includegraphics[width=0.86\textwidth]{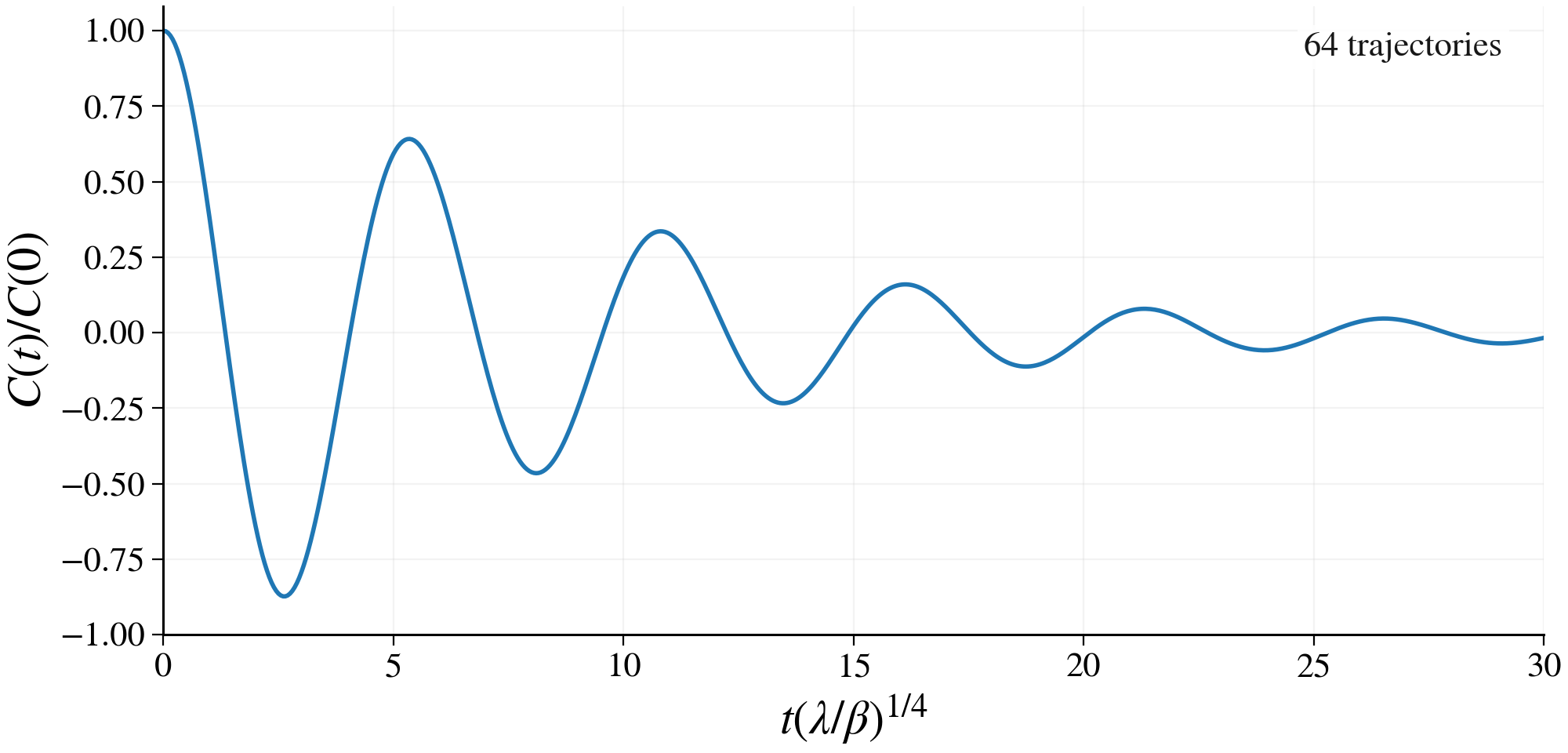}
\caption{Normalized real-time correlator in the classical two-matrix model at $N=16$ and $\lambda/(\beta m^4)=1$, computed using Monte Carlo simulation. The correlator exhibits oscillations with an approximately exponential envelope over the displayed time range.}
\label{fig:two_matrix_realtime}
\end{figure}

\subsection{Correlation functions}
Next we consider correlation functions in multi-matrix models, specializing to the simplest nontrivial case of two matrices $X$ and $Y$. In addition to the adjoint $SU(N)$ symmetry, the model has a global symmetry which rotates $X$ into $Y$. It is useful to diagonalize this $U(1)$ action by introducing
\begin{align}
Z=X+iY,\qquad \overline{Z}=X-iY.
\end{align}
The charged two-point functions vanish in the thermal ensemble. We will study the neutral correlator
\begin{align}
C(t)=\frac{1}{N^2}\left\langle \text{Tr}\bigl[\overline{Z}(t)Z(0)\bigr]\right\rangle_\beta,
\end{align}
which is displayed in Figure~\ref{fig:two_matrix_realtime}. The observed exponential ringdown behavior is characteristic of chaotic systems.

The moment calculation is a direct generalization of the one-matrix construction, 
\begin{align}
\mu_{2n}=\frac{(-1)^n}{N^2}\left\langle \text{Tr}\left(\overline{Z}\mathcal{L}^{2n}Z\right)\right\rangle_\beta ,
\end{align}
where $\mathcal{L}^{2n} Z$ is a linear combination of noncommutative words in $Z,\overline{Z},$ and the conjugate momenta. The difference from the single-matrix case is that $X$ and $Y$ do not commute, except in the strong-coupling limit~\cite{BerensteinHanadaHartnoll2009}. Nevertheless one can solve the matrix integral exactly in the planar limit for any coupling~\cite{KazakovKostovNekrasov1999}. Indeed, since the path integral is Gaussian in $Y$, we can integrate it out exactly. Diagonalizing $X$, one then finds that the planar eigenvalue density of $X$ satisfies an integral equation \cite{BerensteinHanadaHartnoll2009},
\begin{align}
\beta m^2 x
=
2\,\operatorname{P}\!\int ds\,\rho_X(s)
\Bigl[
\frac{1}{x-s}
-
\frac{\frac{\lambda}{2N}(x-s)}
{m^2+\frac{\lambda}{2N}(x-s)^2}
\Bigr],
\end{align}
with the normalization $\int dx\, \rho_X(x)=N$. Here P denotes the principal value. This equation admits an exact solution in terms of elliptic functions, but can also be solved numerically to high accuracy. Computing up to $\mu_{18}$ and Padé approximating then gives an estimate $\beta_0m\approx 5.6$ for $\lambda/(\beta m^4)=1$. As a consistency check, we computed the maximal Lyapunov exponent $\lambda_L\approx 0.12 m$, finding that the chaos bound (\ref{chaosbound}) is satisfied.

\begin{figure}[t]
\centering
\includegraphics[width=\textwidth]{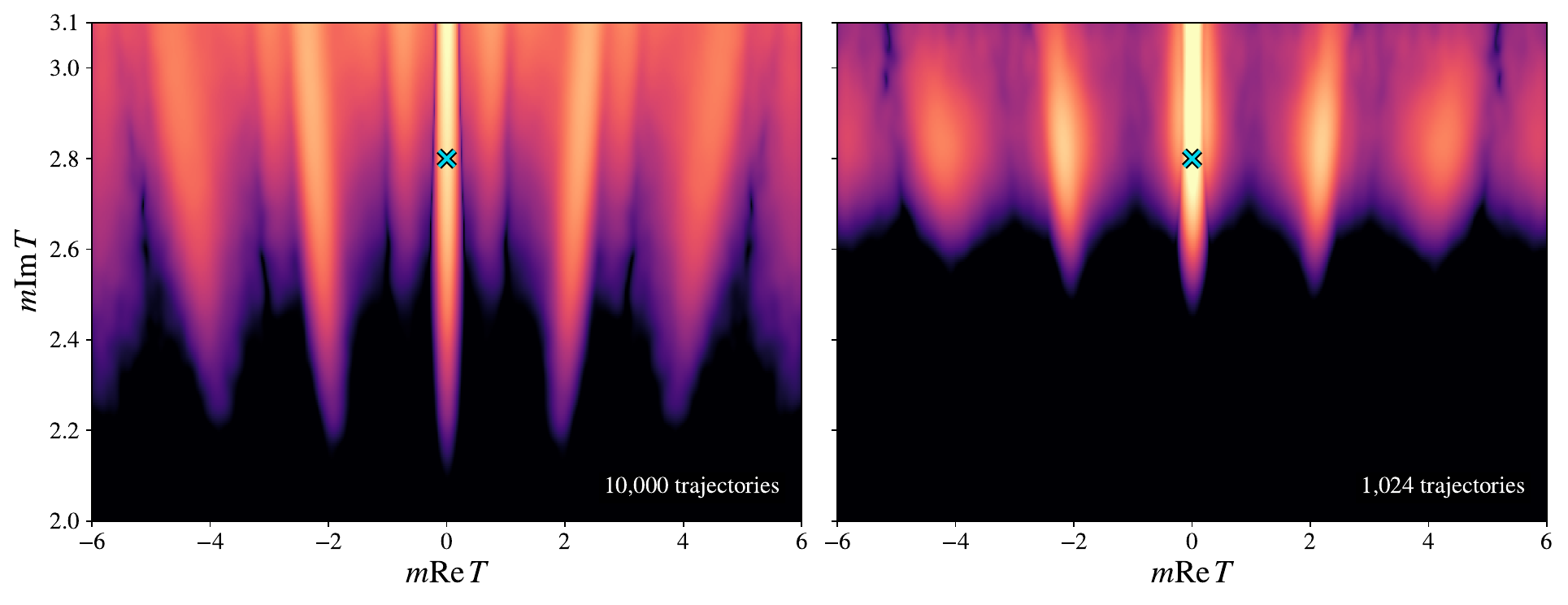}
\caption{Reduced scattering kernel $\tilde{S}_A(T)$ for the correlator $\langle\text{Tr}(\overline{Z}(t)Z(0))\rangle_\beta$ in the two-matrix model at $\lambda/(\beta m^4)=1$, with $N=4$ (left) and $N=32$ (right). We only include singularities on the first sheet. Cyan crosses mark the planar estimate for the leading singularity, with $m\beta_0\approx5.6$\,.}
\label{fig:two_matrix_weighted_scattering_kernel}
\end{figure}

\begin{figure}[t]
\centering
\includegraphics[width=\textwidth]{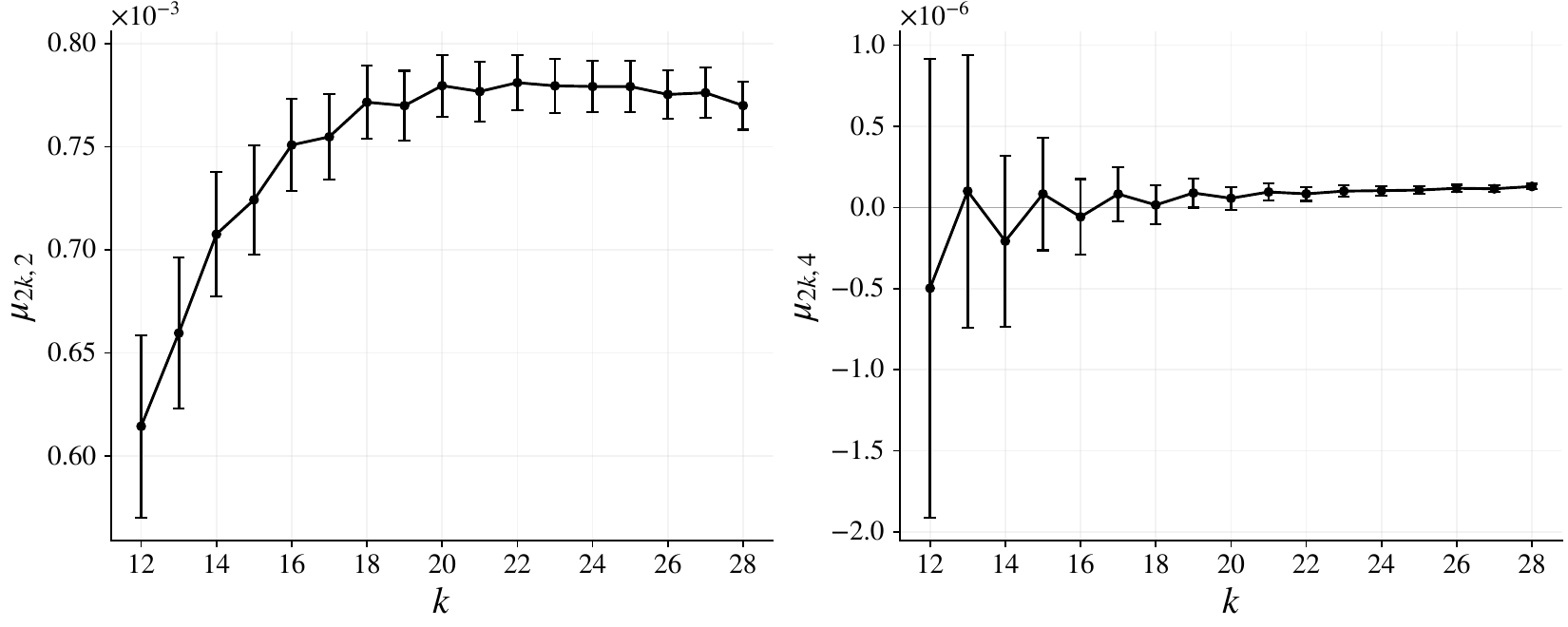}
\caption{Nonplanar coefficients $\mu_{2k,2}$ (left) and $\mu_{2k,4}$ (right) in the two-matrix model at $\lambda/(\beta m^4)=1$, estimated from Monte Carlo moments for $N=24,32,40,48,64,80,96$. The genus expansion (\ref{muplanar}) through $1/N^4$ is fitted independently at each $k$, with the planar moment also fitted. }
\label{fig:two_matrix_nonplanar_moments}
\end{figure}

Since the complexity of long operators grows faster in this model than in the single-matrix model, we were not able to estimate off-axis singularities of correlation functions from the moment method. Instead, we will rely on the scattering kernel approach. Figure~\ref{fig:two_matrix_weighted_scattering_kernel} shows the expected localization of the scattering kernel at large $N$ onto isolated points, with the $\beta_0$ singularity matching the moment method. Performing a Monte Carlo simulation of the moments, we find that the planar limit again breaks down at a scale $k^4\sim N^2$ (see Figure~\ref{fig:two_matrix_nonplanar_moments}). This points to a universal mechanism behind the singularity resolution, which we leave as an open question.
\FloatBarrier
\subsection{Monodromy and high-sheet universality}

\begin{figure}[t]
\centering
\includegraphics[width=\textwidth]{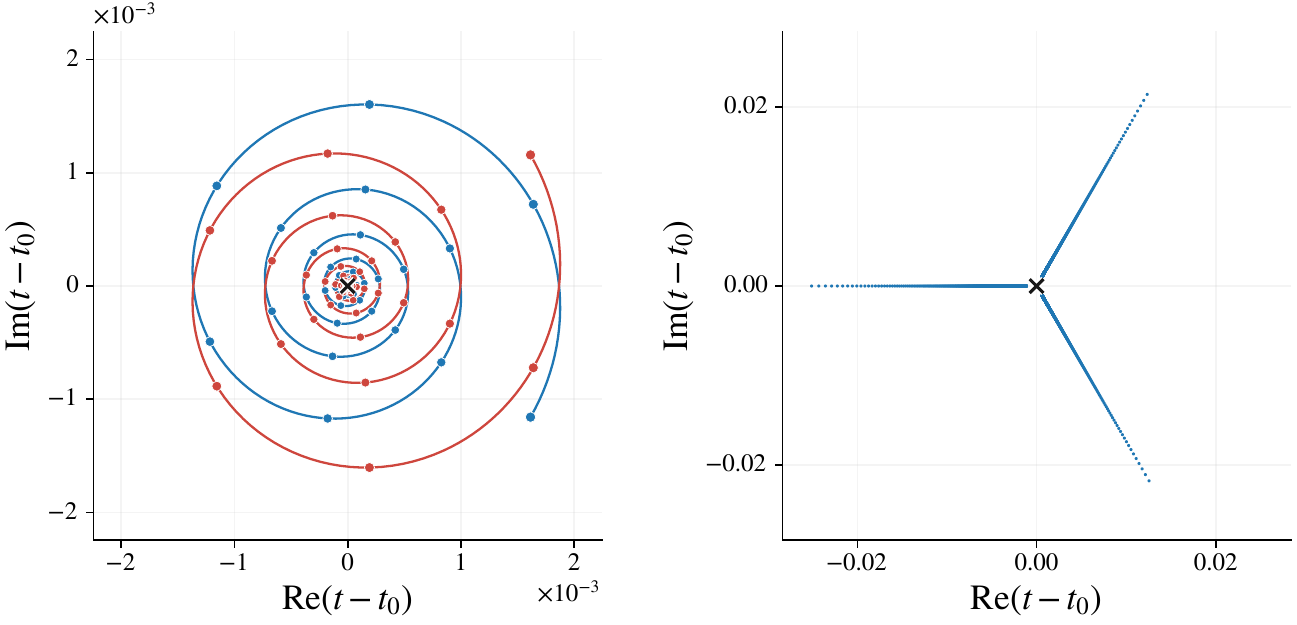}
\par
\begin{subfigure}[t]{0.49\textwidth}
\centering
\caption{}\label{fig:high-sheet-d2-singularities}
\end{subfigure}\hfill
\begin{subfigure}[t]{0.49\textwidth}
\centering
\caption{}\label{fig:high-sheet-nahm-singularities}
\end{subfigure}
\caption{High-sheet singularity patterns projected onto the complex time plane, with $u_n=1$. (a) The spiral family~\eqref{eq:high-sheet-d2-locations} and its clockwise counterpart, for $m\ge 60$. We use the illustrative resonances $r_\pm=6\pm0.6i$ to make successive turns visible; the physical values $(3\pm i\sqrt{7})/2$ give much faster contractions. (b) The three radial families~\eqref{eq:high-sheet-nahm-locations} for selected windings $|K|\ge 40$.}
\label{fig:high-sheet-matrix-singularities}
\end{figure}

Since classical trajectories in chaotic systems have branch cuts, the scattering kernel depends on the Riemann sheets of the initial and final times. Winding around a branch point therefore acts non-trivially on the scattering kernel. We now investigate the action of the monodromy group on classical trajectories.

Consider a singularity with scattering data $(t_n,\mathbf{r}_n)$. The resonances $\mathbf{r}_n$ transform under monodromy around $t_n$.
For example, the shape-mode deformation in $d=2$ takes the form
\begin{align}\label{shapemode}
\delta X^{\pm}_1=\epsilon^{r_\pm-1}B_{\pm}T_1,\hspace{10 mm}
\delta X^{\pm}_2=-\epsilon^{r_\pm-1}B_{\pm}T_2.
\end{align}
Taking $\epsilon\mapsto e^{2\pi i}\epsilon$ induces a transformation
\begin{align}
B_{\pm}\mapsto -e^{\mp\pi \sqrt{7}}B_{\pm}.
\end{align}
Winding many times exponentially amplifies $B_-$, while suppressing $B_+$. We therefore expect the solution to simplify on high sheets.

On high Riemann sheets, the parameter $u\equiv \epsilon^{r_-}$ can become large even though $|\epsilon|$ is small. After winding $K$ times, $B_-$ is amplified by $e^{\pi\sqrt{7}K}$,  so the shape mode competes with the leading pole when $e^{\pi \sqrt{7}K}|u|$ becomes order one. This suggests a resummation of the form
\begin{align}
X_1(t)=\frac{1}{\epsilon}x(u)T_1,\hspace{10 mm}X_2=\frac{1}{\epsilon}y(u)T_2.
\end{align}
Identifying $X_a$ with a gauge field, the resulting equations of motion describe the spatially homogeneous sector of classical $SU(2)$ Yang-Mills theory~\cite{Savvidy1984}. We find
\begin{align}\label{highsheeteqd2}
r_-^2u^2x''-4ux'+2x&=-\frac{\lambda}{2N}xy^2,\\
r_-^2u^2y''-4uy'+2y&=-\frac{\lambda}{2N}x^2y.
\end{align}

In order to pick out the relevant trajectory for the resummation, we make the analytic ansatz
\begin{align}\label{merosol}
x=\sum_{n=0}^{\infty}x_nu^n,\hspace{10 mm}y= \sum_{n=0}^{\infty}y_n u^n,
\end{align}
with the initial conditions
\begin{align}
x_0=y_0=2i\sqrt{\frac{N}{\lambda}},\hspace{10 mm}x_1=-y_1=B_-.
\end{align}
The equations of motion reduce to the recursion relation
\begin{align}
\begin{pmatrix}r_-^2n(n-1)-4n&-4\\-4&r_-^2n(n-1)-4n\end{pmatrix}\begin{pmatrix}x_n\\y_n\end{pmatrix}=-\frac{\lambda}{2N}\sum\limits_{\substack{i+j+k=n\\0\le i,j,k<n}} \begin{pmatrix}
 x_i y_j y_k
\\
x_i x_j y_k
\end{pmatrix}.
\end{align}
For $n>1$, the matrix on the left-hand side is invertible, so $x_n$ and $y_n$ are uniquely determined by the lower-order coefficients. This resummation captures the behavior of the solution after winding $K$ times counterclockwise around $t_n$. The clockwise analysis follows by replacing $r_-$ with $r_+$.

Let $u_n$ be the singularities of the resummed trajectory. Since the solution is analytic in $u$, a fixed singularity $u_n$ leads to an infinite family
\begin{align}\label{eq:high-sheet-d2-locations}
t_{nm}\approx t_0+e^{2\pi i m/r_-}u_n^{1/r_-},\hspace{10 mm}m\to \infty.
\end{align}
The singularities in this scaling limit are completely universal, except for the single undetermined complex number $B_-$ which sets the overall scale and rotational orientation. The projection of $t_{nm}$ down to the complex plane forms the spiral shape shown in Figure~\ref{fig:high-sheet-d2-singularities}, which is invariant under a discrete complex scaling transformation~\cite{ChangEtAl1983, FournierLevineTabor1988}. One can think of this as a complex renormalization group transformation with scaling functions $x(u)$ and $y(u)$. The high-sheet behavior then describes a renormalization group fixed point, exhibiting self-similarity under complex scaling~\cite{LevineTabor1988}.

We can perform a similar resummation on high sheets near the Nahm pole in $d\ge3$. The relevant expansion parameter in this case is 
\begin{align}
u=\epsilon^{3/2}\log \epsilon,
\end{align}
so the resummation becomes important in the regime $|\epsilon|^{3/2}K\sim 1$. The leading pole satisfies the Nahm equations, which govern the self-dual sector of Yang-Mills theory \cite{MasonWoodhouse1996},
\begin{align}
\dot{X}_a=\frac{1}{2}\sqrt{\frac{\lambda}{2N}}\epsilon_{abc}[X_b,X_c].
\end{align}
These equations are known to be integrable~\cite{Hitchin1983Monopoles}, whereas the full equations of motion are chaotic.

In fact, on high sheets $K\gg 1$ the linearized deformation (\ref{nahmdef}) continues to satisfy this equation to leading order in $1/K$. We therefore look for a nonlinear self-dual resummation of the form 
\begin{align}
X_a=\frac{1}{\epsilon}\sum_{n=0}^{\infty}X_{a,n}u^n,
\end{align}
with the initial conditions
\begin{align}
X_{a,0}=i\sqrt{\frac{2N}{\lambda}}T_a,\hspace{10 mm}X_{a,1}=\begin{pmatrix}0&b_a\\c_a&0\end{pmatrix}.
\end{align}
The resulting recursion relation is 
\begin{align}
\left(\frac{3n}{2}-1\right)X_{a,n}-i\epsilon_{abc}[T_b,X_{c,n}]=\sqrt{\frac{\lambda}{8N}}\sum_{m=1}^{n-1}\epsilon_{abc}[X_{b,m},X_{c,n-m}].
\end{align}
The left-hand side defines a linear operator acting on $X_a$. It is straightforward to check that this operator is invertible for $n>1$, so that the equation admits a consistent self-dual resummation on high sheets. The projection of a given singularity down to the complex plane is
\begin{align}\label{eq:high-sheet-nahm-locations}
t_{njK}\approx t_0-e^{2\pi i j/3}\left(\frac{u_n}{2\pi K}\right)^{2/3},\hspace{10 mm} |K|\to \infty,\hspace{10 mm}j=0,1,2,
\end{align}
which is displayed in Figure~\ref{fig:high-sheet-nahm-singularities}.

We have seen that classical trajectories often simplify after winding many times around a singularity, sometimes even becoming integrable like in the Nahm example. The expectation is that the high-sheet behavior of the classical solutions controls the correlation function on high sheets. Assuming that this is the case, the winding procedure admits a simple bulk interpretation. In the holographic limit, the correlator is computed using a spacelike geodesic with turning point \cite{Fidkowski2004}
\begin{align}
r_{\text{min}}\sim (t-t_0)^{2/(D-3)},
\end{align}
where $t_0$ is the location of the bouncing singularity in the correlator. The high-sheet limit is therefore controlled by a null geodesic which winds many times around the black hole singularity. Our results suggest that the geometry simplifies in this regime. It would be interesting to understand the bulk mechanism underlying this simplification. The recently identified gravitational dual of the self-dual sector of $\mathcal{N}=4$ SYM \cite{SharmaSkinner2026Chiral} might be relevant for this question.
\FloatBarrier
\begin{figure}[t]
\centering
\includegraphics[width=\textwidth]{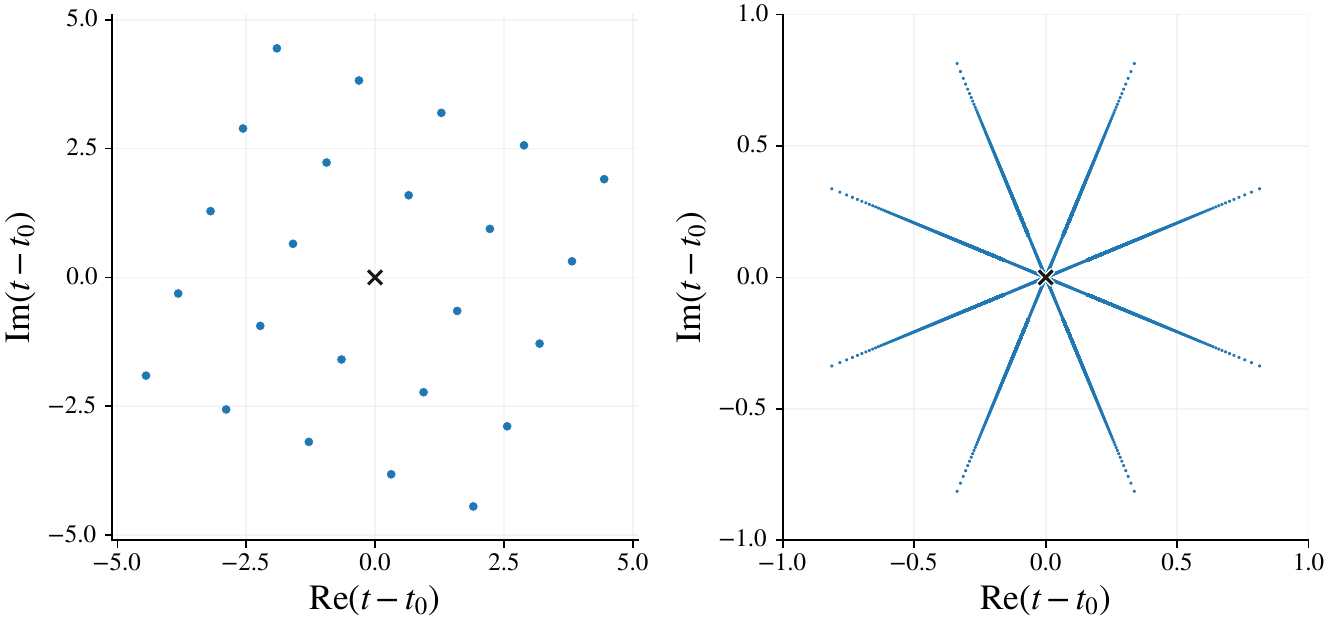}
\caption{High-sheet descendant singularities obtained from the leading-logarithm resummation. Left: the lattice of descendant singularities in the shifted coordinate $t-t_0$ on a fixed high sheet, shown here for $K=20$. Right: the closest poles from the asymptotic lattice formula (\ref{latticeep}), projected onto the complex time plane for selected windings $|K|\ge 6$, showing their accumulation at $t=t_0$.}
\label{fig:syk_high_sheet_k20}
\end{figure}

\subsection{Emergent integrability in SYK}\label{sykint}

In order to rigorously demonstrate that singularities on high sheets of trajectories correspond to actual singularities of the correlation function, one would have to deform the $T_i$ and $T_f$ contours in (\ref{cint}) as $t$ winds around a branch point of $C(t)$. This is beyond the scope of the present work. As a proof of concept, let us instead discuss the SYK model, where we can explicitly study $C(t)$ on high sheets. For the four-body SYK model, the Schwinger-Dyson equation for the two-point function of $\sqrt{2}\psi(t)$ at infinite temperature takes the form \cite{Maldacena:2016hyu}
\begin{align}\label{sykvolterra}
C'(t)=-\frac{\mathcal{J}^2}{2}\int_0^t dt'\, C(t-t')^3C(t'),\qquad C(0)=1.
\end{align}
This equation admits an unambiguous analytic continuation to complex time, provided the contour stays on a fixed sheet. This gives a concrete nonlinear problem whose singularity structure can be studied directly.

Let us first analyze an isolated singularity at $t=t_0$, following \cite{Buric:2026qsp}. Making the ansatz
\begin{align}
C(t)\sim \frac{A}{\epsilon ^{\gamma}},
\end{align}
one finds that the $t'=0$ endpoint of the convolution contributes $\epsilon^{1-3\gamma}$, while the derivative contributes $\epsilon^{-1-\gamma}$. Matching the powers gives $\gamma=1$, together with the coefficient equation
\begin{align}
A=\pm\frac{2i}{\mathcal{J}}.
\end{align}
Thus the leading singularity is a simple pole. The prediction for the residue matches the numerical result derived from Padé approximation of the moment expansion.

We can now systematically compute the series expansion of the correlator around the singularity using an endpoint analysis of (\ref{sykvolterra}). Near the endpoints, one factor in the integrand is regular, and one is singular. For the regular term we substitute the moment expansion 
\begin{align}
C(t)=\sum_{n=0}^{\infty}\mu_{2n}\frac{(it)^{2n}}{(2n)!}.
\end{align}
For the singular term we make a Laurent expansion ansatz for $C(t)$ around $t=t_0$, allowing for potential logarithms. Solving for the expansion coefficients then gives 
\begin{align}\label{seriesexp}
C(t)\sim \pm \frac{2i}{\mathcal{J}\epsilon}\left[1+\frac{\mathcal{J}^2}{6}\epsilon^2+B\epsilon^3+\frac{5\mathcal{J}^4}{24}\epsilon^4(\log \epsilon+D)+\cdots\right].
\end{align} 
The terms of order $1/\epsilon,\epsilon$, and $\epsilon^3\log \epsilon$ yield nonanalytic endpoint contributions, and can therefore be fixed by a local analysis. The other coefficients require a global matching. Just like in the models studied earlier in this section, the solution is pole-like at leading order but not globally meromorphic.

In the scaling limit where $\epsilon^4\log \epsilon$ is held fixed at small $|\epsilon|$, the equation of motion (\ref{sykvolterra}) is dominated by the lower endpoint, and simplifies to 
\begin{align}\label{lowersimp}
C''(t)=-\frac{\mathcal{J}^2}{2}C(t)^3,
\end{align}
which is the equation for a particle in a quartic potential. Although the full equations of motion admit no obvious constant of motion, the reduced equation (\ref{lowersimp}) is integrable, with an approximately conserved energy emerging on high sheets of the Riemann surface. The energy can be determined by expanding around $t=t_0$, which yields
\begin{align}
C'^2+\frac{\mathcal{J}^2}{4}C^4=\frac{50\pi i K\mathcal{J}^2}{3}.
\end{align}
We are finally left with the complete leading-logarithm resummation
\begin{align}\label{snresum}
C(t)=\pm \frac{2 i\left(\frac{25\pi i K}{6}\right)^{1/4}}{\text{sn}\left[\left(\frac{25\pi i K}{6}\right)^{1/4}\mathcal{J}\epsilon,-1\right]}.
\end{align}
 In particular, the singularities on the $K$th Riemann sheet are on a regular lattice
\begin{align}\label{latticeep}
t_{mnK}=t_0+\left(\frac{96}{25\pi i K}\right)^{1/4}\frac{(m+in)K(-1)}{\mathcal{J}},\hspace{10 mm}|m|,|n|\ll |K|^{1/4}.
\end{align}

One can repeat the same analysis for $p>4$ SYK. Curiously, in this case one finds neither logarithms nor complex resonances in the expansion, so there is no emergent high-sheet limit. It would be interesting to understand what makes $p=4$ special.

\FloatBarrier
\section{Beyond high temperatures}
\label{sec:beyondhighT}
\begin{figure}[t]
    \centering
\begin{tikzpicture}[x=1cm,y=0.45cm,font=\small]
  \draw[->] (-1.65,0) -- (5.9,0) node[right] {$\tau$};

  \draw[dashed,gray] (-1,0) -- (-1,6.3);
  \draw[dashed,gray] (5,0) -- (5,6.3);
\draw[dotted,gray] (0,0) -- (0,6.3);
\draw[dotted,gray] (4,0) -- (4,6.3);
  \draw[densely dashed,blue!50] (2,0) -- (2,6.3);

  \draw[<->,thick,blue!70!black,
        domain=-0.72:4.72,samples=121]
    plot (\x,{0.25+1/(1-((\x-2)/3)^2)});
  \node at (0.2,4.5) {$C(-i\tau)$};
  \fill[blue!70!black] (2,1.25) circle (1.7pt);
  \node[above right,blue!70!black,inner sep=1.5pt]
    at (2.1,2.25) {$u=0$};
  \foreach \t/\lab in {
    -1/{-\tau_0},0/{0},2/{\beta/2},
    4/{\beta},5/{\beta+\tau_0}} {
    \draw (\t,0.12) -- (\t,-0.12);
    \node[below] at (\t,-0.12) {$\lab$};
  }
  \draw[<->] (-1,-2) -- (5,-2)
    node[midway,fill=white,inner sep=2pt]
    {$\beta_0=\beta+2\tau_0$};
\end{tikzpicture}
    \caption{Definitions of $\tau_0$ and $\beta_0$. $\tau_0$ is the radius of convergence of the two-point function $C(-i\tau)$ around $\tau=0$. The first singularity is on the Euclidean time axis at some negative location $\tau = -\tau_0$. We define $\beta_0 = \beta+ 2 \tau_0$. %
    The two-sided correlator $C_{\text{LR}}(-iu)$ starts at $u=0$, the reflection-symmetric point. The first singularity it encounters in Euclidean time is at $u= \pm (\beta/2 +\tau_0)$. }
    \label{fig:placeholder}
\end{figure}
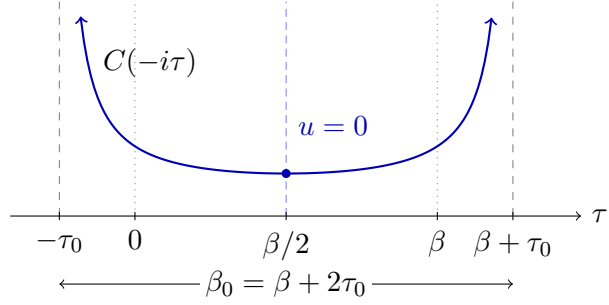

Most of our discussion so far has been strictly in the high-temperature limit, where classical statistical mechanics suffices. The reader might be worried that the complex-time singularities we have identified are an artifact of this limit. Away from high temperatures, the problem is fully quantum and requires new tools (for the one-matrix quantum mechanics and for SYK, one can also analytically analyze the theory at zero temperature, as we did in Section \ref{zero-temp}, but these seem to be rather special cases). One promising non-perturbative method available at finite temperature is the thermal bootstrap~\cite{Fawzi:2023fpg, ChoGabaiSandorYin2024, ChoEtAl2025Euclidean}. In order to give a proof of principle that the bootstrap can be used to extract complex-time singularities of correlation functions, we will study the zero-temperature problem~\cite{HanHartnollKruthoff2020, Lin2023D0, LinZheng2025Ground}. We will also present results on the finite-temperature case, but leave the full treatment to future work \cite{fouriermethod}.

Concretely, we propose two bootstrap approaches for the zero-temperature problem: Section \ref{sec:hybrid-moment-expansion} discusses a method based on the moment expansion discussed in Section \ref{sec:momentexpansion}, and Sections \ref{sec:time-bootstrap} and \ref{sec:frequency-bootstrap} present rigorous bounds on the leading complex-time singularity by bootstrapping the Euclidean two-point function directly \cite{ChoEtAl2025Euclidean}.  
Besides bootstrap, Section \ref{sec:wigner} and Appendix \ref{sec:wigner2} discuss the corrections to the high-temperature expansion. 

In what follows we will find it convenient to work with the two-sided correlator $C_{\text{LR}}(t) = C(-i \beta/2+t)$, which arises naturally in the thermofield double state.
This correlator has the property that it is even in $t$ and real. Note that in the high-temperature limit  $\beta \to 0$, there is no difference between the one-sided correlator $C(t)$ and the two-sided correlator $C_{\text{LR}}(t)$. However, the classical statistical mechanics approximation conveniently has the property that it leads to real and symmetric correlators.
Its singularities are thus at $t=\pm i\beta_0/2$. Away from high temperatures, the moment expansion of $C_{\text{LR}}$ takes the form 
\begin{align}
C_{\text{LR}}(t) = C\left(-\frac{i\beta}{2}+t\right)
=\sum_{n=0}^\infty\frac{\bar\mu_{2n}}{(2n)!}(it)^{2n}.
\label{eq:wk-centered-expansion}
\end{align}
At small $\beta$, we can Taylor expand $C(t) = C_{\text{LR}}(t) + i \beta C'_{\text{LR}}(t)/2  -  \beta^2 C''_{\text{LR}}(t)/8$. This implies %
\begin{align}
\bar\mu_{2n}
=\mu_{2n}-\frac{\beta^2}{8}\mu_{2n+2}
+O(\beta^4\mu_{2n+4}).
\label{eq:wk-midpoint-conversion}
\end{align}
Only the leading-order (classical) value of $\mu_{2n+2}$ is needed in the second term.
For the quartic model, the omitted terms are suppressed by an additional power of $\beta^3$.

\subsection{Hybrid moment expansion approach}
\label{sec:hybrid-moment-expansion}

The moment expansion method determines $\beta_0$ via \eqref{muasympt}. In the high-temperature case, the problem of calculating the moments reduces to computing $\langle \text{Tr}(X^k)\rangle$ using the analytic solution of the matrix integral. In the case of zero temperature, we again have an analytic formula available for computing $\langle \text{Tr}(X^k)\rangle$ \cite{bipz}. Note that there is a slight difference from the high-temperature case for words with both $X$ and $P$: instead of integrating out $P$, we can rewrite them as words with only $X$ using the gauge singlet condition {on the vacuum state} and the condition $\langle [H,\mathcal{O}]\rangle=0$.  Furthermore, the odd moments vanish in the high-temperature limit, but this is not true in general at finite temperature.
\begin{figure}[t]
    \centering
\includegraphics[width=\linewidth]{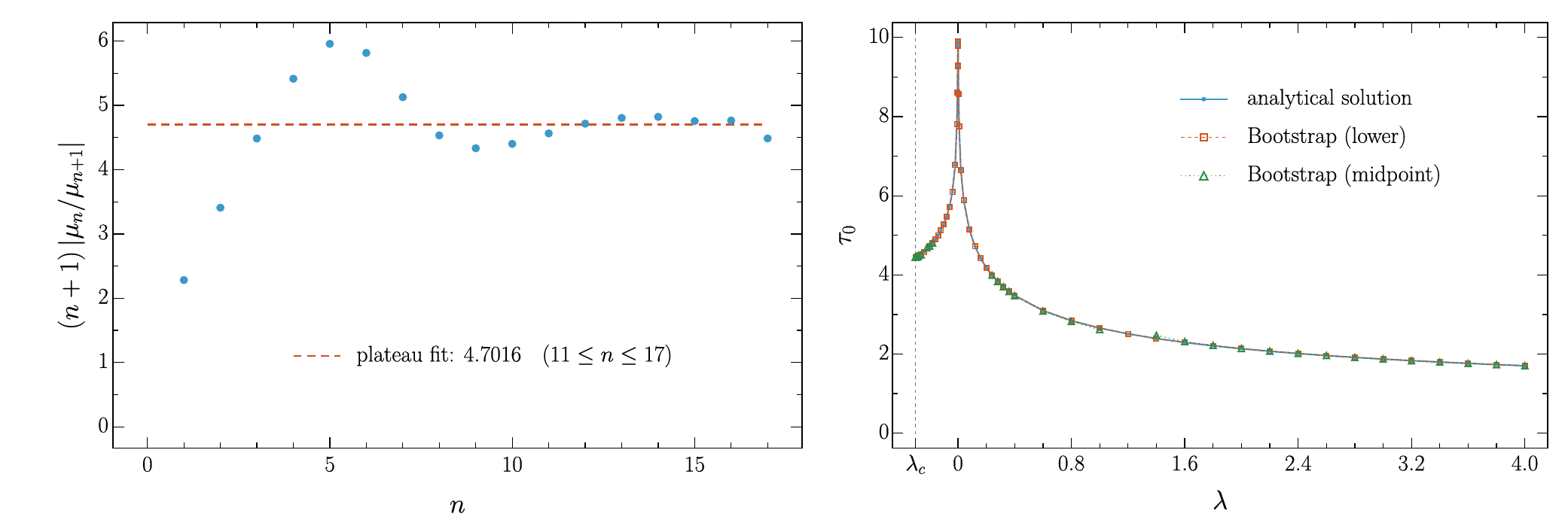}
    \caption{Left: Rescaled moment ratios $(n+1)|\mu_{n}/\mu_{n+1}|$ as a function of $n$ at $\lambda=-0.2$ and $m=1$ in the one-matrix quantum mechanics. We show the estimate of $\tau_0(\lambda) \approx 4.70$ from fitting the location of the converging plateau using the last seven ratios. The moments displayed here are numerically computed from the solution in \cite{bipz}. Right: The zero-temperature function $\tau_0(\lambda)$, obtained by Pad\'e approximation at different couplings. We show the results from computing moments via analytic solution (solid curve, see \eqref{tau0general}) and matrix bootstrap (orange and green dots, corresponding to the lower bounds and midpoints of the bootstrap, respectively). }
\label{fig:beta0zerotemp}
\end{figure}

Let $t=i\tau_0$ be the closest singularity to $t=0$ at zero temperature.  From \eqref{muasympt}, the rescaled moment ratio $(n+1) |\mu_n/\mu_{n+1}|$ approaches $\tau_0$ as $n \rightarrow \infty$. Since the computational cost of $\mu_n$ increases with $n$, it is useful to truncate to finite $n$. In practice, we found that the rescaled moment ratio already stabilized to an approximately constant value at some finite $n \gtrsim 10$. The left panel of Figure \ref{fig:beta0zerotemp} plots the rescaled moment ratio as a function of $n$ at $\lambda=-0.2$, numerically computed using the planar solution \cite{bipz}. {Here we are working in units $m=1$.} This shows a typical scenario of a converging plateau already at finite $n$. We obtain $\tau_0 \approx 4.70$ at this coupling by a horizontal fit with the last few ratios. Alternatively, the Pad\'e approximation gives the first pole at $\tau_0 \approx 4.73$.

Although an analytic solution is available in our current zero-temperature case, we demonstrate that $\tau_0$ can be extracted using moments computed with matrix bootstrap as a proof of principle. Specifically, the matrix bootstrap gives two-sided bounds on $\langle \text{Tr} (X^k)\rangle$, and the bounds become tighter as the bootstrap level $\Lambda$ increases. We can then estimate the moments using, for example, the average of the lower and upper bounds. In many cases, we found that the lower bounds almost converge as $\Lambda$ increases, and therefore arguably provide a better proxy for the actual value. 

We show in the right panel of Figure \ref{fig:beta0zerotemp} the function $\tau_0(\lambda)$ using the Pad\'e approximation. The solid curve shows the analytic formula \eqref{tau0general}, and the orange and green dots show the results using the lower bounds and midpoints of the bootstrap, respectively. For all values of $\lambda$, we compute $\mu_n$ up to $n=14$ and perform a horizontal fit for the last three rescaled moment ratios. The divergence at $\lambda=0$ is consistent with the non-chaotic behavior in the free model. The bootstrapped $\tau_0(\lambda)$ exhibits good agreement with the analytic formula. In particular, we obtain $\tau_0\approx 4.4$ at the critical coupling $\lambda_c = -2\sqrt{2}/(3\pi)$, where the analytic prediction gives $\tau_0(\lambda_c) = \sqrt{2}\pi$. This agreement provides confidence in extending this method to the finite-temperature case, where an analytic expression is not available and moments have to be computed numerically. Lastly, we also check that our analytic formula for $\tau_0(\lambda)$ is consistent with the results obtained from using the moments numerically computed from \cite{bipz}.

\begin{figure}[t]
    \centering
\includegraphics[width=0.7\linewidth]{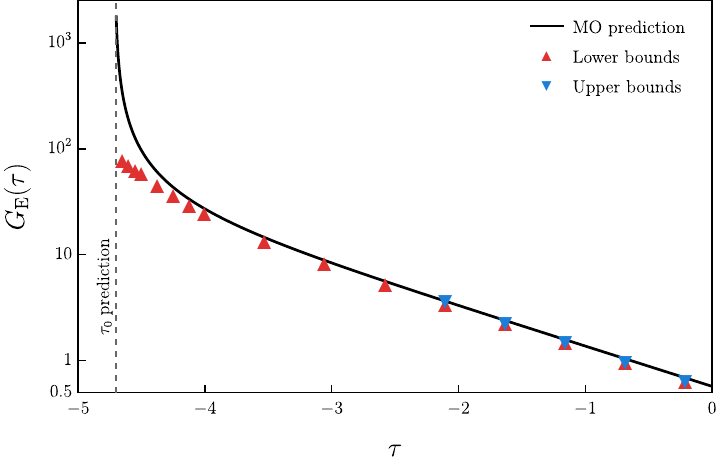}
    \caption{Bootstrap bounds on $G_E(\tau)$ for one-matrix quantum mechanics with the  potential in \eqref{eq:mqmhamiltonian} with $\lambda=-0.2$ from the time-dependent bootstrap in Section \ref{sec:time-bootstrap}, at zero temperature. The figure is obtained at level $\Lambda=10$ and spline degree $d=14$. The black curve shows the two-point function from numerically solving the Marchesini--Onofri equation. The moment expansion gives $\tau_0 \approx 4.7$. The most negative Euclidean time for which we obtained an upper bound is at $\tau = -2.1$, which gives a rigorous lower bound $\tau_0(\lambda =-0.2)\geq 2.1$.}
\label{fig:negativetimebootstrap}
\end{figure}

\subsection{Time-dependent bootstrap}
\label{sec:time-bootstrap}
Let us now turn to an alternative numerical approach for extracting $\tau_0$, by applying the time-dependent bootstrap that directly bounds the two-point correlators at negative Euclidean time. This bootstrap technique was first used in \cite{ChoEtAl2025Euclidean} to bound the (positive) Euclidean time correlators, $G_E(\tau)$, in the quartic matrix quantum mechanics, in both the ground state and the thermal state. Here, we extend the formalism to bound the negative-time correlators. As $\tau_0(\lambda)$ indicates a singularity in the two-point function, we expect both the lower and upper bounds to diverge as $\tau \rightarrow -\tau_0$. An additional bootstrap parameter in the time-dependent bootstrap formalism is the \textit{spline degree} $d$ that is used to parametrize a Lagrange multiplier for time-dependent constraints. We find that at fixed $\Lambda$ and $d$, the bootstrap upper bounds cease to exist for some $\tau \leq -\tau_0^{\Lambda,d}$. Moreover,  $\tau_0^{\Lambda,d}$ increases as either $\Lambda$ or $d$ increases. We can therefore interpret $\tau_0^{\Lambda,d}$ as a \textit{lower bound} on the true $\tau_0$, as the existence of a finite upper bound up to this point indicates the two-point function has not yet diverged. We expect $\tau_0^{\Lambda,d} \rightarrow \tau_0$ as both $\Lambda, d \rightarrow \infty$. 

Figure \ref{fig:negativetimebootstrap} shows the time-dependent bootstrap result at $\lambda=-0.2$ using $\Lambda=10$ and $d=14$. We also show the black curve obtained by numerically solving the Marchesini--Onofri equation, which serves as a benchmark for the time-dependent bootstrap. The disappearing upper bound gives $\tau_0^{10,14} \approx 2.1$. As a comparison, the moment expansion method gives $\tau_0 \approx 4.7$. We see that $\tau_0^{10,14}$ indeed bounds the true $\tau_0$ from below, although it is not tight at this level and spline degree. We expect future developments in the bootstrap method can further improve our results. 

The time-dependent bootstrap can also be applied in the finite-temperature case, where the initial conditions are not analytically available. In this case, we input both the constraints from the equal-time correlators and the finite-time correlators. Here, we expect a singularity in $G_E(\tau)$ at $\beta+\tau_0$. We can similarly apply the time-dependent bootstrap and examine the location where the bounds disappear. Figure \ref{fig:fourierMethod-thermal} shows the time-dependent bootstrap results  in the massless model at $\beta=1$ and $\lambda=4$. We were able to obtain rigorous upper bounds until $\tau \approx 1.5$, translating to $\beta_0 \gtrsim 2$ at this coupling.

\subsection{Spectral positivity}
\label{sec:frequency-bootstrap}

A related but different implementation of the time-dependent bootstrap idea is to work in the frequency domain and exploit the positivity of the spectral measure. The Euclidean correlator and its spectral moments are
\begin{align}
G_E(\tau)
&= \int_0^\infty \left[e^{-\tau\omega} +e^{-(\beta-\tau) \omega}\right]\, p_X(\omega) \d \omega ,
\qquad p_X(\omega) \geq 0,
\label{eq:fourier-spectral-representation}\\
\mu_k
&= \int_0^\infty \omega^k\, p_X(\omega) \d \omega ,
\label{eq:fourier-spectral-moments}
\end{align}
where $p_X(\omega)$ is the spectral measure with respect to the two-point function $\langle \text{Tr} (X(t) X(0))\rangle$. The equations of motion relate these moments to equal-time expectation values, while positivity imposes semidefinite constraints on the moment Gram matrix $\mathcal{M}_{ij}=\langle \omega^{i+j}\rangle$.  At zero temperature, the restriction of the integral to $\omega \ge 0 $ is essentially the statement of ground-state positivity: perturbing the ground state by some operator always increases the energy of the state, and $\omega = E - E_0$ is the energy above the ground-state energy. For more details, we refer the reader to \cite{fouriermethod}.

\begin{figure}[t]
    \centering
    \includegraphics[width=0.98\linewidth]{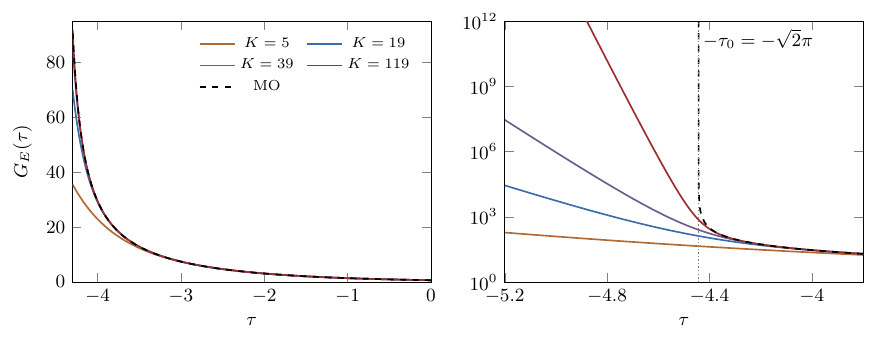}
    \caption{Bootstrap lower bounds on $G_E(\tau)$ for one-matrix quantum mechanics with the potential in \eqref{eq:mqmhamiltonian} with $\lambda=\lambda_c=-2\sqrt{2}/(3\pi)$ from the spectral bootstrap method in Section \ref{sec:frequency-bootstrap} at zero temperature.  The dashed curve shows the $G_E(\tau)$ from numerically solving the Marchesini--Onofri equation.  $K$ corresponds to the maximum degree of the spectral moments used in the bootstrap.}
\label{fig:fourierMethod-1MQM}
\end{figure}

Let us consider this method for the ground-state correlator of the single-matrix model. 
For the ground-state correlators, we deal with exponentials $e^{-\tau \omega}$ by bounding them by polynomials in frequency $P_\text{lower}(\tau \omega) \le e^{-\tau \omega}$. %
For the ground state of the one-matrix model, there is a simple and optimal way to choose $P_\text{lower}$ which is known as ``Gaussian quadrature.''   The idea is that if we know the moments $\mu_0, \mu_1, \ldots ,\mu_K$, we can find a positive measure $\mathfrak{p}(\omega)$ that is a sum of weighted delta functions that reproduces these moments: %
\begin{align} \label{eq:w_omega} \mathfrak{p}(\omega) = \sum_{i=1}^{n} w_i \delta(\omega - \omega_i), \quad 
\sum_{i=1}^{n}w_i\omega_i^k=\mu_k, \quad w_i \ge 0,
\quad k=0,\ldots,K,
\end{align}
where $K=2n-1$.
For $\tau<0$, consider the polynomial $P_{\text{lower}}(\omega) = \sum_{k} c_k \omega^k $ of degree at most $K$ that matches both the value and slope of $e^{-\tau\omega}$ at each $\omega_i$. This polynomial satisfies the bound $P_{\text{lower}}(\omega) \le e^{- \omega \tau}$ on $\omega\geq0$. Then,
\begin{align}
G_E(\tau)
\geq \int_0^\infty P_{\text{lower}}(\omega)\, p_X(\omega) \, \d \omega
= \sum_{i=1}^{n}w_i e^{-\tau\omega_i}. \label{geBDmoments}
\end{align}
Thus, given knowledge of the moments, we simply solve \eqref{eq:w_omega} for the $w_i$ and $\omega_i$ and plot the resulting lower bound \eqref{geBDmoments} (see Figure~\ref{fig:fourierMethod-1MQM}).
In fact the Gaussian quadrature method gives an {\it optimal} bound since the measure $\mathfrak{p}(\omega) = \sum_i w_i \delta(\omega - \omega_i)$ actually saturates the inequality \eqref{geBDmoments} and is (by construction) consistent with all the moments.  %

Note that in the frequency-domain approach, these moment constraints do not yield an upper bound for $\tau<0$. In particular, a finite-degree polynomial cannot control the exponentially growing contribution from arbitrarily high frequencies.  

\begin{figure}[t]
    \centering    \includegraphics[width=0.98\linewidth]{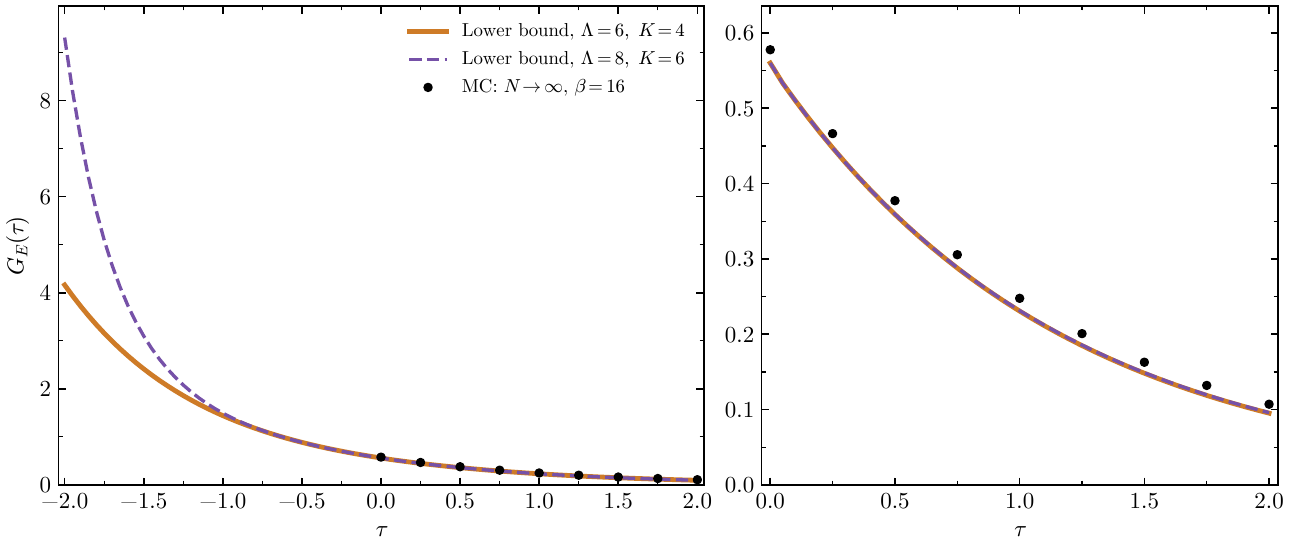}
    \caption{Lower bounds on the ground-state correlator $G_E(\tau)$ for the two-matrix quantum mechanics with $m^2=0$, obtained with the spectral bootstrap. Black points show independent finite-temperature Monte Carlo data at $\beta=16$ and $N_\tau=128$, extrapolated to $N\to\infty$ using a leading $1/N^2$ fit.}%
\label{fig:fourierMethod-O2}
\end{figure}

\begin{figure}[t]
    \centering
    \includegraphics[width=\linewidth]{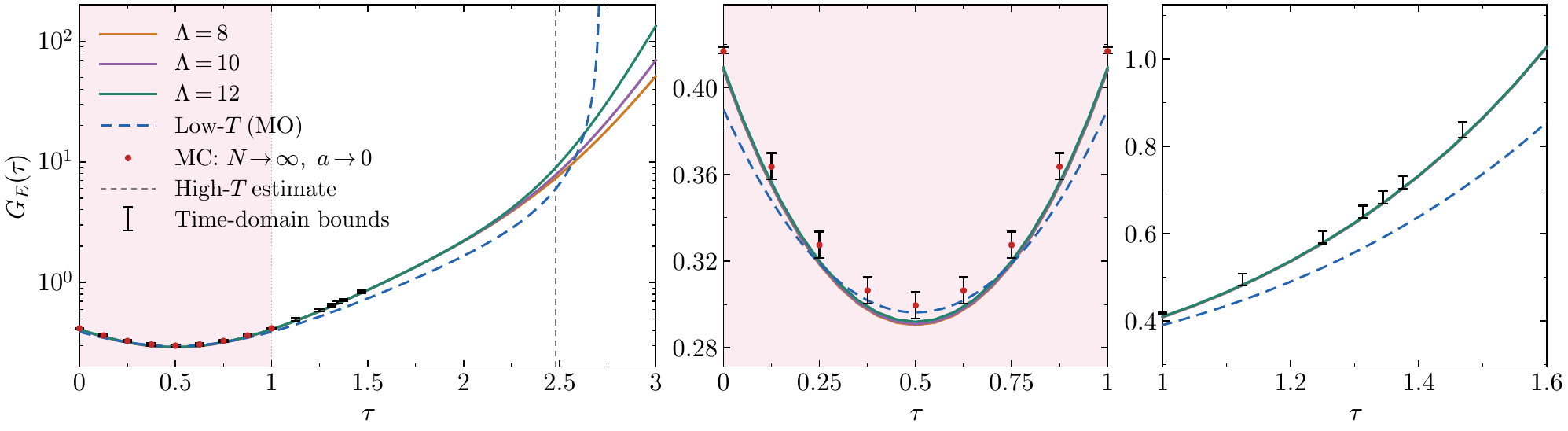}%
    \caption{Bootstrap bounds on $G_E(\tau)$ for one-matrix quantum mechanics at finite temperature $\beta=1$ for the potential $m=0, \lambda=4$. The lower bounds from the spectral bootstrap are shown in solid curves. The upper and lower bounds from the time-domain bootstrap are depicted in black brackets (see the rightmost panel for a close-up). The blue dashed curve shows the low-temperature approximation from the Marchesini--Onofri equation. We also show finite-temperature Monte Carlo data in red dots, after performing a large-$N$ and continuum extrapolation. Here $\Lambda$ is the maximum level of the operator that enters in the bootstrap problem, and the associated values of $K$ (the maximum degree of the polynomial in frequency, see \eqref{eq:fourier-spectral-moments}) associated with $\Lambda = \{8,10,12\}$ were $K = \{7,9,11\}$. %
    }
    \label{fig:fourierMethod-thermal}
\end{figure}

For the ground-state correlator of the two-matrix model, the moments are not known analytically. Hence we have to use a slightly more complicated procedure to obtain rigorous bounds; we first introduce some fixed polynomial $P_\text{lower}(\omega)$. We will call this polynomial an ``approximant'' although it is a rigorous lower bound.  
This introduces more parameters including $R$, the number of frequency bins (with one semi-infinite bin), and $K$, the maximum degree of the polynomial for $P_\text{lower}$. %
Figure~\ref{fig:fourierMethod-O2} uses approximants of degrees $4$ and $6$ at $L=6$ and $8$, respectively. 
In Figure~\ref{fig:fourierMethod-thermal}, the KMS envelopes and the objective envelopes inside the thermal strip $0<\tau<\beta$ are quadratic, whereas the negative-time approximants have degrees $7,9,11$. These degrees are limited by the available moments and are distinct from the spline degree used in the time-domain bootstrap. 

For systems with several operators, we retain matrix-valued spectral moments and impose positivity together with the equations of motion and symmetry constraints. In the two-matrix calculation of Figure \ref{fig:fourierMethod-O2}, $L$ limits the total weighted degree of a spectral matrix entry:
\begin{align}
k+w(A)+w(B)\leq L,
\label{eq:fourier-weighted-truncation}
\end{align}
where $k$ is the frequency-moment order, coordinates have weight $1$, and momenta have weight $2$. The operator basis has weight at most $L/2$. By contrast, the thermal one-matrix calculations in Figure~\ref{fig:fourierMethod-thermal} combine zero-time levels $L=8,10,12$ with selected spectral families: the $X$ moments extend through orders $7,9,11$, respectively, while the $X^2$ moments extend through order $3$.

In the case of finite temperature, the KMS condition takes the form of a detailed balance equation for the spectral density:
\begin{equation}
p_X(-\omega) = e^{-\beta \omega} p_X(\omega).
\end{equation}
This allows us to write the Euclidean correlator with support on only the positive-frequency axis\footnote{For finite temperature, the zero-time moments include both positive and negative frequency transitions, 
\begin{align}
M_k
&\equiv (-1)^k G_E^{(k)}(0) =\int_{0}^{\infty}
\omega^k\left[1+(-1)^k e^{-\beta\omega}\right]
p_X(\omega)\,d\omega.
\end{align}}. Then, we similarly construct a temperature-dependent polynomial $P_{\rm lower}$ that provides a lower bound for the corresponding kernel that involves the $e^{-\beta \omega}$ factor. 
We apply this in the case of one-matrix quantum mechanics. The lower bounds obtained are included in Figure \ref{fig:fourierMethod-thermal}, for the massless model at $\beta=1$ and $\lambda=4$. The bounds are consistent with the time-domain methods in Section \ref{sec:time-bootstrap} and Monte Carlo simulations.

\FloatBarrier
\subsection{The Wigner distribution}
\label{sec:wigner}
To go beyond the high-temperature approximation, it is convenient to use the Wigner distribution. This is the natural formalism for the moment problem, as we need to compute moments of operators that involve both $X$ and $P$. While in statistical mechanics, one can consider a probability distribution on phase space, this is famously impossible in quantum mechanics. However, the Wigner quasi-probability distribution exists and one can use it for the moment problem; in particular, one can systematically compute small-$\beta$ corrections to the ordinary statistical mechanics Boltzmann distribution.

We give a quick review of this before applying it to matrix quantum mechanics.
Let $\widehat\rho_\beta=e^{-\beta H}$. Inserting a resolution of the
identity between its two factors gives
\begin{align}
\operatorname{Tr}_{\mathcal H}(e^{-\beta H}\widehat O)
&=\int \d q\,\d r\,
\langle q|e^{-\beta H}|r\rangle
\langle r|\widehat O|q\rangle.
\end{align}
Here $\widehat{O}$ is any quantum operator (not necessarily symmetrically ordered in $\hat{X}$ and $\hat{P}$).
Let us write $q=X+Y/2$ and $r=X-Y/2$. Here $X$ is the ``midpoint'' coordinate, whereas $Y$ is a shift.
\begin{align}
\operatorname{Tr}_{\mathcal H}(\widehat\rho_\beta\widehat O)
&=\int \d X\,\d Y\,\rho_\beta(X,Y)
\left\langle X-\tfrac Y2\right|\widehat O
\left|X+\tfrac Y2\right\rangle,
\end{align}
where
\begin{align}
\rho_\beta(X,Y)
&=\left\langle X+\tfrac{Y}{2}\right|e^{-\beta H}
\left|X-\tfrac{Y}{2}\right\rangle.
\label{eq:rhobeta}
\end{align}
The Weyl symbol $O_{\text{W}}$\footnote{
The Weyl symbol means that one expands the operator $\widehat O$ in a basis of operators that are symmetric under all permutations of the canonical
ordering. For example if $\widehat O= \widehat X \widehat P$, we write $\widehat X \widehat P =  ( \widehat X \widehat P + \widehat P \widehat X)/2 + i /2$. Then $O_{\text{W}} = XP + i/2$. Note that this map is invertible. Equivalently, the inverse (known as Weyl quantization) maps
$e^{i(a\cdot X+b\cdot P)} \to 
e^{i(a\cdot\widehat X+b\cdot\widehat P)} = e^{i(a\cdot\widehat X / 2)} e^{i b\cdot\widehat P} e^{i(a\cdot\widehat X / 2)}$;
expanding these exponentials generates the symmetrized monomials. Note that $\bra{X-Y/2}e^{i(a\cdot\widehat X / 2)} e^{i b\cdot\widehat P} e^{i a\cdot\widehat X / 2} \ket{X+Y/2}  = e^{i a\cdot  X } \delta(Y-b)$ in agreement with \eqref{weylOrder}.
} satisfies 
\begin{align}
\left\langle X-\tfrac Y2\right|\widehat O
\left|X+\tfrac Y2\right\rangle 
&= \left\langle X\right|e^{-i \hat{P} \cdot Y/2} \widehat O e^{-i \hat{P}\cdot Y/2}
\left|X\right\rangle\\
&=\int\frac{\d P}{(2\pi)^{N^2}}\,
e^{-iP\cdot Y}O_{\text{W}}(X,P). \label{weylOrder}
\end{align}
Here $O_{\text{W}}(X,P)$ is a function of commuting variables.
Thus, defining the unnormalized Wigner distribution
\begin{align}
W_\beta(X,P)
&=\int \d Y\,e^{-iP\cdot Y}\rho_\beta(X,Y),\label{eq:wk-wigner-definition}
\end{align}
we have
\begin{align}
\langle\widehat O\rangle_\beta
&=\frac{\int \d X\,\d P\,W_\beta(X,P)O_{\text{W}}(X,P)}
{\int \d X\,\d P\,W_\beta(X,P)}.
\label{eq:wignerAvg}
\end{align}

At leading order at high temperatures, $W(X,P)$ is just given by the classical Boltzmann distribution. However, this formalism is useful because it allows us to systematically compute a high-temperature expansion for the moments. 
In particular, we can compute $W_\beta(X,P)$ using the Euclidean path integral in the limit where $\beta$ is small in a one-loop approximation; the details of this computation are reported in Appendix \ref{app:1loopWigner}. We obtain
\begin{align}
W_\beta(X,P)&=e^{-\beta H(X,P)}\bigl[1+\mathcal A(X,P)+\cdots\bigr],
\label{eq:wk-weight-main}\\
\mathcal A(X,P)
&=-\frac{\beta^2}{8}\Delta_XV
+\frac{\beta^3}{24}
\left[\sum_aV_a^2+\sum_{a,b}P_aV_{ab}P_b\right],
\label{eq:wk-A}
\end{align}
where $V_a=\partial_aV$, $V_{ab}=\partial_a\partial_b V$, and $\Delta_X V = \sum_a V_{aa}$.
For the massless quartic model with $\lambda=1$, the Hessian acts on a matrix $P$ as
$(X^2P+XPX+PX^2)/N$.  In a basis where $X$ has eigenvalues $x_i$,
\begin{align}
\langle P_{ij}P_{kl}\rangle_{X,\mathrm W}
=\frac{\delta_{il}\delta_{jk}}{\beta}
\left[1+\frac{\beta^2}{12N}(x_i^2+x_ix_j+x_j^2)+\cdots\right].
\label{eq:wk-momentum-covariance}
\end{align}
We now consider the even moments of the real-time correlator $C(t)$ in \eqref{ct_moment_expansion}. 
To average an operator with $W_\beta$, we first use the canonical commutation relations to write it in terms of symmetrized products\footnote{
Symmetrization concerns the canonical components; the matrix-index
contractions are preserved.  We retain the first ordering correction
as well as the corrections to the thermal weights.} of operators.  
Replacing those symmetrized products by
 commuting variables gives its Weyl symbol $O_{\text{W}}(X,P)$. We then use \eqref{eq:wignerAvg} to compute its average.
For example, consider the expectation value of $\text{Tr}(\widehat X^2\widehat P\widehat X)$. Using the canonical commutation relations and the cyclicity of the $SU(N)$ trace,
\begin{align}
\text{Tr}(\widehat X^2\widehat P\widehat X)=\frac12\text{Tr}(\widehat X^3\widehat P+\widehat P\widehat X^3)+\frac{i}{2}(\text{Tr}\widehat X)^2.
\end{align}
Both terms on the right are Weyl ordered in the canonical components, so the symbol to integrate against $W_\beta(X,P)$ is $\text{Tr}(X^3P)+i(\text{Tr}X)^2/2$.

The calculation therefore reduces to momentum Wick contractions \eqref{MQMwick} and moments of the one-matrix integral \eqref{MQMbipz}. %
Note that the Heisenberg equations of motion $\dot{X} = P$, and $\dot{P} = - X^3/N$ are exact as quantum operator equations and do not receive any finite-temperature corrections (in particular, we should {\it not} use the effective potential $\dot{P} = -V'_\text{eff}(X)$).  
The first few results are %
\begin{align}
\beta^{1/2}\mu_0&=\frac4{3\sqrt3}+\frac{\beta^{3/2}}{36}+O(\beta^3),\\
\beta\mu_2&=1+\frac{2\beta^{3/2}}{9\sqrt3}+O(\beta^3),\\
\beta^{3/2}\mu_4&=\frac8{3\sqrt3}+\frac{10\beta^{3/2}}{27}+O(\beta^3).
\label{eq:wk-first-moments}
\end{align}

Using \eqref{eq:wk-midpoint-conversion}, we obtain the
moments $\bar\mu_{k}$ up to $k\leq 28$. To include the one-loop corrections $O(\beta^{3/2})$ for $\bar\mu_k$, we need the classical moments up to $\mu_{30}$, since the shift by $\mu_{k+2}$ in \eqref{eq:wk-midpoint-conversion} is of the same order:
\begin{align}
\bar\mu_k
&=\mu_k^{\mathrm{cl}}+\delta\mu_k
-\frac{\beta^{3/2}}{8}(\beta^{1/2}\mu_{k+2}^{\mathrm{cl}})+\cdots.
\end{align}
We then construct the Pad\'e approximant in the variable
$z=(it)^2/\sqrt\beta$. The correlator is then  %
\begin{align}
\sqrt\beta\,C_{\text{LR}}(t) %
=\sum_{n=0}^{\infty}\frac{\beta^{(n+1)/2}\bar\mu_{2n}}{(2n)!}z^n.
\end{align}
Let $Q_0(z)+\beta^{3/2}Q_1(z)+\cdots$ be the denominator of a
Pad\'e approximant, normalized to one at $z=0$.  Its coefficients are
found by matching the moment series, first at classical order and then
at order $\beta^{3/2}$.  If its positive pole is
$z_0+\beta^{3/2}z_1+\cdots$, then
\begin{align}
z_1=-\frac{Q_1(z_0)}{Q_0'(z_0)},\qquad
\frac{\beta_0}{\beta^{1/4}}=b_0+\beta^{3/2}b_1+\cdots,\qquad
b_0=2\sqrt{z_0},\quad b_1=\frac{z_1}{\sqrt{z_0}}.
\label{eq:wk-pole-shift}
\end{align}
This extracts the first correction directly from the moment series.
Representative results are shown in Table~\ref{tab:wk-pade}. This gives
\begin{align}
\beta_0
&\simeq
\left(\frac{\beta}{\lambda}\right)^{1/4}
\left[5.60452+0.175\,(\lambda\beta^3)^{1/2}+\cdots\right].
\label{eq:wk-beta0-result}
\end{align}

\begin{table}[t]
\centering
\begin{tabular}{ccc}
\toprule
Pad\'e order $[L/M]$ & $b_0$ & $b_1$ \\
\midrule
$[6/6]$ & $5.6053455$ & $0.1750456$ \\
$[7/6]$ & $5.6046520$ & $0.1748159$ \\
$[6/7]$ & $5.6045124$ & $0.1746989$ \\
$[7/7]$ & $5.6045710$ & $0.1748021$ \\
\bottomrule
\end{tabular}
\caption{Estimates of the constant term and first correction in
$\beta_0/\beta^{1/4}$ from the centered moment expansion.
The orders refer to Pad\'e approximation in $z=u^2/\sqrt\beta$.
The displayed digits allow comparison between approximants and do not
specify the uncertainty.}
\label{tab:wk-pade}
\end{table}

\FloatBarrier
\setcounter{topnumber}{4}
\setcounter{totalnumber}{5}
\section{Conclusion}

In this paper we have identified signatures of the black hole singularity that persist far beyond the regime where one might expect a classical geometric description to apply. The essential remaining problem is to reconstruct a bulk description from the complex-time singularities of boundary correlators. In particular, we would like to formulate precisely what it means to ask about the fate of the black hole horizon and singularity in string theory and quantum gravity, where a local spacetime metric need not be an adequate description. Our complex scattering kernel provides a natural starting point for this reconstruction.

A curious fact is that the bouncing geodesic singularity in the $SL(2,\mathbf{R})/U(1)$ black hole solution is already resolved by $\alpha'$ effects~\cite{BuricGusevParnachev2026Resolution,Itzhaki:2018rld}. This is in tension with our results, since we found that quantum gravity played an essential role in smoothing the singularities. However, there is no direct contradiction, since the $SL(2,\mathbf{R})/U(1)$ black hole is asymptotically flat, so that there is no obvious quantum-mechanical boundary dual. It would be interesting to generalize the mechanism of \cite{BuricGusevParnachev2026Resolution,Itzhaki:2018rld} to the AdS black hole context, where the usual holographic dictionary applies, in order to identify a scenario where stringy corrections are sufficient for singularity resolution.

Eventually one would like to understand the bouncing geodesic singularities in higher-dimensional field theories. In particular, is the $N^{-1/2}$ smoothing scale universal? In contrast to quantum mechanics, quantum field theories do not become classical at high temperatures, due to the ultraviolet catastrophe. Nevertheless one can obtain a classical effective theory for the low-energy degrees of freedom by integrating out the hard modes~\cite{Iancu1998Classical,Bodeker1998Soft}. It would be interesting to use this approach to analyze the complex-time singularities in a realistic system like QCD. If we consider a compact spatial manifold, the situation is improved; for example, we can consider $\mathcal{N}=4$ SYM on $T^3 \times S^1_\beta$ where the characteristic size of $T^3$ is $R$. There are then two dimensionless parameters, $\beta/R$ and $\lambda$. In the ``double-scaling'' limit $\beta/R \to \infty$ with $\lambda (\beta/R)^3 = \tilde\lambda$ held fixed, the system reduces to the BFSS matrix quantum mechanics with effective 't Hooft coupling $\tilde\lambda$.

Ultimately, one wants to understand what differentiates black holes from generic large-$N$ systems. We hope that the scattering formalism developed here provides a concrete framework for answering this question.

\section*{Acknowledgements}
We are grateful to Ahmed Almheiri, Gauri Batra, Daniel Jafferis, Aidan Herderschee, Igor Klebanov, Zohar Komargodski, Juan Maldacena, Stephen Shenker, Eva Silverstein, Atul Sharma, Ahmed Sheta,  Douglas Stanford, Zechuan Zheng, Alexander Zhiboedov, and Winston Zoullas for helpful discussions. We would like to thank Gauri Batra, Stephen Shenker, Ahmed Sheta, and Winston Zoullas for comments on the draft. We acknowledge the use of OpenAI’s Codex and Anthropic's Claude. Numerical computations used the FASRC Cannon cluster, supported by the FAS Division of Science Research Computing Group at Harvard University. MD's work is supported by DOE grant DE-SC/0007870 and the Frankel-Goldfield Research Fund. MD thanks the Aspen Center for Physics for its hospitality during the 2026 summer program. This work was performed in part at the Aspen Center for Physics, which is supported by National Science Foundation grant PHY-2210452.

\appendix
\section{\texorpdfstring{$\tau_0$}{tau0} at zero temperature}\label{appb}
\subsection{Analytic value of \texorpdfstring{$\tau_0$}{tau0} from Regge eigenfunctions}

\label{sec:regge}
In this appendix, we compute $\tau_0$ and determine the behavior of the correlator near this singularity in one-matrix quantum mechanics at criticality $\lambda = \lambda_c$.
This can be done using the Marchesini--Onofri (MO) equation \cite{Marchesini:1979yq}, which gives the adjoint spectrum:
\begin{equation}
    \Delta_n\Phi_n(x)=
    \operatorname{P}\int_{x_1}^{x_2} \d y\,\rho(y)
    {\frac{\Phi_n(x)-\Phi_n(y)}{(x-y)^2}},
    \qquad
    \rho(x)=\frac{1}{\pi}\sqrt{2\bigl[\mu_F-V(x)\bigr]},
\end{equation}
where $\operatorname{P}$ denotes the principal value. Here $\Phi_n$ and $\Delta_n$ are the adjoint wavefunctions and energies above the ground-state energy.
In \cite{klebanov2026reggetrajectoriesadjointsector}, it was shown that at criticality $\lambda_c =-2\sqrt{2}/(3\pi)$, the MO equation can be written as an eigenvalue equation of the Hamiltonian
\begin{equation}
\label{eq:regge-h}
    {\cal H}(p,\tau)=
    p\coth\left(\frac{\pi p}{\sqrt2}\right)
    +\frac{2}{\pi}\Bigl[\tau\tanh\left(\frac{\tau}{\sqrt2}\right)-\sqrt2\Bigr].
\end{equation}
The matrix elements of $X^J$, the $J$-fold matrix product of $X$, can be computed by projecting onto the MO eigenfunction,
\begin{equation}
    \langle 0|X^J|n\rangle
    =\int_{x_1}^{x_2}\d x\,\rho(x)\,x^J\,\Phi_n(x).
    \label{eq:Mn-def}
\end{equation}
The singularity is governed by high energies $n \gg 1$. In this regime, we can use the WKB approximation for the Hamiltonian \eqref{eq:regge-h} to obtain the matrix elements (restricted to the appropriate $\mathbb{Z}_2$ parity) 
\begin{equation}
\label{eq:MJ-tail}
    \langle 0|X^J|n\rangle
    \simeq
    \frac{4\sqrt2}{\Gamma(J)}
    \left(\frac{3\pi}{\sqrt2}\right)^{(J+1)/2}
    (\Delta_n+c)^{J-3/2}
    \exp\!\left[-\frac{\pi(\Delta_n+c)}{\sqrt2}\right],
\end{equation}
where the eigenvalues are
\begin{equation}
    \Delta_n \simeq \sqrt{2n+\frac{2}{3}}-c,
    \qquad c = \frac{2\sqrt{2}}{\pi}.
    \label{eq:regge-spectrum}
\end{equation}
Inserting a complete set of states, the Euclidean correlator becomes
    $G_J =
    \sum_{n\ge1}| \langle 0|X^J|n\rangle|^2e^{-\Delta_nt_E}$.
So the high-energy tail is either exponentially growing or damped in $n$,
\begin{equation}
   | \langle 0|X^J|n\rangle|^2 e^{-\Delta_n t_E }
    \sim
    \frac{32}{\Gamma(J)^2}
    \left(\frac{3\pi}{\sqrt2}\right)^{J+1}
    e^{ct_E}(\Delta_n+c)^{2J-3}
    \exp\!\left[-(\sqrt2\pi+t_E)(\Delta_n+c)\right],
\end{equation}
depending on the sign of $\sqrt{2}\pi+t_E$. Thus the sum converges until it hits $t_E=-\tau_0$, where
\begin{equation}
\label{tau0crit}
    \tau_0 = \sqrt2\,\pi.
\end{equation}
To determine the behavior of $G$ near this singularity, we may replace the sum over Regge states by an integral; this yields
\begin{equation}
\begin{aligned}
    G_J(t_E)
    &\sim
    \frac{\mathcal{N}_J}{(t_E+\sqrt2\pi)^{2J-1}}
    +\hbox{less singular terms},
    \\
    \mathcal{N}_J
    &=
    16e^{-4}
    \left(\frac{3\pi}{\sqrt2}\right)^{J+1}
    \frac{\Gamma(2J-1)}{\Gamma(J)^2}.
\end{aligned}
    \label{eq:GJ-singularity}
\end{equation}
Thus the singularity is at the same location for every $J$; however, the order of the pole changes with $J$.

\subsection{String interpretation}
\label{app:string}
We offer the following string interpretation of this divergence. The states that are responsible for this divergence are $n \gg 1$ Regge states, which consist of a string stretched by a distance $\propto \sqrt{n}$. As we approach $t_E = -\tau_0$, the typical $n$ diverges and the string is getting extremely long. To be more concrete, consider the linear dilaton background
\begin{equation}
    S\left[\phi\right]=\int d^2 \sigma  \sqrt{- g}\Bigl(
    \frac{1}{4 \pi \alpha^{\prime}}
 \bigl[-(\partial X^0)^2 +(\partial\phi)^2\bigr] + \frac{1}{4 \pi} Q \phi R(g)- \mu e^{2\phi/\sqrt{\alpha'}}\Bigr).
\end{equation}
This has the classical Lorentzian solution \cite{Maldacena:2005hi}
\begin{equation}
    \phi(t,\sigma)=\phi_0'
    -Q\alpha'\log\left[\cosh\left(\frac{t}{\alpha'Q}\right)
    +\cosh\left(\frac{\sigma}{\alpha'Q}\right)\right].
\end{equation}
The tip at $\sigma=0$ obeys
\begin{equation}
    \phi_{\rm tip}(t)
    =\phi_0-2Q\alpha'\log\cosh\left(\frac{t}{2\alpha'Q}\right),
    \qquad Q=\frac{2}{\sqrt{\alpha'}}.
\end{equation}
Thus the nearest complex-time singularity of
this classical tip trajectory is at
\begin{equation}
    t_0 =\pm i\pi\alpha'Q=\pm 2\pi i\sqrt{\alpha'}.
\end{equation}
For the quartic matrix model, our conventions set $\alpha' = 1/2$, so that we recover \eqref{tau0crit}.

\subsection{Away from criticality}
The computation can be generalized to fixed nonzero $\lambda>\lambda_c$ using the high-energy WKB approximation of \cite{Gross:1990md,klebanov2026reggetrajectoriesadjointsector}, which describes ``long strings'' near criticality \cite{Maldacena:2005hi}. Writing $\phi=\rho\Phi$ and introducing the time-of-flight coordinate $\tau=\int_0^x \d y/\bigl[\pi\rho(y)\bigr]$, the approximate MO Hamiltonian is
\begin{equation}
    {\cal H}(p,\tau)\simeq |p|+\eta_\lambda(\tau),
    \qquad
    \eta_\lambda(\tau)=\operatorname{P}\int_{-a}^{a}
    \d y\,\frac{\rho(y)}{\bigl[x(\tau)-y\bigr]^2},
\end{equation}
where $[-a,a]$ is the support of the eigenvalue density and $\phi$ vanishes at the endpoints of the finite time-of-flight interval. 
Defining $A=\lambda a^2/(2+\lambda a^2)$, examining the poles of $x(\tau)$ gives %
\begin{equation}\label{tau0general}
    \tau_0(\lambda)=
    \begin{cases}
        2\sqrt{1-A}\,K(1+A), & -1<A<0,\\[2mm]
        2\sqrt{\dfrac{1-A}{1+A}}\,K\!\left(\dfrac{1}{1+A}\right),
        & 0<A<1
    \end{cases},
\end{equation}
where $K$ is the complete elliptic integral of the first kind, see the right panel of Figure \ref{fig:beta0zerotemp}. This reduces to $\tau_0=\sqrt{2}\pi$ as $\lambda\to\lambda_c$ and diverges logarithmically as $\lambda\to0$. As in the critical computation, the WKB phase cancels two powers of the pole in $x^Jx'$, leaving a pole of order $J$. Thus, on the allowed parity subsequence, $\langle0|X^J|n\rangle\sim\Delta_n^{J-1}e^{-\tau_0\Delta_n/2}$. Since the high-energy levels are now asymptotically evenly spaced, we obtain
\begin{equation}
    G_J(t_E)\sim
    \int^\infty \d\Delta\,\Delta^{2J-2}
    e^{-(t_E+\tau_0)\Delta}
    \sim \frac{\mathcal{N}_J(\lambda)}
    {(t_E+\tau_0)^{2J-1}},
    \qquad t_E\to-\tau_0^+.
\end{equation}
We checked this computation by numerically solving the MO equation and extracting the overlaps from the MO wavefunctions.
\section{Wigner at one loop}\label{app:1loopWigner}
\subsection{One-loop correction to the Wigner distribution}
\label{sec:wigner2}
We will compute $W_\beta(X,P)$ using the Euclidean path integral in the approximation where $\beta$ is small.
Expand an arbitrary Euclidean trajectory between the two endpoints as
\begin{align}
X_a(\tau)=X_a+\left(\frac{\tau}{\beta}-\frac12\right)Y_a
+\eta_a(\tau),
\qquad \eta_a(0)=\eta_a(\beta)=0.
\end{align}
The Euclidean action is
\begin{align}
S_E
&=\int_0^\beta d\tau\,
\left[\frac12 \dot X_a(\tau)^2+V(X(\tau))\right]
\nonumber\\
&=\frac{1}{2\beta} Y_a^2
+\int_0^\beta d\tau\,
\Bigl[
\frac12 \dot\eta_a(\tau)^2
+V\!\left[X+\left(\frac{\tau}{\beta}-\frac12\right)Y
+\eta(\tau)\right]
\Bigr],
\end{align}
where the $a$ indices are summed over. Then writing $X_{l}(\tau)
=X+\left(\frac{\tau}{\beta}-\frac12\right)Y$,
\begin{align}
S_E
&=\frac{1}{2\beta}Y_a^2
+\int_0^\beta d\tau\,\Bigl[
\frac12 \dot\eta_a(\tau)^2
+V(X_{l}(\tau))
\nonumber\\
&\qquad
+V_a(X_{l}(\tau))\,\eta_a(\tau)
+\frac12 V_{ab}(X_{l}(\tau))\,
\eta_a(\tau)\eta_b(\tau)
\Bigr]
+O(\eta^3),
\end{align}
where $V_a=\partial_a V$ and $V_{ab}=\partial_a\partial_b V$. The kinetic term gives the $\eta$ propagator 
\begin{align}
\langle\eta_a(\tau)\eta_b(\tau')\rangle
=\delta_{ab}g(\tau,\tau'),
\qquad
g(\tau,\tau')=\min(\tau,\tau')-\frac{\tau\tau'}{\beta}.
\end{align}
Note that $\langle{\eta \eta}\rangle \sim \beta$, so at high temperatures the fluctuations are suppressed.  Expanding in these fluctuations and ``bringing them down'' from the exponent gives %
\begin{align}
    \rho_\beta(X,Y) &= \int D \eta \, \exp \Bigl(-{\frac{1}{2\beta}Y_a^2
-\int_0^\beta d\tau\, \Bigl[
\frac12 \dot\eta_a(\tau)^2
+V(X)} \Bigr] \Bigr) \times \\
 &\Bigl[1 + \frac{1}{2}\int d \tau \, d \tau' \,(V_a)^2 g(\tau, \tau')  - \frac{V_{aa}}{2} \int  g(\tau, \tau) d \tau - \frac{V_{ab} Y_a Y_b}{2} \int d \tau\,  \left(\frac\tau\beta-\frac12\right)^2 \Bigr]. \nonumber
\end{align}
Doing the Euclidean time integrals, we obtain
\begin{align}
\rho_\beta(X,Y)
={}&\frac{e^{- Y_a^2/(2\beta)-\beta V(X)}}
{(2\pi\beta)^{N^2/2}}
\left[
1-\frac{\beta^2}{12}V_{aa}
-\frac{\beta}{24}Y_aV_{ab}Y_b
+\frac{\beta^3}{24}V_a^2+\cdots
\right].
\end{align}
Fourier transforming in $Y$ gives
\begin{align}
\frac{\int \d Y\,Y_aY_b\,
e^{-Y_c^2/(2\beta)-iP_cY_c}}
{\int \d Y\,e^{-Y_c^2/(2\beta) -iP_cY_c}}
=\beta\delta_{ab}-\beta^2P_aP_b.
\end{align}

Combining the terms gives
\begin{align}
W_\beta(X,P)&=e^{-\beta H(X,P)}\bigl[1+\mathcal A(X,P)+\cdots\bigr],
\label{eq:wk-weight-app}\\
\mathcal A(X,P)
&=-\frac{\beta^2}{8}\Delta_XV
+\frac{\beta^3}{24}
\left[\sum_aV_a^2+\sum_{a,b}P_aV_{ab}P_b\right].
\label{eq:wk-A-app}
\end{align}
This phase-space weight reproduces the leading Wigner--Kirkwood
correction~\cite{Wigner1932,Kirkwood1933}.
Integrating over $P$, we obtain a positive measure for $X$. This can be used to define an effective potential $V_\text{eff}(X)$, via
\begin{align}
    \int \d P \, W(X,P)\propto e^{-\beta \,  V_\text{eff}(X)}.
\end{align}
In the large-$N$ massless model with $\lambda=1$,
\begin{align}
V_{\mathrm{eff}}(X)
=\frac1{4N}\text{Tr}(X^4)
+\frac{\beta}{6}\text{Tr}(X^2)
-\frac{\beta^2}{24N^2}\text{Tr}(X^6)+\cdots.
\label{eq:wk-main-effective-potential}
\end{align}
This potential determines the distribution of the instantaneous matrix
$X(0)$, or by stationarity $X(t)$, at any fixed real time. Note that it is not the same as the effective potential for the zero mode of $X(t)$; however, the two are related. %

\subsection{The one-loop thermal mass}
\label{app:wk-thermal-mass}

To connect the thermal mass to the density-matrix derivation in
Section~\ref{sec:wigner},
\eqref{eq:wk-wigner-definition}--\eqref{eq:wk-A}, we compute
$\langle\text{Tr}(X(t)^2)\rangle_\beta$ at leading order in $1/N$.
Thermal stationarity makes this equal-time observable independent of
$t$, so it suffices to evaluate it at $t=0$. Its zero-mode and
nonzero-mode contributions can then be computed separately.

Separate the Euclidean path into its zero mode and its nonzero modes:
\begin{align}
X(\tau)=X_0+\eta(\tau),\qquad
X_0=\frac1\beta\int_0^\beta X(\tau)\,d\tau,\qquad
\int_0^\beta\eta(\tau)\,d\tau=0.
\end{align}
The term linear in $\eta$ integrates to zero.  The quadratic action is
\begin{align}
S_E=\beta V(X_0)
+\frac12\int_0^\beta d\tau\,
\left[\sum_a\dot\eta_a^2
+\sum_{a,b}\eta_aV_{ab}(X_0)\eta_b\right]+\cdots.
\end{align}
Integrating out the nonzero modes gives a weight $e^{-\beta V_0(X_0)}$,
where, up to a field-independent constant,
\begin{align}
V_0(X_0)
&=V(X_0)+\frac1{2\beta}\sum_{n\ne0}
\log\det\left[\delta_{ab}+\frac{V_{ab}(X_0)}{\omega_n^2}\right]
+\cdots,
&\omega_n&=\frac{2\pi n}{\beta}.
\end{align}
The determinant acts on the $N^2$ canonical components.
Expanding the logarithm and using
$\sum_{n\ne0}\omega_n^{-2}=\beta^2/12$ gives
\begin{align}
V_0(X_0)=V(X_0)+\frac{\beta}{24}\Delta_XV(X_0)+\cdots.
\label{eq:wk-zero-mode-potential}
\end{align}
For the quartic matrix potential,
\begin{align}
\Delta_XV=2\text{Tr}(X^2)+\frac1N(\text{Tr}X)^2,
\end{align}
and we can neglect the double-trace term at infinite $N$. Hence, %
\begin{align}
V_0(X_0)
=\frac1{4N}\text{Tr}(X_0^4)
+\frac{\beta}{12}\text{Tr}(X_0^2).
\end{align}
To relate this effective potential of the zero mode to the effective potential of the instantaneous $X(t)$, write $X(t) = \bar{X} + \eta(t)$.
 To the order needed here, the nonzero modes are Gaussian with the propagator
 \begin{align}
\langle\eta_a(0)\eta_b(0)\rangle
 =\frac{\delta_{ab}}{\beta}\sum_{n\ne0}\frac1{\omega_n^2}
 =\frac{\beta}{12}\delta_{ab}+\cdots.
 \label{eq:wk-nonzero-variance}
 \end{align}
 Note that $\eta$ is independent of the zero mode at this order. %
 Then the probability distribution for the instantaneous $X(t)$ is
\begin{align}
p_X(X(t))
&=\int \d X_0\,\d\eta\,
p_0(X_0)\,p(\eta)\,
\delta(X-X_0-\eta)\\
&\propto
\int \d\eta\,
e^{-6\mathrm{Tr} (\eta^2)/\beta}
e^{-\beta V_0(X(t)-\eta)}.
\end{align}
Using \eqref{eq:wk-zero-mode-potential} and
$\Delta_Xe^{-\beta V}
=e^{-\beta V}\left[-\beta\Delta_XV+\beta^2\sum_aV_a^2\right]$,
we obtain
\begin{align}
p_X(X)
&\propto
e^{-\beta V(X)}
\left[
1-\frac{\beta^2}{12}\Delta_XV
+\frac{\beta^3}{24}\sum_aV_a^2+\cdots
\right]
= e^{- \beta V_\text{eff}(X)}. %
\end{align}
Thus the zero-mode calculation reproduces the effective potential obtained in \eqref{eq:wk-main-effective-potential}.
Said differently, one can freely translate the one-loop computation of the effective potential for the zero mode into the one-loop computation of the Wigner--Kirkwood effective potential \cite{Wigner1932, Kirkwood1933}, giving a ``path integral'' derivation of their formulas.

\bibliographystyle{JHEP}
\bibliography{refs}

\providecommand{\href}[2]{#2}\begingroup\raggedright\begin{thebibliography}{100}

\bibitem{Fidkowski2004}
L.~Fidkowski, V.~Hubeny, M.~Kleban and S.~Shenker, \emph{{The black hole
  singularity in AdS/CFT}},
  \href{http://dx.doi.org/10.1088/1126-6708/2004/02/014}{\emph{JHEP} {\bf 02}
  (2004) 014}, [\href{http://arxiv.org/abs/hep-th/0306170}{{\tt
  hep-th/0306170}}].

\bibitem{FestucciaLiu2006}
G.~Festuccia and H.~Liu, \emph{{Excursions beyond the horizon: black hole
  singularities in Yang-Mills theories. I}},
  \href{http://dx.doi.org/10.1088/1126-6708/2006/04/044}{\emph{JHEP} {\bf 04}
  (2006) 044}, [\href{http://arxiv.org/abs/hep-th/0506202}{{\tt
  hep-th/0506202}}].

\bibitem{Kraus:2002iv}
P.~Kraus, H.~Ooguri and S.~Shenker, \emph{{Inside the horizon with AdS/CFT}},
  \href{http://dx.doi.org/10.1103/PhysRevD.67.124022}{\emph{Phys. Rev. D} {\bf
  67} (2003) 124022}, [\href{http://arxiv.org/abs/hep-th/0212277}{{\tt
  hep-th/0212277}}].

\bibitem{AfkhamiJeddi2025Imprint}
N.~Afkhami-Jeddi, S.~Caron-Huot, J.~Chakravarty and A.~Maloney, \emph{{Imprint
  of the black hole singularity on thermal two-point functions}},
  \href{http://dx.doi.org/10.1103/1j3s-fzjl}{\emph{Phys. Rev. D} {\bf 114}
  (2026) 025005}, [\href{http://arxiv.org/abs/2510.21673}{{\tt 2510.21673}}].

\bibitem{Chakravarty:2025ncy}
J.~Chakravarty, \emph{{Imprint of black hole interior on thermal four-point
  correlators}}, \href{http://dx.doi.org/10.1007/JHEP09(2026)034}{\emph{JHEP}
  {\bf 09} (2026) 034}, [\href{http://arxiv.org/abs/2512.10912}{{\tt
  2512.10912}}].

\bibitem{Ceplak2026Fooling}
N.~{\v C}eplak, H.~Liu, A.~Parnachev and S.~Valach, \emph{{Fooling the censor:
  going beyond inner horizons with the OPE}},
  \href{http://dx.doi.org/10.1007/JHEP05(2026)001}{\emph{JHEP} {\bf 05} (2026)
  001}, [\href{http://arxiv.org/abs/2511.09638}{{\tt 2511.09638}}].

\bibitem{Jia2025ThermalAsymptotics}
H.~F. Jia and M.~Rangamani, \emph{{Thermal spectral function asymptotics and
  black hole singularity in holography}},
  \href{http://arxiv.org/abs/2512.15114}{{\tt 2512.15114}}.

\bibitem{Jia:2026ryl}
H.~F. Jia and M.~Rangamani, \emph{{Exact holographic thermal spectral
  functions: OPE, non-perturbative corrections, and black hole singularity}},
  \href{http://arxiv.org/abs/2604.10803}{{\tt 2604.10803}}.

\bibitem{Hartnoll:2026vhu}
S.~A. Hartnoll and A.~Zhiboedov, \emph{{Can one hear the shape of a black hole
  singularity?}},  \href{http://arxiv.org/abs/2609.07514}{{\tt 2609.07514}}.

\bibitem{Amado2008ReflectingCavities}
I.~Amado and C.~Hoyos-Badajoz, \emph{{AdS black holes as reflecting cavities}},
  \href{http://dx.doi.org/10.1088/1126-6708/2008/09/118}{\emph{JHEP} {\bf 09}
  (2008) 118}, [\href{http://arxiv.org/abs/0807.2337}{{\tt 0807.2337}}].

\bibitem{Horowitz:2023ury}
G.~T. Horowitz, H.~Leung, L.~Queimada and Y.~Zhao, \emph{{Boundary signature of
  singularity in the presence of a shock wave}},
  \href{http://dx.doi.org/10.21468/SciPostPhys.16.2.060}{\emph{SciPost Phys.}
  {\bf 16} (2024) 060}, [\href{http://arxiv.org/abs/2310.03076}{{\tt
  2310.03076}}].

\bibitem{Grozdanov:2026cut}
S.~Grozdanov, S.~Valach and M.~Vrbica, \emph{{Bouncing geodesics, black hole
  singularities, and singularities of thermal correlators}},
  \href{http://dx.doi.org/10.1007/JHEP08(2026)022}{\emph{JHEP} {\bf 08} (2026)
  022}, [\href{http://arxiv.org/abs/2603.15598}{{\tt 2603.15598}}].

\bibitem{JiaKulaxizi2026PhaseShift}
Y.~Jia and M.~Kulaxizi, \emph{{Bulk phase shift and singularity}},
  \href{http://arxiv.org/abs/2602.06558}{{\tt 2602.06558}}.

\bibitem{AliAhmad:2026wem}
S.~Ali~Ahmad, A.~Almheiri and S.~Lin, \emph{{Continuing past the inner horizon
  using WKB}},  \href{http://arxiv.org/abs/2601.02354}{{\tt 2601.02354}}.

\bibitem{Araya:2026shz}
I.~J. Araya, C.~Esper, Y.~Jia, M.~Kulaxizi and A.~Parnachev, \emph{{Bulkcone
  singularities and complex geodesics}},
  \href{http://arxiv.org/abs/2602.12893}{{\tt 2602.12893}}.

\bibitem{Grozdanov:2026ktq}
S.~Grozdanov, V.~Movrin and S.~Valach, \emph{{Bouncing geodesics,
  singularities, and the cavity thermal product formula in asymptotically flat
  and de Sitter black holes}},  \href{http://arxiv.org/abs/2606.11297}{{\tt
  2606.11297}}.

\bibitem{ArnaudoWithers2026Bouncing}
P.~Arnaudo and B.~Withers, \emph{{Bouncing singularities in Schwarzschild: a
  geometric origin of the QNM convergence region}},
  \href{http://arxiv.org/abs/2605.16489}{{\tt 2605.16489}}.

\bibitem{ArnaudoWithers2026Fourier}
P.~Arnaudo and B.~Withers, \emph{{Analytic structure of holographic thermal
  correlators from Fourier series}},
  \href{http://dx.doi.org/10.1007/JHEP06(2026)205}{\emph{JHEP} {\bf 06} (2026)
  205}, [\href{http://arxiv.org/abs/2603.13469}{{\tt 2603.13469}}].

\bibitem{Ceplak:2024bja}
N.~{\v{C}}eplak, H.~Liu, A.~Parnachev and S.~Valach, \emph{{Black hole
  singularity from OPE}},
  \href{http://dx.doi.org/10.1007/JHEP10(2024)105}{\emph{JHEP} {\bf 10} (2024)
  105}, [\href{http://arxiv.org/abs/2404.17286}{{\tt 2404.17286}}].

\bibitem{Sachdev:1992fk}
S.~Sachdev and J.~Ye, \emph{{Gapless spin fluid ground state in a random,
  quantum Heisenberg magnet}},
  \href{http://dx.doi.org/10.1103/PhysRevLett.70.3339}{\emph{Phys. Rev. Lett.}
  {\bf 70} (1993) 3339}, [\href{http://arxiv.org/abs/cond-mat/9212030}{{\tt
  cond-mat/9212030}}].

\bibitem{Kitaev2015Talks}
A.~Kitaev, ``{A simple model of quantum holography}.'' Talks at KITP,
  \href{https://online.kitp.ucsb.edu/online/entangled15/kitaev/}{April 7} and
  \href{https://online.kitp.ucsb.edu/online/entangled15/kitaev2/}{May 27},
  2015.

\bibitem{Polchinski:2016xgd}
J.~Polchinski and V.~Rosenhaus, \emph{{The spectrum in the Sachdev-Ye-Kitaev
  model}}, \href{http://dx.doi.org/10.1007/JHEP04(2016)001}{\emph{JHEP} {\bf
  04} (2016) 001}, [\href{http://arxiv.org/abs/1601.06768}{{\tt 1601.06768}}].

\bibitem{Maldacena:2016hyu}
J.~Maldacena and D.~Stanford, \emph{{Remarks on the Sachdev-Ye-Kitaev model}},
  \href{http://dx.doi.org/10.1103/PhysRevD.94.106002}{\emph{Phys. Rev. D} {\bf
  94} (2016) 106002}, [\href{http://arxiv.org/abs/1604.07818}{{\tt
  1604.07818}}].

\bibitem{Dodelson2025Stringy}
M.~Dodelson, C.~Iossa and R.~Karlsson, \emph{{Bouncing off a stringy
  singularity}}, \href{http://dx.doi.org/10.1007/JHEP07(2026)161}{\emph{JHEP}
  {\bf 07} (2026) 161}, [\href{http://arxiv.org/abs/2511.09616}{{\tt
  2511.09616}}].

\bibitem{Dodelson:2024atp}
M.~Dodelson, \emph{{Ringdown in the SYK model}},
  \href{http://dx.doi.org/10.21468/SciPostPhys.19.3.081}{\emph{SciPost Phys.}
  {\bf 19} (2025) 081}, [\href{http://arxiv.org/abs/2408.05790}{{\tt
  2408.05790}}].

\bibitem{DodelsonEtAl2026Temperature}
M.~Dodelson, O.~Gupta, M.~Mezei and D.~Wang, \emph{{On the temperature
  dependence of quasinormal modes in SYK and holography}},
  \href{http://arxiv.org/abs/2606.22679}{{\tt 2606.22679}}.

\bibitem{Buric:2026qsp}
I.~Buri{\'c}, C.-M. Chang, I.~Gusev, E.~Helfenberger, A.~Parnachev and
  M.~Rangamani, \emph{{Thermal two-point functions in SYK and complex-time
  singularities}},  \href{http://arxiv.org/abs/2607.05258}{{\tt 2607.05258}}.

\bibitem{Hubeny2007BulkCone}
V.~E. Hubeny, H.~Liu and M.~Rangamani, \emph{{Bulk-cone singularities \&
  signatures of horizon formation in AdS/CFT}},
  \href{http://dx.doi.org/10.1088/1126-6708/2007/01/009}{\emph{JHEP} {\bf 01}
  (2007) 009}, [\href{http://arxiv.org/abs/hep-th/0610041}{{\tt
  hep-th/0610041}}].

\bibitem{Dodelson2024BulkCone}
M.~Dodelson, C.~Iossa, R.~Karlsson, A.~Lupsasca and A.~Zhiboedov, \emph{{Black
  hole bulk-cone singularities}},
  \href{http://dx.doi.org/10.1007/JHEP07(2024)046}{\emph{JHEP} {\bf 07} (2024)
  046}, [\href{http://arxiv.org/abs/2310.15236}{{\tt 2310.15236}}].

\bibitem{Batra2026NonPrompt}
G.~Batra, \emph{{A non-prompt null ray cures its own singularity}},
  \href{http://arxiv.org/abs/2610.22478}{{\tt 2610.22478}}.

\bibitem{DodelsonOoguri2021}
M.~Dodelson and H.~Ooguri, \emph{{Singularities of thermal correlators at
  strong coupling}},
  \href{http://dx.doi.org/10.1103/PhysRevD.103.066018}{\emph{Phys. Rev. D} {\bf
  103} (2021) 066018}, [\href{http://arxiv.org/abs/2010.09734}{{\tt
  2010.09734}}].

\bibitem{Gesteau2023Ergodic}
E.~Gesteau, \emph{{Emergent spacetime and the ergodic hierarchy}},
  \href{http://arxiv.org/abs/2310.13733}{{\tt 2310.13733}}.

\bibitem{Ouseph2024Poincare}
S.~Ouseph, K.~Furuya, N.~Lashkari, K.~L. Leung and M.~Moosa, \emph{{Local
  Poincar{\'e} algebra from quantum chaos}},
  \href{http://dx.doi.org/10.1007/JHEP01(2024)112}{\emph{JHEP} {\bf 01} (2024)
  112}, [\href{http://arxiv.org/abs/2310.13736}{{\tt 2310.13736}}].

\bibitem{GesteauLiu2024Horizons}
E.~Gesteau and H.~Liu, \emph{{Toward stringy horizons}},
  \href{http://arxiv.org/abs/2408.12642}{{\tt 2408.12642}}.

\bibitem{Belin2025Mirage}
A.~Belin, S.~Bintanja, A.~Castro and W.~Knop, \emph{{Symmetric product orbifold
  universality and the mirage of an emergent spacetime}},
  \href{http://dx.doi.org/10.1007/JHEP05(2025)190}{\emph{JHEP} {\bf 05} (2025)
  190}, [\href{http://arxiv.org/abs/2502.01734}{{\tt 2502.01734}}].

\bibitem{Herderschee2025Stringy}
A.~Herderschee and J.~Kudler-Flam, \emph{{Stringy algebras, stretched horizons,
  and quantum-connected wormholes}},
  \href{http://arxiv.org/abs/2510.01556}{{\tt 2510.01556}}.

\bibitem{Dodelson2025Chaos}
M.~Dodelson, \emph{{Black holes from chaos}},
  \href{http://arxiv.org/abs/2501.06170}{{\tt 2501.06170}}.

\bibitem{Newton1982Scattering}
R.~G. Newton,
  \emph{{\href{https://doi.org/10.1007/978-3-642-88128-2}{Scattering Theory of
  Waves and Particles}}}.
\newblock Springer, 2nd~ed., 1982.

\bibitem{Yaffe1982}
L.~G. Yaffe, \emph{{Large $N$ limits as classical mechanics}},
  \href{http://dx.doi.org/10.1103/RevModPhys.54.407}{\emph{Rev. Mod. Phys.}
  {\bf 54} (1982) 407}.

\bibitem{JevickiSakita1980}
A.~Jevicki and B.~Sakita, \emph{{The quantum collective field method and its
  application to the planar limit}},
  \href{http://dx.doi.org/10.1016/0550-3213(80)90046-2}{\emph{Nucl. Phys. B}
  {\bf 165} (1980) 511--527}.

\bibitem{GopakumarGross1995}
R.~Gopakumar and D.~J. Gross, \emph{{Mastering the master field}},
  \href{http://dx.doi.org/10.1016/0550-3213(95)00340-X}{\emph{Nucl. Phys. B}
  {\bf 451} (1995) 379--415}, [\href{http://arxiv.org/abs/hep-th/9411021}{{\tt
  hep-th/9411021}}].

\bibitem{Kawahara:2007ib}
N.~Kawahara, J.~Nishimura and S.~Takeuchi, \emph{{High temperature expansion in
  supersymmetric matrix quantum mechanics}},
  \href{http://dx.doi.org/10.1088/1126-6708/2007/12/103}{\emph{JHEP} {\bf 12}
  (2007) 103}, [\href{http://arxiv.org/abs/0710.2188}{{\tt 0710.2188}}].

\bibitem{AsplundBerensteinDzienkowski2013}
C.~T. Asplund, D.~Berenstein and E.~Dzienkowski, \emph{{Large $N$ classical
  dynamics of holographic matrix models}},
  \href{http://dx.doi.org/10.1103/PhysRevD.87.084044}{\emph{Phys. Rev. D} {\bf
  87} (2013) 084044}, [\href{http://arxiv.org/abs/1211.3425}{{\tt 1211.3425}}].

\bibitem{GurAriHanadaShenker2016}
G.~Gur-Ari, M.~Hanada and S.~H. Shenker, \emph{{Chaos in classical D0-brane
  mechanics}}, \href{http://dx.doi.org/10.1007/JHEP02(2016)091}{\emph{JHEP}
  {\bf 02} (2016) 091}, [\href{http://arxiv.org/abs/1512.00019}{{\tt
  1512.00019}}].

\bibitem{Komargodski2026Bouncing}
Z.~Komargodski, \emph{{Fate of bouncing singularities at finite $G_N$}},
  \href{http://arxiv.org/abs/2609.31309}{{\tt 2609.31309}}.

\bibitem{Batra:2026gbq}
G.~Batra, A.~Levine and S.~H. Shenker, \emph{{A backreacted bouncing geodesic
  in 2D}},  \href{http://arxiv.org/abs/2609.32071}{{\tt 2609.32071}}.

\bibitem{Cvitanovic2020Chaos}
P.~Cvitanovi{\'c}, R.~Artuso, R.~Mainieri, G.~Tanner and G.~Vattay,
  \emph{{\href{https://chaosbook.org/version16/chapters/ChaosBook.pdf}{Chaos:
  Classical and Quantum}}}.
\newblock Niels Bohr Institute, Copenhagen, 16.4.8~ed., 2020.

\bibitem{Gaspard1998}
P.~Gaspard, \emph{{\href{https://doi.org/10.1017/CBO9780511628856}{Chaos,
  Scattering and Statistical Mechanics}}}, vol.~9 of \emph{Cambridge Nonlinear
  Science Series}.
\newblock Cambridge University Press, 1998.

\bibitem{BowenRuelle1975}
R.~Bowen and D.~Ruelle, \emph{{The ergodic theory of Axiom A flows}},
  \href{http://dx.doi.org/10.1007/BF01389848}{\emph{Invent. Math.} {\bf 29}
  (1975) 181--202}.

\bibitem{Hille1976}
E.~Hille, \emph{{Ordinary Differential Equations in the Complex Domain}}.
\newblock Pure and Applied Mathematics. Wiley-Interscience, New York, 1976.

\bibitem{FrischMorf1981}
U.~Frisch and R.~Morf, \emph{{Intermittency in nonlinear dynamics and
  singularities at complex times}},
  \href{http://dx.doi.org/10.1103/PhysRevA.23.2673}{\emph{Phys. Rev. A} {\bf
  23} (1981) 2673--2705}.

\bibitem{Hone2009Painleve}
A.~N.~W. Hone,
  \emph{{\href{https://doi.org/10.1007/978-3-540-88111-7_8}{Painlev{\'e} tests,
  singularity structure and integrability}}},  in \emph{{Integrability}} (A.~V.
  Mikhailov, ed.), vol.~767 of \emph{Lecture Notes in Physics}, pp.~245--277.
\newblock Springer, 2009.
\newblock \href{http://arxiv.org/abs/nlin/0502017}{{\tt nlin/0502017}}.

\bibitem{Lorenz1963}
E.~N. Lorenz, \emph{{Deterministic nonperiodic flow}},
  \href{http://dx.doi.org/10.1175/1520-0469(1963)020<0130:DNF>2.0.CO;2}{\emph{J.
  Atmos. Sci.} {\bf 20} (1963) 130--141}.

\bibitem{TaborWeiss1981}
M.~Tabor and J.~Weiss, \emph{{Analytic structure of the Lorenz system}},
  \href{http://dx.doi.org/10.1103/PhysRevA.24.2157}{\emph{Phys. Rev. A} {\bf
  24} (1981) 2157--2167}.

\bibitem{LevineTabor1988}
G.~Levine and M.~Tabor, \emph{{Integrating the nonintegrable: analytic
  structure of the Lorenz system revisited}},
  \href{http://dx.doi.org/10.1016/S0167-2789(98)90018-5}{\emph{Physica D} {\bf
  33} (1988) 189--210}.

\bibitem{Landau1959}
L.~D. Landau, \emph{{On analytic properties of vertex parts in quantum field
  theory}}, \href{http://dx.doi.org/10.1016/0029-5582(59)90154-3}{\emph{Nucl.
  Phys.} {\bf 13} (1959) 181--192}.

\bibitem{AdlerVanMoerbeke1980}
M.~Adler and P.~van Moerbeke, \emph{{Linearization of Hamiltonian systems,
  Jacobi varieties and representation theory}},
  \href{http://dx.doi.org/10.1016/0001-8708(80)90008-0}{\emph{Adv. Math.} {\bf
  38} (1980) 318--379}.

\bibitem{Toda1967}
M.~Toda, \emph{{Vibration of a chain with nonlinear interaction}},
  \href{http://dx.doi.org/10.1143/jpsj.22.431}{\emph{J. Phys. Soc. Jap.} {\bf
  22} (1967) 431--436}.

\bibitem{KacVanMoerbeke1975}
M.~Kac and P.~van Moerbeke, \emph{{A complete solution of the periodic Toda
  problem}}, \href{http://dx.doi.org/10.1073/pnas.72.8.2879}{\emph{Proc. Nat.
  Acad. Sci. U.S.A.} {\bf 72} (1975) 2879--2880}.

\bibitem{Pollicott1985}
M.~Pollicott, \emph{{On the rate of mixing of Axiom A flows}},
  \href{http://dx.doi.org/10.1007/BF01388579}{\emph{Invent. Math.} {\bf 81}
  (1985) 413--426}.

\bibitem{Ruelle1986}
D.~Ruelle, \emph{{Resonances of chaotic dynamical systems}},
  \href{http://dx.doi.org/10.1103/PhysRevLett.56.405}{\emph{Phys. Rev. Lett.}
  {\bf 56} (1986) 405--407}.

\bibitem{HenonHeiles1964}
M.~H{\'e}non and C.~Heiles, \emph{{The applicability of the third integral of
  motion: some numerical experiments}},
  \href{http://dx.doi.org/10.1086/109234}{\emph{Astron. J.} {\bf 69} (1964)
  73--79}.

\bibitem{MacKayMeissPercival1984}
R.~S. MacKay, J.~D. Meiss and I.~C. Percival, \emph{{Transport in Hamiltonian
  systems}},
  \href{http://dx.doi.org/10.1016/0167-2789(84)90270-7}{\emph{Physica D} {\bf
  13} (1984) 55--81}.

\bibitem{MeissOtt1985}
J.~D. Meiss and E.~Ott, \emph{{Markov-tree model of intrinsic transport in
  Hamiltonian systems}},
  \href{http://dx.doi.org/10.1103/PhysRevLett.55.2741}{\emph{Phys. Rev. Lett.}
  {\bf 55} (1985) 2741--2744}.

\bibitem{tHooft1974}
G.~'t~Hooft, \emph{{A planar diagram theory for strong interactions}},
  \href{http://dx.doi.org/10.1016/0550-3213(74)90154-0}{\emph{Nucl. Phys. B}
  {\bf 72} (1974) 461}.

\bibitem{bipz}
E.~Br{\'e}zin, C.~Itzykson, G.~Parisi and J.~B. Zuber, \emph{{Planar
  diagrams}}, \href{http://dx.doi.org/10.1007/BF01614153}{\emph{Commun. Math.
  Phys.} {\bf 59} (1978) 35}.

\bibitem{BruschiCalogero2000Matrix}
M.~Bruschi and F.~Calogero, \emph{{On the integrability of certain matrix
  evolution equations}},
  \href{http://dx.doi.org/10.1016/S0375-9601(00)00482-5}{\emph{Phys. Lett. A}
  {\bf 273} (2000) 167--172}.

\bibitem{BruschiCalogero2004Quartic}
M.~Bruschi and F.~Calogero, \emph{{Integrable systems of quartic oscillators.
  II}}, \href{http://dx.doi.org/10.1016/j.physleta.2004.05.039}{\emph{Phys.
  Lett. A} {\bf 327} (2004) 320--326},
  [\href{http://arxiv.org/abs/nlin/0403011}{{\tt nlin/0403011}}].

\bibitem{ParkerEtAl2019}
D.~E. Parker, X.~Cao, A.~Avdoshkin, T.~Scaffidi and E.~Altman, \emph{{A
  universal operator growth hypothesis}},
  \href{http://dx.doi.org/10.1103/PhysRevX.9.041017}{\emph{Phys. Rev. X} {\bf
  9} (2019) 041017}, [\href{http://arxiv.org/abs/1812.08657}{{\tt
  1812.08657}}].

\bibitem{Lin:2023trc}
H.~W. Lin and D.~Stanford, \emph{{A symmetry algebra in double-scaled SYK}},
  \href{http://dx.doi.org/10.21468/SciPostPhys.15.6.234}{\emph{SciPost Phys.}
  {\bf 15} (2023) 234}, [\href{http://arxiv.org/abs/2307.15725}{{\tt
  2307.15725}}].

\bibitem{AvdoshkinDymarsky2020}
A.~Avdoshkin and A.~Dymarsky, \emph{{Euclidean operator growth and quantum
  chaos}},
  \href{http://dx.doi.org/10.1103/PhysRevResearch.2.043234}{\emph{Phys. Rev.
  Res.} {\bf 2} (2020) 043234}, [\href{http://arxiv.org/abs/1911.09672}{{\tt
  1911.09672}}].

\bibitem{GuKitaevZhang2022}
Y.~Gu, A.~Kitaev and P.~Zhang, \emph{{A two-way approach to out-of-time-order
  correlators}}, \href{http://dx.doi.org/10.1007/JHEP03(2022)133}{\emph{JHEP}
  {\bf 03} (2022) 133}, [\href{http://arxiv.org/abs/2111.12007}{{\tt
  2111.12007}}].

\bibitem{Witten1996}
E.~Witten, \emph{{Bound states of strings and p-branes}},
  \href{http://dx.doi.org/10.1016/0550-3213(95)00610-9}{\emph{Nucl. Phys. B}
  {\bf 460} (1996) 335--350}, [\href{http://arxiv.org/abs/hep-th/9510135}{{\tt
  hep-th/9510135}}].

\bibitem{Bachas1996}
C.~Bachas, \emph{{D-brane dynamics}},
  \href{http://dx.doi.org/10.1016/0370-2693(96)00238-9}{\emph{Phys. Lett. B}
  {\bf 374} (1996) 37--42}, [\href{http://arxiv.org/abs/hep-th/9511043}{{\tt
  hep-th/9511043}}].

\bibitem{Lin:2013jra}
Y.-H. Lin, S.-H. Shao, Y.~Wang and X.~Yin, \emph{{A low temperature expansion
  for matrix quantum mechanics}},
  \href{http://dx.doi.org/10.1007/JHEP05(2015)136}{\emph{JHEP} {\bf 05} (2015)
  136}, [\href{http://arxiv.org/abs/1304.1593}{{\tt 1304.1593}}].

\bibitem{TracyWidom1994}
C.~A. Tracy and H.~Widom, \emph{{Level spacing distributions and the Airy
  kernel}}, \href{http://dx.doi.org/10.1007/BF02100489}{\emph{Commun. Math.
  Phys.} {\bf 159} (1994) 151--174},
  [\href{http://arxiv.org/abs/hep-th/9211141}{{\tt hep-th/9211141}}].

\bibitem{DeiftGioev2007}
P.~Deift and D.~Gioev, \emph{{Universality at the edge of the spectrum for
  unitary, orthogonal, and symplectic ensembles of random matrices}},
  \href{http://dx.doi.org/10.1002/cpa.20164}{\emph{Commun. Pure Appl. Math.}
  {\bf 60} (2007) 867--910}, [\href{http://arxiv.org/abs/math-ph/0507023}{{\tt
  math-ph/0507023}}].

\bibitem{BerensteinMaldacenaNastase2002}
D.~Berenstein, J.~Maldacena and H.~Nastase, \emph{{Strings in flat space and pp
  waves from $\mathcal{N}=4$ super Yang Mills}},
  \href{http://dx.doi.org/10.1088/1126-6708/2002/04/013}{\emph{JHEP} {\bf 04}
  (2002) 013}, [\href{http://arxiv.org/abs/hep-th/0202021}{{\tt
  hep-th/0202021}}].

\bibitem{KristjansenEtAl2002}
C.~Kristjansen, J.~Plefka, G.~W. Semenoff and M.~Staudacher, \emph{{A new
  double-scaling limit of $\mathcal{N}=4$ super Yang-Mills theory and PP-wave
  strings}}, \href{http://dx.doi.org/10.1016/S0550-3213(02)00749-6}{\emph{Nucl.
  Phys. B} {\bf 643} (2002) 3--30},
  [\href{http://arxiv.org/abs/hep-th/0205033}{{\tt hep-th/0205033}}].

\bibitem{ConstableEtAl2002}
N.~R. Constable, D.~Z. Freedman, M.~Headrick, S.~Minwalla, L.~Motl,
  A.~Postnikov et~al., \emph{{PP-wave string interactions from perturbative
  Yang-Mills theory}},
  \href{http://dx.doi.org/10.1088/1126-6708/2002/07/017}{\emph{JHEP} {\bf 07}
  (2002) 017}, [\href{http://arxiv.org/abs/hep-th/0205089}{{\tt
  hep-th/0205089}}].

\bibitem{Berenstein2013Collective}
D.~Berenstein, \emph{{Giant gravitons: a collective coordinate approach}},
  \href{http://dx.doi.org/10.1103/PhysRevD.87.126009}{\emph{Phys. Rev. D} {\bf
  87} (2013) 126009}, [\href{http://arxiv.org/abs/1301.3519}{{\tt 1301.3519}}].

\bibitem{Ferrari2014GaugeBranes}
F.~Ferrari, \emph{{Gauge theories, D-branes and holography}},
  \href{http://dx.doi.org/10.1016/j.nuclphysb.2014.01.007}{\emph{Nucl. Phys. B}
  {\bf 880} (2014) 247--289}, [\href{http://arxiv.org/abs/1310.6788}{{\tt
  1310.6788}}].

\bibitem{Ferrari2014MatrixProbes}
F.~Ferrari, \emph{{D-brane probes in the matrix model}},
  \href{http://dx.doi.org/10.1016/j.nuclphysb.2013.12.017}{\emph{Nucl. Phys. B}
  {\bf 880} (2014) 290--320}, [\href{http://arxiv.org/abs/1311.4520}{{\tt
  1311.4520}}].

\bibitem{Marchesini:1979yq}
G.~Marchesini and E.~Onofri, \emph{{Planar limit for SU($N$) symmetric quantum
  dynamical systems}}, \href{http://dx.doi.org/10.1063/1.524532}{\emph{J. Math.
  Phys.} {\bf 21} (1980) 1103}.

\bibitem{Karczmarek2009Adjoint}
J.~L. Karczmarek, \emph{{Scattering in the adjoint sector of the c = 1 Matrix
  Model}}, \href{http://dx.doi.org/10.1088/1126-6708/2009/02/011}{\emph{JHEP}
  {\bf 02} (2009) 011}, [\href{http://arxiv.org/abs/0809.3543}{{\tt
  0809.3543}}].

\bibitem{BenedettiGurau2018}
D.~Benedetti and R.~Gurau, \emph{{2PI effective action for the SYK model and
  tensor field theories}},
  \href{http://dx.doi.org/10.1007/JHEP05(2018)156}{\emph{JHEP} {\bf 05} (2018)
  156}, [\href{http://arxiv.org/abs/1802.05500}{{\tt 1802.05500}}].

\bibitem{Berkooz:2018jqr}
M.~Berkooz, M.~Isachenkov, V.~Narovlansky and G.~Torrents, \emph{{Towards a
  full solution of the large N double-scaled SYK model}},
  \href{http://dx.doi.org/10.1007/JHEP03(2019)079}{\emph{JHEP} {\bf 03} (2019)
  079}, [\href{http://arxiv.org/abs/1811.02584}{{\tt 1811.02584}}].

\bibitem{Cotler:2016fpe}
J.~S. Cotler, G.~Gur-Ari, M.~Hanada, J.~Polchinski, P.~Saad, S.~H. Shenker
  et~al., \emph{{Black Holes and Random Matrices}},
  \href{http://dx.doi.org/10.1007/JHEP05(2017)118}{\emph{JHEP} {\bf 05} (2017)
  118}, [\href{http://arxiv.org/abs/1611.04650}{{\tt 1611.04650}}].

\bibitem{Savvidy1984}
G.~K. Savvidy, \emph{{Classical and quantum mechanics of non-Abelian gauge
  fields}}, \href{http://dx.doi.org/10.1016/0550-3213(84)90298-0}{\emph{Nucl.
  Phys. B} {\bf 246} (1984) 302--334}.

\bibitem{MandalMahatoMorita2010}
G.~Mandal, M.~Mahato and T.~Morita, \emph{{Phases of one dimensional large $N$
  gauge theory in a $1/D$ expansion}},
  \href{http://dx.doi.org/10.1007/JHEP02(2010)034}{\emph{JHEP} {\bf 02} (2010)
  034}, [\href{http://arxiv.org/abs/0910.4526}{{\tt 0910.4526}}].

\bibitem{BanksEtAl1997}
T.~Banks, W.~Fischler, S.~H. Shenker and L.~Susskind, \emph{{M theory as a
  matrix model: a conjecture}},
  \href{http://dx.doi.org/10.1103/PhysRevD.55.5112}{\emph{Phys. Rev. D} {\bf
  55} (1997) 5112--5128}, [\href{http://arxiv.org/abs/hep-th/9610043}{{\tt
  hep-th/9610043}}].

\bibitem{MaldacenaMilekhin2018}
J.~Maldacena and A.~Milekhin, \emph{{To gauge or not to gauge?}},
  \href{http://dx.doi.org/10.1007/JHEP04(2018)084}{\emph{JHEP} {\bf 04} (2018)
  084}, [\href{http://arxiv.org/abs/1802.00428}{{\tt 1802.00428}}].

\bibitem{ItzhakiEtAl1998}
N.~Itzhaki, J.~M. Maldacena, J.~Sonnenschein and S.~Yankielowicz,
  \emph{{Supergravity and the large $N$ limit of theories with sixteen
  supercharges}},
  \href{http://dx.doi.org/10.1103/PhysRevD.58.046004}{\emph{Phys. Rev. D} {\bf
  58} (1998) 046004}, [\href{http://arxiv.org/abs/hep-th/9802042}{{\tt
  hep-th/9802042}}].

\bibitem{BiggsMaldacena2023}
A.~Biggs and J.~Maldacena, \emph{{Scaling similarities and quasinormal modes of
  D0 black hole solutions}},
  \href{http://dx.doi.org/10.1007/JHEP11(2023)155}{\emph{JHEP} {\bf 11} (2023)
  155}, [\href{http://arxiv.org/abs/2303.09974}{{\tt 2303.09974}}].

\bibitem{Maldacena2023BlackHole}
J.~Maldacena, \emph{{A simple quantum system that describes a black hole}},
  \href{http://arxiv.org/abs/2303.11534}{{\tt 2303.11534}}.

\bibitem{Nahm1980}
W.~Nahm, \emph{{A simple formalism for the BPS monopole}},
  \href{http://dx.doi.org/10.1016/0370-2693(80)90961-2}{\emph{Phys. Lett. B}
  {\bf 90} (1980) 413--414}.

\bibitem{GaiottoWitten2009Boundary}
D.~Gaiotto and E.~Witten, \emph{{Supersymmetric boundary conditions in
  $\mathcal{N}=4$ super Yang-Mills theory}},
  \href{http://dx.doi.org/10.1007/s10955-009-9687-3}{\emph{J. Stat. Phys.} {\bf
  135} (2009) 789--855}, [\href{http://arxiv.org/abs/0804.2902}{{\tt
  0804.2902}}].

\bibitem{BerensteinHanadaHartnoll2009}
D.~E. Berenstein, M.~Hanada and S.~A. Hartnoll, \emph{{Multi-matrix models and
  emergent geometry}},
  \href{http://dx.doi.org/10.1088/1126-6708/2009/02/010}{\emph{JHEP} {\bf 02}
  (2009) 010}, [\href{http://arxiv.org/abs/0805.4658}{{\tt 0805.4658}}].

\bibitem{KazakovKostovNekrasov1999}
V.~A. Kazakov, I.~K. Kostov and N.~Nekrasov, \emph{{D-particles, matrix
  integrals and KP hierarchy}},
  \href{http://dx.doi.org/10.1016/S0550-3213(99)00393-4}{\emph{Nucl. Phys. B}
  {\bf 557} (1999) 413--442}, [\href{http://arxiv.org/abs/hep-th/9810035}{{\tt
  hep-th/9810035}}].

\bibitem{ChangEtAl1983}
Y.~F. Chang, J.~M. Greene, M.~Tabor and J.~Weiss, \emph{{The analytic structure
  of dynamical systems and self-similar natural boundaries}},
  \href{http://dx.doi.org/10.1016/0167-2789(83)90317-2}{\emph{Physica D} {\bf
  8} (1983) 183--207}.

\bibitem{FournierLevineTabor1988}
J.~D. Fournier, G.~Levine and M.~Tabor, \emph{{Singularity clustering in the
  Duffing oscillator}},
  \href{http://dx.doi.org/10.1088/0305-4470/21/1/014}{\emph{J. Phys. A} {\bf
  21} (1988) 33--54}.

\bibitem{MasonWoodhouse1996}
L.~J. Mason and N.~M.~J. Woodhouse,
  \emph{{\href{https://doi.org/10.1093/oso/9780198534983.001.0001}{Integrability,
  Self-Duality, and Twistor Theory}}}, vol.~15 of \emph{London Mathematical
  Society Monographs, New Series}.
\newblock Oxford University Press, Oxford, 1996.

\bibitem{Hitchin1983Monopoles}
N.~J. Hitchin, \emph{{On the construction of monopoles}},
  \href{http://dx.doi.org/10.1007/BF01211826}{\emph{Commun. Math. Phys.} {\bf
  89} (1983) 145--190}.

\bibitem{SharmaSkinner2026Chiral}
A.~Sharma and D.~Skinner, \emph{{Chiral holography}},
  \href{http://dx.doi.org/10.1007/JHEP08(2026)202}{\emph{JHEP} {\bf 08} (2026)
  202}, [\href{http://arxiv.org/abs/2512.04152}{{\tt 2512.04152}}].

\bibitem{Fawzi:2023fpg}
H.~Fawzi, O.~Fawzi and S.~O. Scalet, \emph{{Certified algorithms for
  equilibrium states of local quantum Hamiltonians}},
  \href{http://dx.doi.org/10.1038/s41467-024-51592-3}{\emph{Nature Commun.}
  {\bf 15} (2024) 7394}, [\href{http://arxiv.org/abs/2311.18706}{{\tt
  2311.18706}}].

\bibitem{ChoGabaiSandorYin2024}
M.~Cho, B.~Gabai, J.~Sandor and X.~Yin, \emph{{Thermal bootstrap of matrix
  quantum mechanics}},
  \href{http://dx.doi.org/10.1007/JHEP04(2025)186}{\emph{JHEP} {\bf 04} (2025)
  186}, [\href{http://arxiv.org/abs/2410.04262}{{\tt 2410.04262}}].

\bibitem{ChoEtAl2025Euclidean}
M.~Cho, B.~Gabai, H.~W. Lin, J.~Yeh and Z.~Zheng, \emph{{Bootstrapping
  Euclidean two-point correlators}},
  \href{http://dx.doi.org/10.1007/JHEP08(2026)147}{\emph{JHEP} {\bf 08} (2026)
  147}, [\href{http://arxiv.org/abs/2511.08560}{{\tt 2511.08560}}].

\bibitem{HanHartnollKruthoff2020}
X.~Han, S.~A. Hartnoll and J.~Kruthoff, \emph{{Bootstrapping matrix quantum
  mechanics}},
  \href{http://dx.doi.org/10.1103/PhysRevLett.125.041601}{\emph{Phys. Rev.
  Lett.} {\bf 125} (2020) 041601}, [\href{http://arxiv.org/abs/2004.10212}{{\tt
  2004.10212}}].

\bibitem{Lin2023D0}
H.~W. Lin, \emph{{Bootstrap bounds on D0-brane quantum mechanics}},
  \href{http://dx.doi.org/10.1007/JHEP06(2023)038}{\emph{JHEP} {\bf 06} (2023)
  038}, [\href{http://arxiv.org/abs/2302.04416}{{\tt 2302.04416}}].

\bibitem{LinZheng2025Ground}
H.~W. Lin and Z.~Zheng, \emph{{Bootstrapping ground state correlators in matrix
  theory. Part I}},
  \href{http://dx.doi.org/10.1007/JHEP01(2025)190}{\emph{JHEP} {\bf 01} (2025)
  190}, [\href{http://arxiv.org/abs/2410.14647}{{\tt 2410.14647}}].

\bibitem{fouriermethod}
H.~W. Lin, J.~Yeh and Z.~Zheng. In preparation.

\bibitem{BuricGusevParnachev2026Resolution}
I.~Buri{\'c}, I.~Gusev and A.~Parnachev, \emph{{Stringy resolution of bouncing
  singularities}},  \href{http://arxiv.org/abs/2610.12311}{{\tt 2610.12311}}.

\bibitem{Itzhaki:2018rld}
N.~Itzhaki and L.~Liram, \emph{{A stringy glimpse into the black hole
  horizon}}, \href{http://dx.doi.org/10.1007/JHEP04(2018)018}{\emph{JHEP} {\bf
  04} (2018) 018}, [\href{http://arxiv.org/abs/1801.04939}{{\tt 1801.04939}}].

\bibitem{Iancu1998Classical}
E.~Iancu, \emph{{Effective theory for real-time dynamics in hot gauge
  theories}},
  \href{http://dx.doi.org/10.1016/S0370-2693(98)00772-2}{\emph{Phys. Lett. B}
  {\bf 435} (1998) 152--158}, [\href{http://arxiv.org/abs/hep-ph/9710543}{{\tt
  hep-ph/9710543}}].

\bibitem{Bodeker1998Soft}
D.~B{\"o}deker, \emph{{On the effective dynamics of soft non-Abelian gauge
  fields at finite temperature}},
  \href{http://dx.doi.org/10.1016/S0370-2693(98)00279-2}{\emph{Phys. Lett. B}
  {\bf 426} (1998) 351--360}, [\href{http://arxiv.org/abs/hep-ph/9801430}{{\tt
  hep-ph/9801430}}].

\bibitem{klebanov2026reggetrajectoriesadjointsector}
I.~R. Klebanov, H.~W. Lin and P.~Meshcheriakov, \emph{{Regge trajectories from
  the adjoint sector of matrix quantum mechanics}},
  \href{http://arxiv.org/abs/2603.04522}{{\tt 2603.04522}}.

\bibitem{Maldacena:2005hi}
J.~M. Maldacena, \emph{{Long strings in two dimensional string theory and
  non-singlets in the matrix model}},
  \href{http://dx.doi.org/10.1088/1126-6708/2005/09/078}{\emph{JHEP} {\bf 09}
  (2005) 078}, [\href{http://arxiv.org/abs/hep-th/0503112}{{\tt
  hep-th/0503112}}].

\bibitem{Gross:1990md}
D.~J. Gross and I.~R. Klebanov, \emph{{Vortices and the nonsinglet sector of
  the c = 1 matrix model}},
  \href{http://dx.doi.org/10.1016/0550-3213(91)90363-3}{\emph{Nucl. Phys. B}
  {\bf 354} (1991) 459--474}.

\bibitem{Wigner1932}
E.~P. Wigner, \emph{{On the quantum correction for thermodynamic equilibrium}},
  \href{http://dx.doi.org/10.1103/PhysRev.40.749}{\emph{Phys. Rev.} {\bf 40}
  (1932) 749--760}.

\bibitem{Kirkwood1933}
J.~G. Kirkwood, \emph{{Quantum statistics of almost classical assemblies}},
  \href{http://dx.doi.org/10.1103/PhysRev.44.31}{\emph{Phys. Rev.} {\bf 44}
  (1933) 31--37}.

\end{thebibliography}\endgroup

\end{document}